\documentclass[review,12pt]{elsarticle} 

\usepackage[colorlinks, citecolor =blue]{hyperref}
\hypersetup{linkcolor=blue}

\usepackage[english]{babel}
\usepackage{amsmath}
\usepackage{graphicx} 
\usepackage{color}
\usepackage{multirow}
\usepackage{hyperref}
\usepackage{booktabs}
\usepackage{morefloats}
\usepackage{textcomp}
\usepackage{gensymb}
\usepackage{epstopdf}
\usepackage{geometry}
\usepackage{caption}
\usepackage{subfig} 
\usepackage{amsmath}
\usepackage{textcomp}
\usepackage{amsfonts}
\usepackage{mathtools}
\usepackage{booktabs}
\usepackage{array}
\usepackage[table,xcdraw]{xcolor}
\usepackage{colortbl}
\usepackage{float}
\usepackage{makecell}
\usepackage{color}
\usepackage{ragged2e}
\usepackage{enumitem} 
\usepackage{longtable}
\usepackage{adjustbox}

\usepackage{cleveref} 
\usepackage{subfig}
\biboptions{sort&compress}

\crefname{figure}{}{}
\crefrangelabelformat{figure}{#3#1#4--#5\crefstripprefix{#1}{#2}#6}
\crefmultiformat{figure}{~#2#1\xdef\mycreffirstarg{#1}#3}%
{ and~#2{\crefstripprefix{\mycreffirstarg}{#1}}#3}{, #2#1#3}{ and~#2#1#3}
 
\justifying
\usepackage{physics}
\usepackage[left]{lineno}
\usepackage[none]{hyphenat}
\usepackage[nodisplayskipstretch]{setspace}

\makeatletter
\newcommand\tabfill[1]{%
    \dimen@\linewidth
    \advance\dimen@\@totalleftmargin
    \advance\dimen@-\dimen@curtab
    \parbox[t]\dimen@{#1\ifhmode\strut\fi}
}\makeatother
\begin{document}
\linenumbers
\begin{frontmatter}
\title{Modelling of nucleate pool boiling on coated substrates using machine learning and empirical approaches}

\author{Vijay Kuberan\corref{CB}}
\ead{vijaykuberan98@gmail.com}
\author{Sateesh Gedupudi}
\cortext[CB]{Corresponding author}
 
\address{Heat Transfer and Thermal Power Laboratory, Department of Mechanical Engineering, Indian Institute of Technology Madras, Chennai-600036, India}

\vspace{-2cm}
\begin{abstract}
Surface modification results in substantial improvement in pool boiling heat transfer. Thin film-coated and porous-coated substrates, through different materials and techniques, significantly boost heat transfer through increased nucleation due to the presence of micro-cavities on the surface. The existing models and empirical correlations for boiling on these coated surfaces are constrained by specific operating conditions and parameter ranges and are hence limited by their prediction accuracy. This study focuses on developing an accurate and reliable Machine Learning (ML) model by effectively capturing the actual relationship between the influencing variables. Various ML algorithms have been evaluated on the thin film-coated and porous-coated datasets amassed from different studies. The CatBoost model demonstrated the best prediction accuracy after cross-validation and hyperparameter tuning. For the optimized CatBoost model, SHAP analysis has been carried out to identify the prominent influencing parameters and interpret the impact of parameter variation on the target variable. This model interpretation clearly justifies the decisions behind the model predictions, making it a robust model for the prediction of nucleate boiling Heat Transfer Coefficient (HTC) on coated surfaces. Finally, the existing empirical correlations have been assessed, and new correlations have been proposed to predict the HTC on these surfaces with the inclusion of influential parameters identified through SHAP interpretation.
\end{abstract}


\end{frontmatter}

\clearpage    

\section*{Nomenclature}
\begin{tabbing}
    xxxxxxxxxxxxxxxxxx \= \kill

    $C_{pl}$ \> Specific heat of the liquid corresponding to $T_{film}$ (kJ/kgK) \\
    $C_{pv}$ \> Specific heat of the vapor corresponding to $T_{sat}$ (kJ/kgK) \\
    $C_{s}$\> Fluid surface coefficient in the Pioro correlation \\
    $C_{sf}$\> Fluid surface coefficient in the Rohsenow correlation \\
    $h$ \> Heat transfer coefficient (kW/m$^2$K) \\
    $h_l$ \> Specific enthalpy of liquid (kJ/kg) \\
    $h_{lv}$ \> Latent heat of vapourisation (kJ/kg) \\
    $h_v$ \> Specific enthalpy of vapor (kJ/kg) \\
    $k_{co}$ \> Thermal conductivity of the coating (W/mK) \\
    $k_{eff}$ \> Effective thermal conductivity of the porous coating (W/mK) \\
    $k_{w}$ \> Thermal conductivity of the substrate (W/mK) \\
    $L_c$ \> Boiling length scale ($\mu$m) \\
    $m$ \> Experimental constant in Rohsenow correlation\\
    $M_{w}$ \> Molecular mass of water (kg/kmol) \\
    $n$ \> Constant in Pioro correlation\\
    $P_{cr}$ \> Critical pressure (bar) \\
    $P_{film}$ \> Saturation pressure corresponding to $T_{film}$ (bar) \\
    $P_{op}$ \> Operating pressure (bar) \\
    $P_{red}$, $P_{r}$  \> Reduced pressure (bar) \\
    $R^2$ \> Coefficient of determination \\
    $r_{cav}$ \> Cavity radius required for nucleation ($\mu$m) \\
    $R_{co}$ \> Thermal resistance of the coating (m$^2$K/W) \\
    $R_q$ \> Surface roughness ($\mu$m) \\
    $R_{sll}$ \> Thermal resistance of superheated liquid layer (m$^2$K/W) \\
    $t$ \> Coating thickness ($\mu$m) \\
    $T_{film}$ \> Film temperature (K) \\
    $T_w$ \> Temperature of the heated surface (K) \\
    $\triangle T$ \> Wall superheat (K) \\
    \\
    Greek symbols \\
    $\alpha$ \> Thermal diffusivity (m$^2$/s) \\
    $\kappa$ \> Thermal conductivity (W/mK) \\
    $\mu$ \> Dynamic viscosity (kg/m s) \\
    $\nu$ \> Kinematic viscosity (m$^2$/s) \\
    $\rho$ \> Density (kg/m$^3$) \\
    $\sigma$ \> Surface tension (N/m) \\
    $\theta$ \> Contact angle (degrees) \\
    $\varepsilon$ \> Volumetric porosity \\
    \\
    Non-dimensional Quantities \\
    $Ja$ \> Jakob number \\
    $Nu$ \> Nusselt number \\
    $Pr$ \> Prandtl number \\
    \\
    Abbreviations \\
    HTC \> Heat Transfer Coefficient \\
    MAE \> Mean Absolute Error \\
    MAPE \> Mean Absolute Percentage Error \\
    ND \> Non-Dimensional \\
    RMSE \> Root Mean Squared Error \\
    SD \> Standard Deviation \\
    \\
    Subscripts \\
    co \> coating \\
    cr \> critical \\
    eff \> effective \\
    l \> liquid \\
    sat \> saturated \\
    sll \> superheated liquid layer \\
    v \> vapor \\
\end{tabbing}

\clearpage


\section{Introduction}\label{sec:intro}
Global energy demand has amplified remarkably in recent years due to the rising population and growth of different industries. Effective cooling through high-heat flux dissipation and efficient thermal management of devices becomes crucial in a wide range of applications, ranging from nuclear reactors to electronic systems. Nucleate boiling heat transfer, a two-phase process, is capable of dissipating a large amount of thermal energy with a small temperature differential. Several techniques have been employed to augment the boiling process, which can be broadly classified into active and passive techniques \cite{Bergles1997}. Active enhancement \cite{Bergles1997} involves the agitation of liquids using external sources such as electric fields, mechanical vibrations, and ultrasonic sources, while passive techniques involve changing the properties of working fluid and modifying the heating surface \cite{Bergles1997}. Due to the limitation of increased energy consumption in active methods, passive methods have been widely employed to increase the Critical Heat Flux (CHF), reduce the Onset of Nucleate Boiling (ONB), and improve heat transfer efficiency. Increased nucleation, improved bubble dynamics, and optimum wettability characteristics attained by surface modification facilitate efficient heat removal from the surface \cite{Moghadasi2021}. These include modifying the surface by roughening, applying coatings, and incorporating extended or structured surfaces \cite{Li2020}.

Micro cavities introduced on the surface by employing thin film coatings and porous coatings alter the surface characteristics and act as re-entrant cavities. It enhances the nucleation site density, improves the capillary pumping effects, establishes optimum wettability, and forms stable vapor traps on the surface, promoting increased heat transfer performance \cite{Li2007,Surtaev2018}. Owing to these advantages, several studies have implemented different coatings on the surface, including  SiO$_2$, TiO$_2$, ZrO$_2$, CuO, different nano-composites, various nanostructures, CNT (Carbon Nanotubes), etc, deposited by diverse techniques such as Electrochemical Deposition, Physical Vapor Deposition, Sintering, Plasma Spraying, Electron Beam Physical Vapor Deposition (EBPVD), and much more. All these surface modifications – thin film-coated \cite{Das2016a, Das2016b, Das2014a, Das2014b, Kim2018a, Kim2018b, Pinni2022, Majumder2022c, Majumder2023, Das2017a, Das2017b, Jaikumar2017, Thangavelu2023, Nithyanandam2020, Gajghate2020, Gajghate2021a, Gajghate2021b, Patel2019, Kumar2024, Deb2020, Laskar2022, Ray2018} and porous-coated \cite{Katarkar2021a, Katarkar2021b, Jun2015, Ahmadi2022, Gupta2020, Gupta2023a, Gupta2023b, Gupta2019a, Gupta2019b, Hsieh1997, Dewangan2016, Ashok2019, Godinez2019, Gupta2018, Majumder2022a, Majumder2022b, Gheitaghy2017, Joseph2019, Pialago2013, Wang2018, Jun2016} - have been found to increase heat transfer efficiency compared to plain surfaces.

Different empirical correlations have been proposed to predict the heat transfer performance. These include correlations by Rohsenow \cite{Rohsenow1952}, Pioro \cite{Pioro1999}, Forster-Zuber \cite{Forster1955}, Borishansky \cite{borishanskii1969correlation}, Kichigin \& Tobilevich \cite{kichigin1955generalization}, Labuntsov \cite{labuntsov1973heat}, Kruzhilin \cite{kruzhilin1947free}, Cooper \cite{COOPER1984157}, Kutateladze \cite{kutateladze1966concise, kutateladze1990heat}, Cornwell-Housten \cite{Cornwell1994}, and Ribatski \& Jabardo \cite{Ribatski2003}. Even though there are various correlations and different boiling models to predict boiling performance, they are constrained by the working fluid, parameter ranges, surface characteristics, and specific operating conditions.

Machine Learning (ML) models are highly efficient and reliable in capturing the underlying mechanism and discovering complex non-linear patterns in the data \cite{Sarker2021}. ML is extensively adopted in various sectors to model systems accurately where the underlying phenomenon is not completely understood. In the field of thermal and fluid engineering, various studies have employed these machine-learning techniques to model thermal systems. Swain and Das \cite{Swain2014} adopted the Artificial Neural Networks (ANN) and Adaptive Neuro-Fuzzy Inference System (ANFIS) to model the flow boiling HTC over tube bundles and showed better predictive capability than conventional correlations. A similar study was done by Scalabrin et al. \cite{Scalabrin2006} to model flow boiling heat transfer inside horizontal tubes using ANN. Furthermore, in a study done by Chang et al. \cite{Chang2018}, heat transfer prediction in supercritical water using ANN exhibited a lower mean error percentage. Another study by Barroso et al. predicted the frictional pressure drop \cite{Barroso2019} and two-phase convective heat transfer coefficient \cite{Barroso2018} for non-azeotropic mixtures with a lower mean relative error. Khosravi et al. \cite{Khosravi2018} also predicted the frictional pressure drop in the two-phase flow of R407C by employing ANN and Support Vector Regressor (SVR). In a study by Alic et al. \cite{Alic2019}, SVR showed improved performance in predicting boiling heat transfer over a horizontal tube compared to ANN and the Decision Tree (DT) algorithm. Similarly, in concentric-tube open thermosyphon, SVR predicted the Critical Heat Flux (CHF) more accurately than ANN \cite{Cai2012}. A study by Bard et al. \cite{Bard2022} on the prediction of HTC in flow boiling of mini/micro-channels inferred that SVR predicts the HTC with lower Mean Absolute Error (MAE). Zhou et al. \cite{Zhou2020} used non-dimensional input parameters and highlighted that Extreme Gradient Boosting (XGBoost) and ANN predicted the HTC for flow condensation in mini/micro-channels with Mean Absolute Error (MAE) of less than 10\%.

The minimum film boiling temperature of quenched substrate rods in distilled pools was predicted by Bahman and Ebrahim  \cite{Bahman2020} by employing a 2-layer ANN model with $R^2$ value of around 0.96. Qiu et al. \cite{QIU2021121607} used Extreme Gradient Boosting (XGBoost), Light Gradient Boosting Machine (LightGBM), K-Nearest Neighbor (KNN), and ANN to predict pressure drop for saturated flow boiling in mini/micro channels using various non-dimensional numbers as input variables. He concluded that XGBoost and ANN performed better than other models. Wen et al. \cite{Wen2024} also compared the ANN, XGBoost, SVR, and Random Forest (RF) models and inferred that ANN and XGBoost predicted the critical heat transfer deterioration points in the prediction of heat transfer characteristics of supercritical carbon dioxide in the pseudo-critical region. A study by Vijay and Gedupudi \cite{K2024} concluded that the XGBoost algorithm showed better prediction in estimating the heat transfer coefficient on plain and roughened surfaces with $R^2$ value of 0.99 in comparison to RF and DT and identified the key influencing parameters in the prediction of HTC through SHAP interpretation. An analysis of the prediction of heat transfer performance in an open pulsating heat pipe by Wu et al. \cite{wu2022heat} illustrated the better predictive performance of the Categorical Boosting (CatBoost) algorithm against XGBoost, LightGBM, and Gradient boosting decision tree (GBDT) models. The enhanced effectiveness of the CatBoost algorithm is also emphasized in the prediction of the boiling crisis inside channels and the thermohydraulic performance of double pipe heat exchangers, as illustrated in studies by Abdurakipov et al. \cite{Abdurakipov2022} and Sammil and Sridharan, \cite{Sammil2024}, respectively. All the studies infer that the machine learning model's performance varies according to the specific application.

The above literature review highlights the importance of employing ML techniques and the limitations of traditional approaches in predicting boiling performance. This study aims to compare various ML models to predict the heat transfer coefficient on thin film-coated and porous-coated substrates and identify the best-performing model in terms of accuracy and reliability, thus expanding its applicability in industries.

\subsection{Objective and contributions of the present study} \label{sec: Objective of the present study}
A complete understanding of the boiling phenomenon remains elusive due to the complex interaction of various parameters such as surface characteristics, operating conditions, liquid and vapor properties, substrate properties, and bubble dynamics. Though several correlations are available, they fall short of accurately predicting HTC on coated surfaces because they fail to identify the complex interplay of all influencing parameters. To navigate these challenges, machine learning is instrumental in understanding these patterns and predicting the heat transfer characteristics accurately. Despite earlier efforts to employ ML in this context, several challenges still persist.

ML models perform well with a large number of data points and a broader range of parameters. Previous studies \cite{Sajjad2021a,Sajjad2021b,Sajjad2021c} have trained these models with a smaller number of datasets, typically around 1000 data points. This study employs 5244 and 5142 data points for thin film-coated and porous-coated surfaces, respectively. To capture the actual phenomena, it is crucial to consider all the critical parameters affecting the boiling process. Compared to earlier studies \cite{Sajjad2021a,Sajjad2021b,Sajjad2021c} ], this study considers a more comprehensive set of parameters, including operating conditions ($\triangle T$, $T_w$, $P_{op}$), surface characteristics ($k_{w}$, $k_{co}$, $R_{co}$, $t$, $\varepsilon$, $R_q$, $\theta$), and thermophysical properties ($P_{film}$, $\rho_l$, $\rho_v$,  $C_{pl}$, $C_{pv}$, $\mu_l$, $\mu_v$,  $k_l$, $k_v$, $\sigma$, $h_{lv}$). Furthermore, this study evaluates nineteen ML models to identify the best model to predict HTC on thin film-coated and porous-coated substrates.

Heat transfer primarily occurs at the liquid-vapor interface. This study employs thermophysical properties at film temperature, which accurately present the actual conditions at the interface compared to prior studies, which primarily use properties at saturation temperature \cite{Sajjad2021a,Sajjad2021b,Sajjad2021c}. Even though the ML model exhibits high predictive accuracy, it is imperative to understand and interpret the model’s predictions. This study uses SHAP (SHapley Additive exPlanations) to identify the important parameters affecting HTC. Unlike previous studies, it explains how the variations in parameters affect HTC and validates the findings with the observed phenomena in existing studies, thus trusting the model predictions. This approach ensures the model’s applicability beyond the trained parametric ranges.

Thus, this research aims to bridge these critical gaps and develop a robust ML model to predict the nucleate boiling heat transfer coefficient on thin film-coated and porous-coated substrates. Furthermore, the study non-dimensionalizes the parameters and carries out a separate analysis to identify the critical non-dimensional parameters affecting the boiling heat transfer. The study further divides the dataset into water and refrigerants to uncover the key parameters influencing these fluid categories. Finally, the study also makes an assessment of the existing empirical correlations and proposes new correlations. 

\section{Methodology}\label{sec:methodology}
The overall methodology followed in the study is illustrated in Fig.\ref{fig: Methodology}. The detailed description of the methodology followed is presented below:

\begin{figure}[H]    
\centering
    \includegraphics[width=18cm]{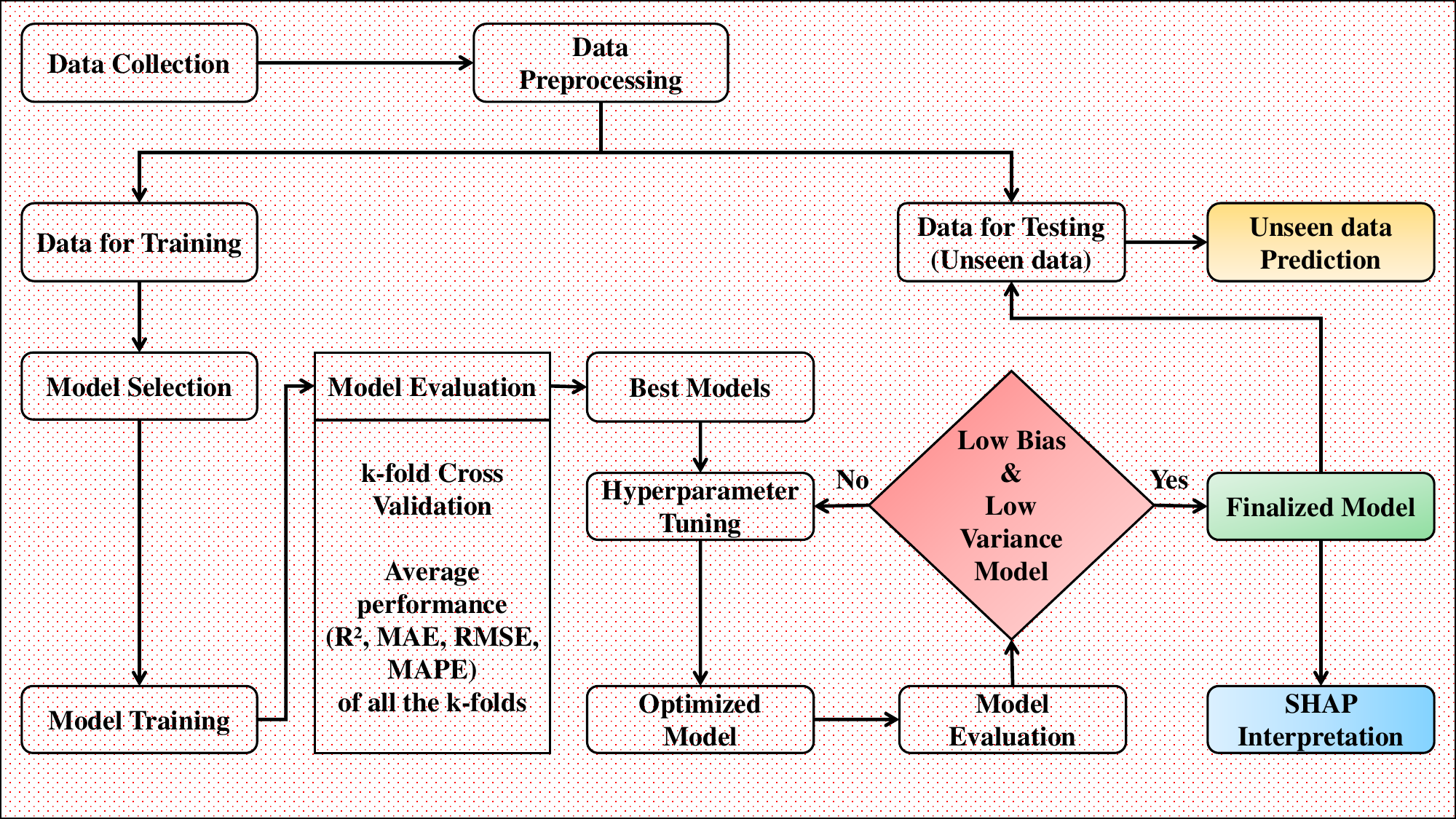}
\caption{Schematic representation of machine learning framework.}
\label{fig: Methodology}
\end{figure}

\subsection{Data collection}
This study compiles a wide range of datasets for nucleate pool boiling on thin film-coated \cite{Das2016a, Das2016b, Das2014a, Das2014b, Kim2018a, Kim2018b, Pinni2022, Majumder2022c, Majumder2023, Das2017a, Das2017b, Jaikumar2017, Thangavelu2023, Nithyanandam2020, Gajghate2020, Gajghate2021a, Gajghate2021b, Patel2019, Kumar2024, Deb2020, Laskar2022, Ray2018} and porous-coated substrates \cite{Katarkar2021a, Katarkar2021b, Jun2015, Ahmadi2022, Gupta2020, Gupta2023a, Gupta2023b, Gupta2019a, Gupta2019b, Hsieh1997, Dewangan2016, Ashok2019, Godinez2019, Gupta2018, Majumder2022a, Majumder2022b, Gheitaghy2017, Joseph2019, Pialago2013, Wang2018, Jun2016}, sourced from various studies.

The collected data includes a broad range of coatings and coating techniques, different fluids, and various substrate materials, as detailed in Table \ref{tab:coating-techniques}. In total, 5244 data points for thin film-coated substrates and 5142 data points for porous-coated substrates have been collected to predict the HTC (target variable) under saturated boiling conditions.

\begin{table}[H]
\centering
\caption{Overview of coating techniques, coatings, fluids, and substrate materials in the dataset.}
\label{tab:coating-techniques}
\begin{adjustbox}{max width=\textwidth}
\begin{tabular}{>{\arraybackslash}p{2.5cm} >{\arraybackslash}p{8cm} >{\arraybackslash}p{8cm}}
\toprule
& \multicolumn{2}{c}{\textbf{Description}}\\
\cmidrule(lr){2-3} 
\vspace{-0.7cm}\hspace{10pt}\textbf{Type} & \hspace{68pt}\textbf{Thin film coating} & \hspace{72pt}\textbf{Porous coating} \\
\midrule

\vspace{1.2cm}\textbf{Coating techniques} & 
Thermal evaporation physical vapor deposition \cite{Pinni2022,Majumder2022c,Majumder2023,Laskar2022}, Electron beam evaporation \cite{Das2016a,Das2014b, Das2017a,Das2017b}, Spin coating \cite{Kim2018a}, Nanofluid boiling nanoparticle deposition \cite{Gajghate2021a}, Dip coating \cite{Gajghate2020,Gajghate2021b,Jaikumar2017}, Glancing angle deposition \cite{Ray2018}, Electrophoretic deposition \cite{Kumar2024}, Sputtering technique \cite{Nithyanandam2020,Thangavelu2023}, Electron beam physical vapor deposition \cite{Das2016b}, Sol-gel spin coating \cite{Patel2019,Deb2020}
& 
Flame spraying \cite{Ashok2019,Dewangan2016,Hsieh1997}, Brazing \cite{Godinez2019,Jun2015}, Electrochemical deposition \cite{Gupta2023b,Gupta2019b,Gupta2023a, Gupta2020,Gupta2018,Gupta2019a}, Sintering \cite{Ahmadi2022}, Electrodeposition method \cite{Katarkar2021a,Katarkar2021b,Gheitaghy2017, Majumder2022b}, Mechanical milling \cite{Majumder2022a,Pialago2013}, Plasma spraying \cite{Hsieh1997}, Sol-gel dip coating \cite{Joseph2019}, Hydrogen bubble template electrodeposition \cite{Wang2018} \\
\midrule

\vspace{2.8cm}\textbf{Coatings} & 
SiO$_2$ - thin film \cite{Nithyanandam2020,Das2016a, Deb2020,Das2016b}, SiO$_2$ - nanoparticles \cite{Das2016b,Das2017b,Das2014b}, Hexagonal boron nitride (h-BN) \cite{Kim2018b}, Aluminium - thin film \cite{Pinni2022,Majumder2022c}, ZnO nanostructures \cite{Majumder2023}, Graphene - Graphene oxide mixture \cite{Jaikumar2017}, Polytetrafluoroethylene (PTFE) \cite{Kim2018a}, TiO$_2$ - thin film \cite{Thangavelu2023,Das2016a,Das2014a}, ZrO$_2$ - thin film \cite{Gajghate2021a}, Graphene - thin film \cite{Gajghate2020}, Graphene poly(3,4-ethylenedioxythiophene) polystyrene sulfonate \cite{Gajghate2021b}, CuO - thin film \cite{Patel2019}, TiO$_2$ - crystalline nanoparticles \cite{Das2017a}, TiO$_2$ \& SiO$_2$ composite \cite{Kumar2024}, Nano Cu - thin film \cite{Laskar2022}, TiO$_2$ nanostructures \cite{Ray2018,Ray2018}
&
Copper powder \cite{Ashok2019,Dewangan2016,Ahmadi2022,Jun2016}, Dendritic copper \cite{Wang2018}, CNT (Carbon nanotube) - Cu composite \cite{Pialago2013}, Cu - Al$_2$O$_3$ nanocomposite \cite{Gupta2018,Gupta2019a}, Nano CuO \cite{Joseph2019}, Al-GNP (Graphite nanoplatelets) \cite{Majumder2022a}, Cu-GNP \cite{Katarkar2021a, Katarkar2021b}, Microporous copper, \cite{Gheitaghy2017}, Microporous aluminium \cite{Majumder2022b}, CuAl$_2$O$_3$ nanoparticle \cite{Gupta2023a}, High-temperature conductive microporous Al \cite{Godinez2019}, High-temperature conductive microporous coating Cu \cite{Jun2015}, Mo coating \cite{Hsieh1997}, Al coating \cite{Hsieh1997}, Cu coating \cite{Hsieh1997}, Zn coating \cite{Hsieh1997}, Cu-alumina \cite{Gupta2023b}, Cu-TiO$_2$ nanocomposite \cite{Gupta2019b,Gupta2020} \\
\midrule

\textbf{Fluids} & \multicolumn{2}{>{\arraybackslash}p{15cm}}{Water, R134A, R141b, R600A, R410A, R407C} \\
\midrule

\textbf{Substrate materials} & \multicolumn{2}{>{\arraybackslash}p{15cm}}{Copper, Aluminium, Silicon wafer, Stainless steel} \\
\bottomrule

\end{tabular}
\end{adjustbox}
\end{table}

\subsection{Feature selection}
Selecting the important features (parameters) affecting the HTC is a critical step in ML. With the aim to completely capture the underlying pattern and illustrate the intricate interactions of various parameters in predicting the HTC, the present study considers all the influencing parameters that affect the nucleate pool boiling on thin film-coated and porous-coated substrates. The coating thickness for thin film-coated substrates ranges from 0.05 $\mu$m to 27 $\mu$m, while for porous-coated substrates, it ranges from 6 $\mu$m to 2000 $\mu$m.

Significant parameters that affect the HTC in this analysis are considered, including different operating conditions ($P_{op}$, $\Delta T$, $T_{w}$, Working fluids), substrate and coating properties ($k_{w}$, $k_{co}$, $R_{co}$, $\varepsilon$ and $t$), surface characteristics ($R_q$, and $\theta$), liquid thermophysical properties ($\sigma$, $\rho_l$, $C_{pl}$, $\mu_l$, $P_{film}$, and $k_l$) corresponding to film temperature ($T_{film}$ = ($T_{w}$ + $T_{sat}$) / 2), and vapor thermophysical properties ($\rho_v$,  $C_{pv}$, $\mu_v$, $k_v$, and $h_{lv}$) corresponding to liquid saturation temperature. Evaluating liquid properties at film temperature captures the accurate thermal interaction between the heating surface and the liquid. It is appropriate to take vapor properties at saturation temperature because the vapor pressure inside the bubbles will be nearly equal to the operating pressure, as the pressure difference across the liquid-vapor interface is marginal.

\begin{table}[h!]
\centering
\caption{Range of parameters for raw data.}
\label{tab: Range of parameters for raw data}
\begin{adjustbox}{max width=\textwidth}
\begin{tabular}{@{}lllllllllll@{}}
\toprule
\multicolumn{1}{l}{} & \hspace{10pt} & \multicolumn{4}{c}{\textbf{Thin film-coated}} & \hspace{10pt} & \multicolumn{4}{c}{\textbf{Porous-coated}} \\ 
\cmidrule(lr){3-6} \cmidrule(lr){8-11}
\hspace{10pt}\textbf{Features} & \hspace{10pt} & \textbf{Min} & \textbf{Max} & \textbf{Mean} & \textbf{SD} & \hspace{10pt} & \textbf{Min} & \textbf{Max} & \textbf{Mean} & \textbf{SD} \\ 
\midrule
\hspace{10pt}$\triangle T$ (K) & & 0.09 & 58.9 & 10.7 & 5.55 & & 0.45 & 24.52 & 7.03 & 3.75 \\
\hspace{10pt}$T_w$ (K) & & 287.3 &  432.05 & 355.42 & 39.42 & & 278.40 & 392.33 & 343.55 & 44.73 \\
\hspace{10pt}$P_{op}$ (bar) & & 1.01 & 2.193 & 2.49 & 4.07 & & 1.01 & 10.90 & 2.42 & 2.46 \\
\hspace{10pt}$k_{w}$ (W/mK) & & 37.66 &  401 & 381.06 & 67.72 & & 22 & 401 & 363.13 & 72.48 \\
\hspace{10pt}$k_{co}$ (W/mK) & & 0.25 & 610 & 146.22 & 211 & & 32.9 & 470 & 326.31 & 91.09 \\
\hspace{10pt}$R_{co}$ (m$^2$K/W) & & $1.3 \times 10^{-10}$ & $3.1 \times 10^{-5}$ & $1.1 \times 10^{-6}$ & $4.4 \times 10^{-6}$ & & $6.0 \times 10^{-8}$ & $1.5 \times 10^{-2}$ & $9.4 \times 10^{-4}$ & $2.5 \times 10^{-3}$ \\
\hspace{10pt}$t$ ($\mu$m) & & 0.05 & 27 & 1.2 & 3.52 & & 6 & 2000 & 190.28 & 315 \\
\hspace{10pt}$\varepsilon$ & & Nil & Nil & Nil & Nil & & 0.03 & 0.89 & 0.44 & 0.22 \\
\hspace{10pt}$R_q$ ($\mu$m) & & 0.01 & 15.1 & 0.54 & 1.87 & & 0.06 & 13.58 & 4.52 & 4.22 \\
\hspace{10pt}$\theta$ ($^\circ$) & & 4 & 133 & 52.85 & 35.14 & & 8 & 131 & 37.43 & 27.61 \\
\hspace{10pt}$P_{film}$ (bar) & & 1.02 & 6.04 & 2.08 & 1.58 & & 1.02 & 12.20 & 2.89 & 2.74 \\
\hspace{10pt}$\rho_l$ (kg/m$^3$) & & 554.9 & 1253.73 & 1024.14 & 155.15 & & 564.68 & 1275.46 & 1040.55 & 161.93 \\
\hspace{10pt}$C_{pl}$ (kJ/kgK) & & 1.16 & 4.26 & 3.25 & 1.34 & & 1.36 & 4.23 & 3.13 & 1.34 \\
\hspace{10pt}$\mu_l$ (kg m$^2$/s) & & $1.6 \times 10^{-4}$ & $3.8 \times 10^{-4}$ & $2.6 \times 10^{-4}$ & $4.1 \times 10^{-5}$ & & $1.4 \times 10^{-4}$ & $2.8 \times 10^{-4}$ & $2.5 \times 10^{-4}$ & $3.8 \times 10^{-5}$ \\
\hspace{10pt}$k_l$ (W/mK) & & 0.08 & 0.68 & 0.47 & 0.28 & & 0.08 & 0.68 & 0.44 & 0.29 \\
\hspace{10pt}$\sigma$ (N/m) & & $8.5 \times 10^{-3}$ & $5.9 \times 10^{-2}$ & $4.2 \times 10^{-2}$ & $2.2 \times 10^{-2}$ & & 0.01 & 0.06 & 0.04 & 0.02 \\
\hspace{10pt}$\rho_{v}$ (kg/m$^3$) & & 0.6 & 29.33 & 6.45 & 9.58 & & 0.60 & 47.25 & 10.01 & 13.07 \\
\hspace{10pt}$C_{pv}$ (kJ/kgK) & & 0.81 & 2.24 & 1.72 & 0.54 & & 0.92 & 2.12 & 1.69 & 0.52 \\
\hspace{10pt}$\mu_v$ (kg m$^2$/s) & & $7 \times 10^{-6}$ & $1.3 \times 10^{-5}$ & $1.2 \times 10^{-5}$ & $1 \times 10^{-6}$ & & $7 \times 10^{-6}$ & $1.3 \times 10^{-5}$ & $1.2 \times 10^{-5}$ & $1 \times 10^{-6}$ \\
\hspace{10pt}$k_v$ (W/mK) & & 0.01 & 0.028 & 0.021 & 0.006 & & 0.012 & 0.026 & 0.020 & 0.006 \\
\hspace{10pt}$h_{lv}$ (kJ/kg) & & 180.71 & 2256.28 & 1536.71 & 967.44 & & 180.79 & 2255.80 & 1431.59 & 1000.46 \\
\hspace{10pt}$h$ (kW/m$^2$K) & & 0.68 & 197.95 & 41.91 & 42.91 & & 0.02 & 413.72 & 84.37 & 85.95 \hspace{2.5em} \\
\bottomrule
\end{tabular}
\end{adjustbox}
\end{table}

\begin{table}[h!]
\centering
\caption{Range of parameters for non-dimensional data.}
\label{tab: Range of parameters for non-dimensional data}
\begin{adjustbox}{max width=\textwidth}
\begin{tabular}{@{}lllllllllll@{}}
\toprule
\multicolumn{1}{l}{} & \hspace{10pt} & \multicolumn{4}{c}{\textbf{Thin film-coated}} & \hspace{10pt} & \multicolumn{4}{c}{\textbf{Porous-coated}} \\ 
\cmidrule(lr){3-6} \cmidrule(lr){8-11}
\hspace{10pt}\textbf{Features} & \hspace{10pt} & \textbf{Min} & \textbf{Max} & \textbf{Mean} & \textbf{SD} & \hspace{10pt} & \textbf{Min} & \textbf{Max} & \textbf{Mean} & \textbf{SD} \\ 
\midrule
\hspace{10pt}$Pr_l$ & & 1.33 & 4.93 & 2.47 & 1.16 & & 1.59 & 4.36 & 2.40 & 0.93 \\
\hspace{10pt}$Pr_v$ & & 0.73 & 1.07 & 0.96 & 0.11 & & 0.79 & 1.17 & 0.98 & 0.10 \\
\hspace{10pt}$C_{pl}/C_{pv}$ & & 1.36 & 2.03 & 1.8 & 0.28 & & 1.26 & 2.03 & 1.77 & 0.30 \\
\hspace{10pt}$\rho_l/\rho_v$ & & 41.55 & 1599.65 & 916.62 & 605.74 & & 23.52 & 1589.77 & 882.19 & 677.59 \\
\hspace{10pt}$\mu_l/\mu_v$ & & 16.14 & 39.83 & 22.67 & 4.93 & & 10.68 & 24.63 & 20.87 & 2.80 \\
\hspace{10pt}$k_l/k_v$ & & 5.49 & 27.56 & 20.05 & 9.63 & & 5.98 & 27.54 & 19.08 & 10.00 \\
\hspace{10pt}$h_l/h_v$ & & 0.16 & 0.56 & 0.29 & 0.17 & & 0.16 & 0.56 & 0.30 & 0.18 \\
\hspace{10pt}$\varepsilon$ & & Nil & Nil & Nil & Nil & & 0.03 & 0.89 & 0.44 & 0.22 \\
\hspace{10pt}$R_q/r_{cav}$ & & $1.2 \times 10^{-4}$ & 52.13 & 0.77 & 2.69 & & 0.007 & 192.52 & 12.17 & 25.21 \\
\hspace{10pt}$Ja_l$ & & $1.7 \times 10^{-4}$ & 0.18 & 0.04 & 0.03 & & 0.001 & 0.18 & 0.03 & 0.03 \\
\hspace{10pt}$\theta/90$ & & 0.04 & 1.48 & 0.59 & 0.39 & & 0.09 & 1.46 & 0.42 & 0.31 \\
\hspace{10pt}$P_{red}$ & & 0.005 & 0.604 & 0.05 & 0.117 & & 0.005 & 0.22 & 0.04 & 0.06 \\
\hspace{10pt}$k_w/k_l$ & & 55.14 & 4825.28 & 1863.5 & 1833.62 & & 32.37 & 4755.36    & 1988.15    & 1841.80 \\
\hspace{10pt}$k_w/k_{co}$ & & 0.66 & 1604 & 246.29 & 442.19 & & 0.67 & 3.46 & 1.19 & 0.48 \\
\hspace{10pt}$R_{co}$ / $R_{sll}$ & & $9.1 \times 10^{-9}$ & $8.6 \times 10^{-3}$ & $3.0 \times 10^{-4}$ & $1.2 \times 10^{-3}$ & & $6.0 \times 10^{-6}$ & $4.24$\hspace{3em}  & $0.25$\hspace{3em}  & $0.68$\hspace{3em} \\
\hspace{10pt}$t/R_q$ & & 0.03 & 15.1 & 2.91 & 3.63 & & 4.86 & 33333.33 & 1272.41 & 5397.03 \\
\hspace{10pt}$P_{film}/P_{op}$ & & 0.12 & 2.62 & 1.16 & 0.24 & & 1.01 & 5.94 & 1.29 & 0.78 \\
\hspace{10pt}$Nu$ & & 8.28 & 725.21 & 169.31 & 146.63 & & 0.24 & 1523.07 & 320.89 & 306.85 \\
\bottomrule
\end{tabular}
\end{adjustbox}
\end{table}

\begin{table}[h!]
\centering
\caption{Features and their expressions.}
\label{tab: Features and their expressions}
\begin{adjustbox}{max width=\textwidth}
\begin{tabular}{ll}
\toprule
\hspace{10pt}\textbf{Features} & \hspace{10pt}\textbf{Expressions} \\
\midrule
\hspace{10pt}$\Delta T$ & \hspace{10pt} $T_w - T_{sat}$ \\ [4pt]
\hspace{10pt}$Pr_l$ &  \hspace{10pt} $\frac{\mu_{l} \cdot C_{pl}}{k_l}$ \\ [4pt]
\hspace{10pt}$Pr_v$ &  \hspace{10pt} $\frac{\mu_v \cdot C_{pv}}{k_v}$ \\ [4pt]
\hspace{10pt}$r_{cav}$ & \hspace{10pt}  $\frac{2\sigma\left(\frac{1}{\rho_v} - \frac{1}{\rho_l}\right)T_{sat}}{\Delta T \cdot h_{lv}}$ \cite{MAHMOUD2021101024} \\ [4pt]
\hspace{10pt}$Ja_l$ &  \hspace{10pt} $\frac{C_{pl}\cdot\Delta T}{h_{lv}}$ \\ [4pt]
\hspace{10pt}$P_{red}$ & \hspace{10pt}  $\frac{P_{op}}{P_{cr}}$ \\ [4pt]
\hspace{10pt}$L_c$ &  \hspace{10pt} $\sqrt{\frac{\sigma}{g(\rho_l - \rho_v)}}$ \cite{Pioro1999} \\ [9pt]
\hspace{10pt}$Nu$ & \hspace{10pt} $\frac{h \cdot L_c}{k_l}$ \\ [3pt]
\bottomrule
\end{tabular}
\end{adjustbox}
\end{table}

Non-dimensional numbers play a pivotal role in analyzing the boiling behavior under different conditions. To understand its influence in pool boiling on thin film-coated and porous-coated substrates, the above parameters have been non-dimensionalized. The non-dimensional features used in the present analysis are $Pr_l$, $Pr_v$, $C_{pl}/C_{pv}$, $\rho_l/\rho_v$, $\mu_l/\mu_v$, $k_l/k_v$, $h_l/h_v$, $R_q/r_{cav}$, $Ja_l$, $\theta/90^\circ$, $P_{red}$, $k_w/k_l$, $k_w/k_{co}$, $R_{co}$/$R_{sll}$, $t/R_q$, $\varepsilon$, and $P_{film}/P_{op}$. For porous-coated surfaces, volumetric porosity ($\varepsilon$) is included in the analysis, while for thin film-coated substrates, $\varepsilon$ is not considered. All other parameters are identical for both raw data and non-dimensional data analysis. Nu is the target variable in the non-dimensional analysis.\\
The thermal resistance of the superheated liquid layer is determined by, 
\begin{equation}
R_{sll} = \frac{L_c}{k_l}
\end{equation}
\vspace{2pt}The resistance of the thin film-coated substrates is expressed as, 
\begin{equation}
R_{co} = \frac{t}{k_{co}}   
\end{equation}
For the porous-coated substrates, the resistance of the coating is represented as, 
\begin{equation}
R_{co} = \frac{t}{k_{eff}}
\end{equation}
\vspace{0.5pt}
\begin{equation}
k_{eff} = k_f^{1-n} \cdot k_s^n
\end{equation}
\vspace{-\baselineskip}
\begin{equation*}
n = 0.280 - 0.757 \cdot \log_{10} (\varepsilon) + 0.057 \cdot \log_{10} \left( \frac{k_f}{k_s} \right)
\end{equation*}
\vspace{1.5pt}The effective thermal conductivity $k_{eff}$ is determined using the Krupiczka model \cite{krupiczka1967analysis}, where $k_s$ and $k_f$ denote the thermal conductivity of the solid and fluid phases of the porous coatings, respectively. The range of parameters for both raw and non-dimensional analyses of both surfaces are specified in Tables  \ref{tab: Range of parameters for raw data} and \ref{tab: Range of parameters for non-dimensional data}. Table \ref{tab: Features and their expressions} presents the expressions used in the analysis. The fluid properties are taken from the National Institute of Standards and Technology (NIST) \cite{lemmon_fluid_systems_chemistry_webbook}, and CoolProp \cite{Bell2014} databases.

\subsection{Data visualization} 
\begin{figure}[H]
	\centering
  \includegraphics[width=18cm, height=24cm]
  {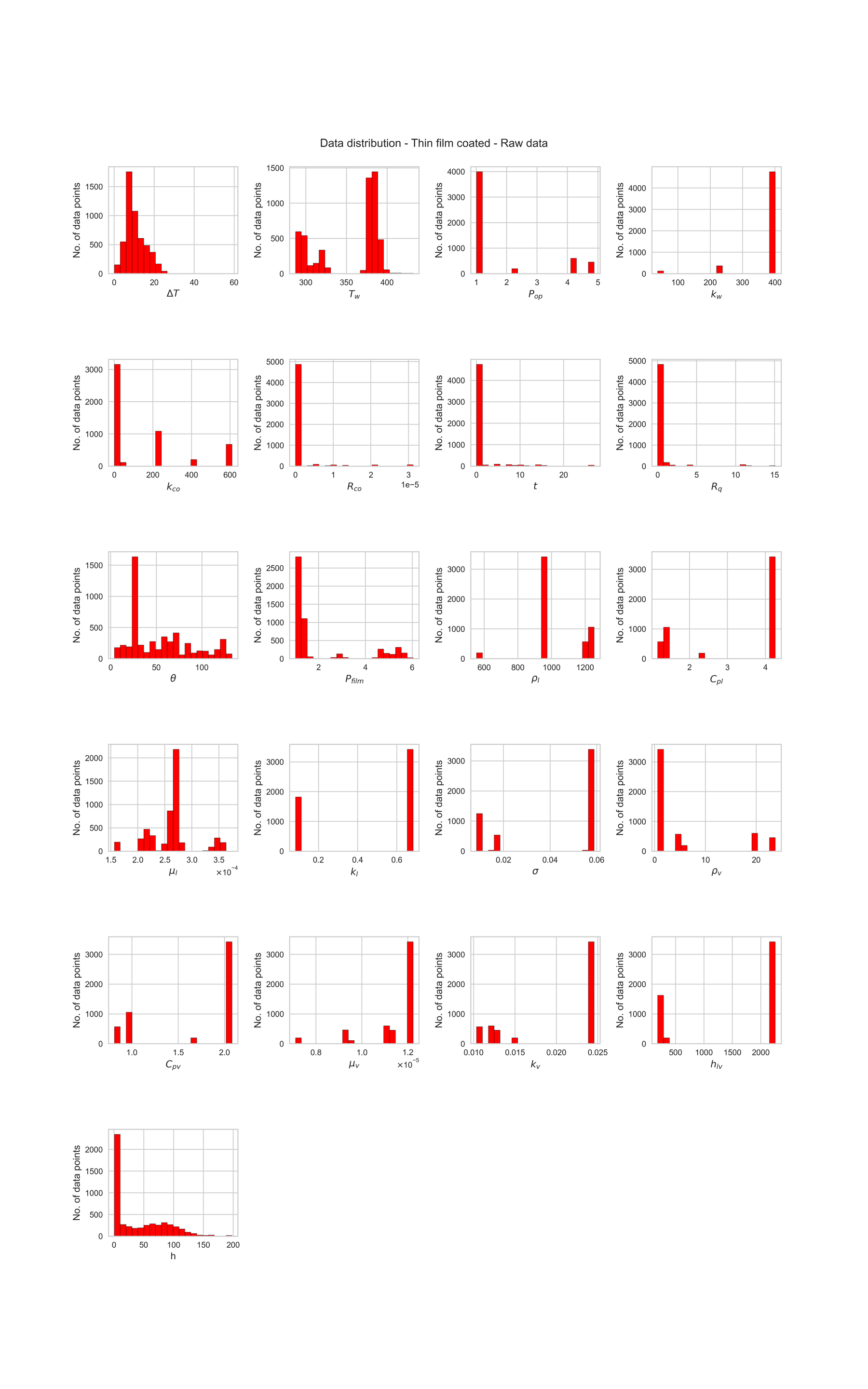}
\caption{Data distribution of raw data (Thin film-coated).}
\label{fig: Data distribution of raw data (Surface Coated).}
\end{figure}

\begin{figure}[H]
	\centering
  \includegraphics[width=18cm, height=24cm]{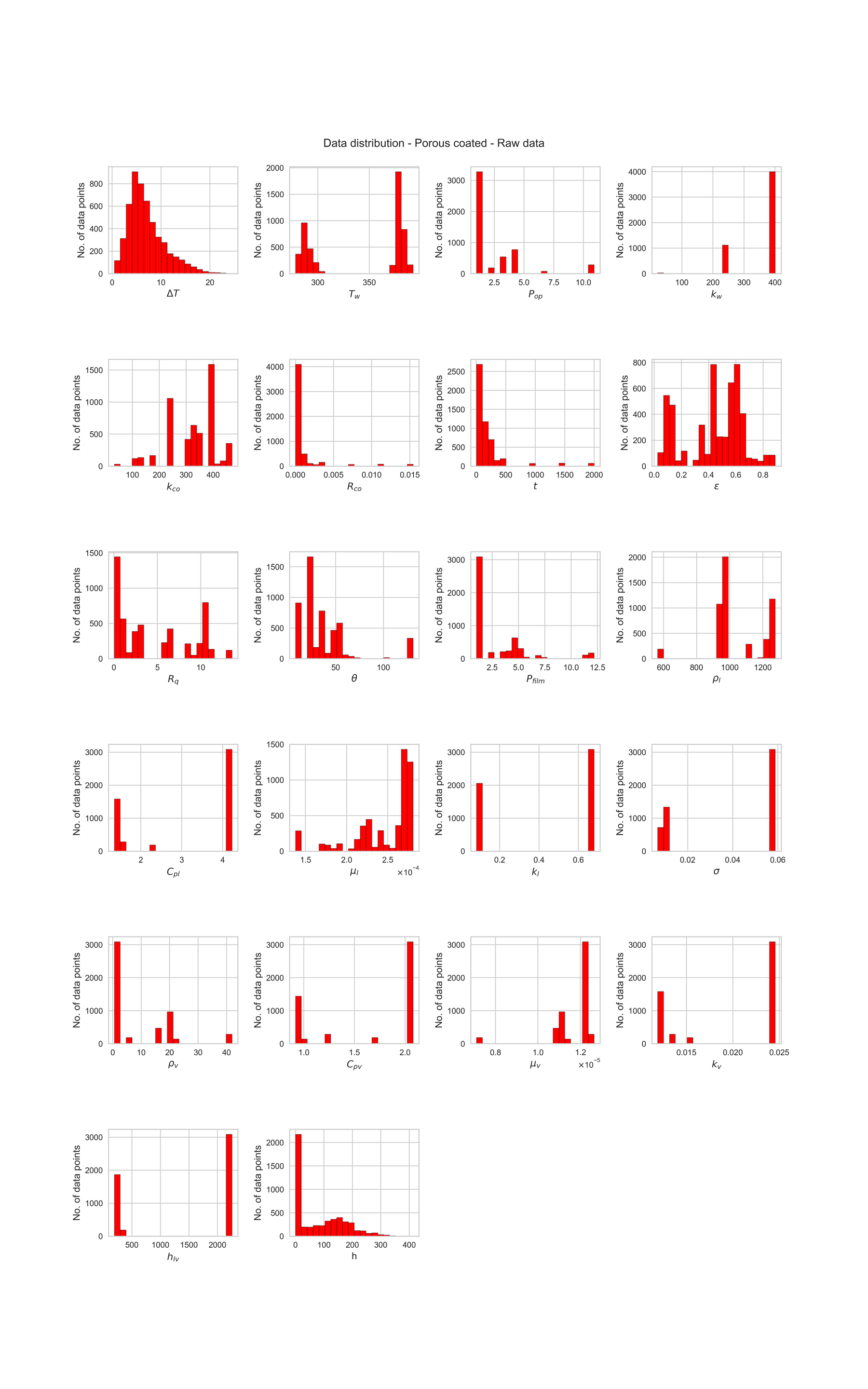}
\caption{Data distribution of raw data (Porous-coated).}
\label{fig: Data distribution of raw data (Porous Coated).}
\end{figure}

\begin{figure}[H]
	\centering
  \includegraphics[width=18cm]{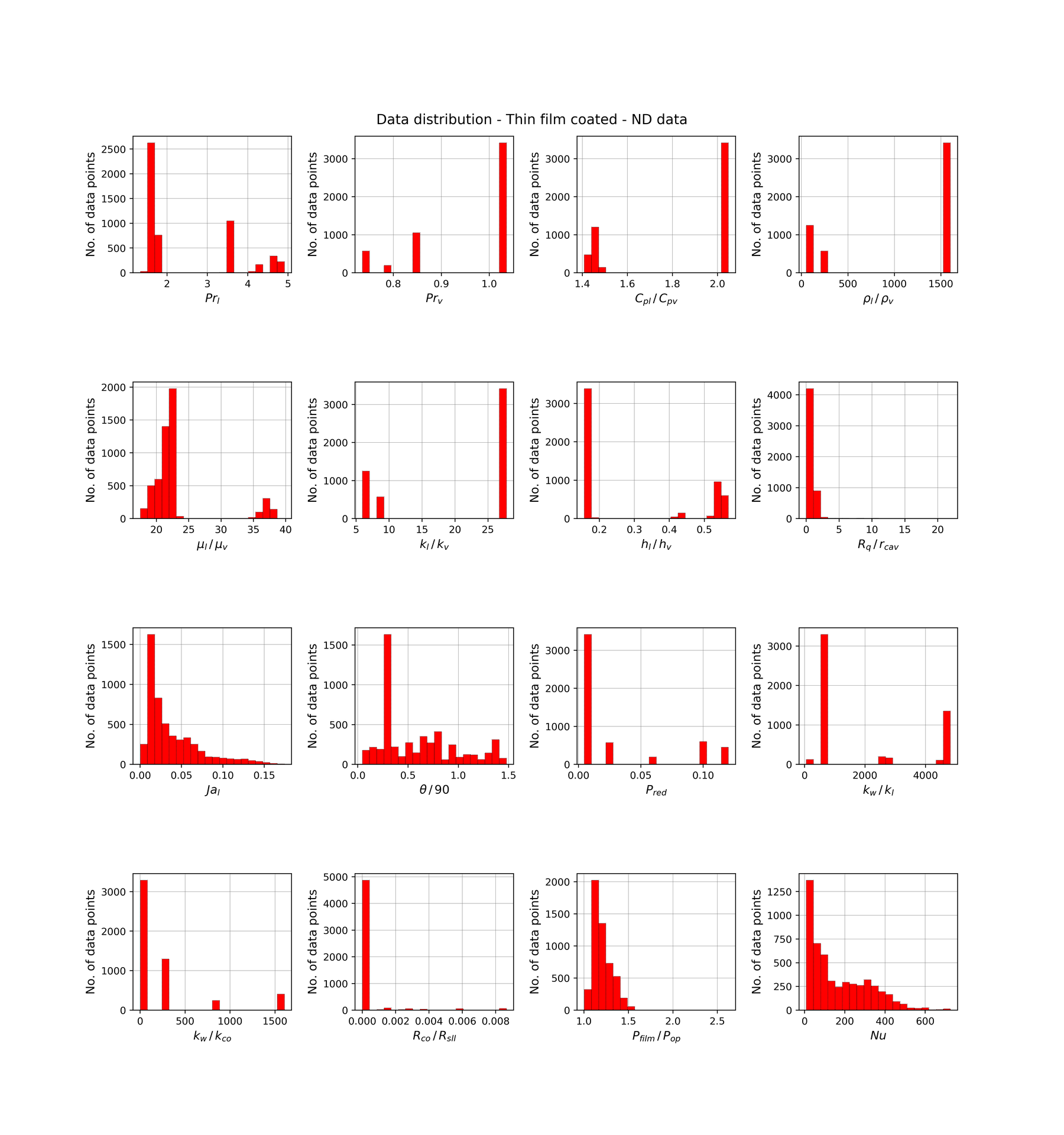}
\caption{Data distribution of ND data (Thin film-coated).}
\label{fig: Data distribution of ND data (Surface Coated).}
\end{figure}

\begin{figure}[H]
	\centering
  \includegraphics[width=18cm]{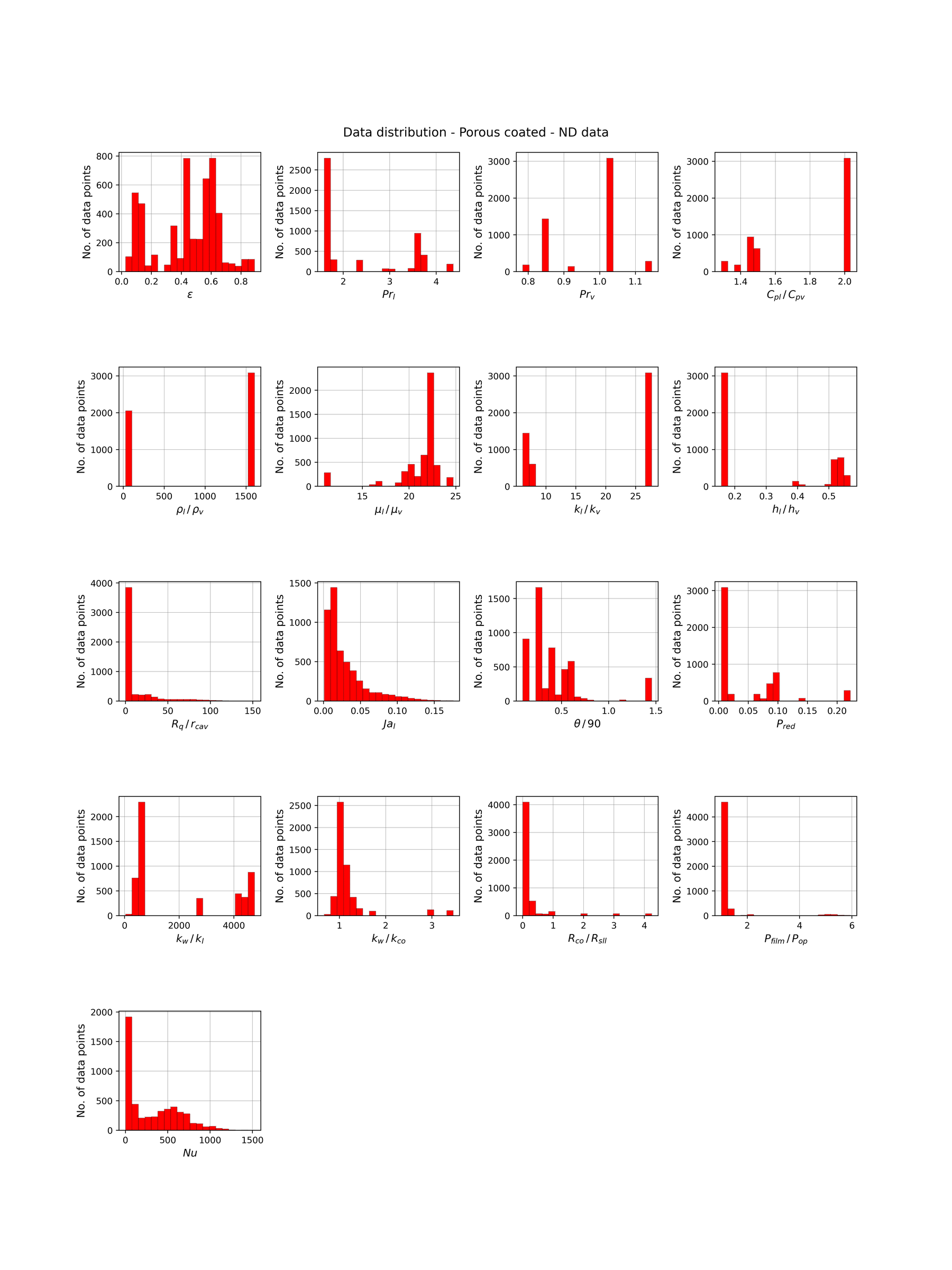}
\caption{Data distribution of ND data (Porous-coated).}
\label{fig: Data distribution of ND data (Porous Coated).}
\end{figure}

\begin{figure}[H]
	\centering
  \includegraphics[width=18cm,height=12cm]{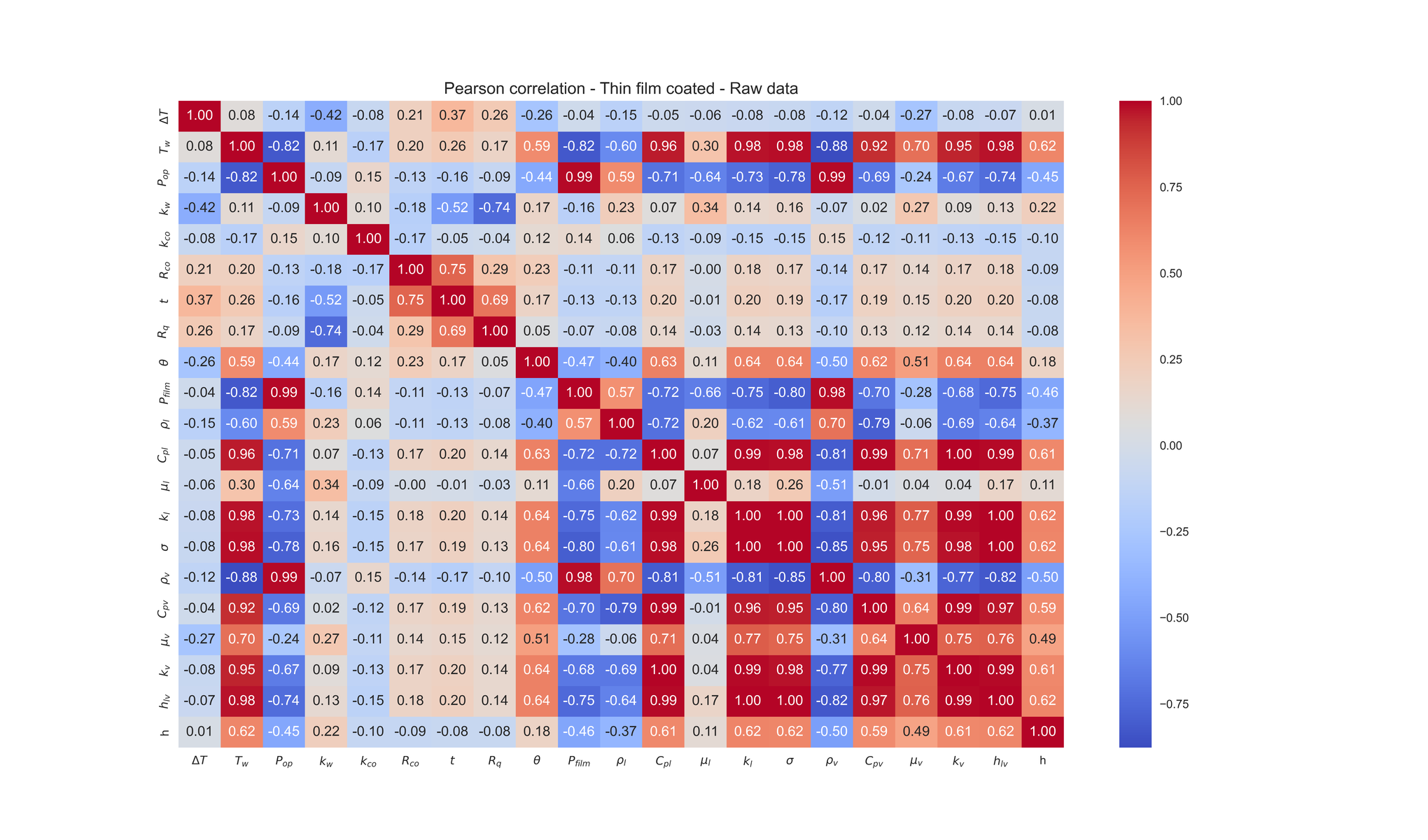}
\caption{Pearson correlation chart for raw data (Thin film-coated).}
\label{fig: Pearson correlation chart for raw data (Surface Coated).}
\end{figure}

\begin{figure}[H]
	\centering
  \includegraphics[width=18cm,height=12cm]{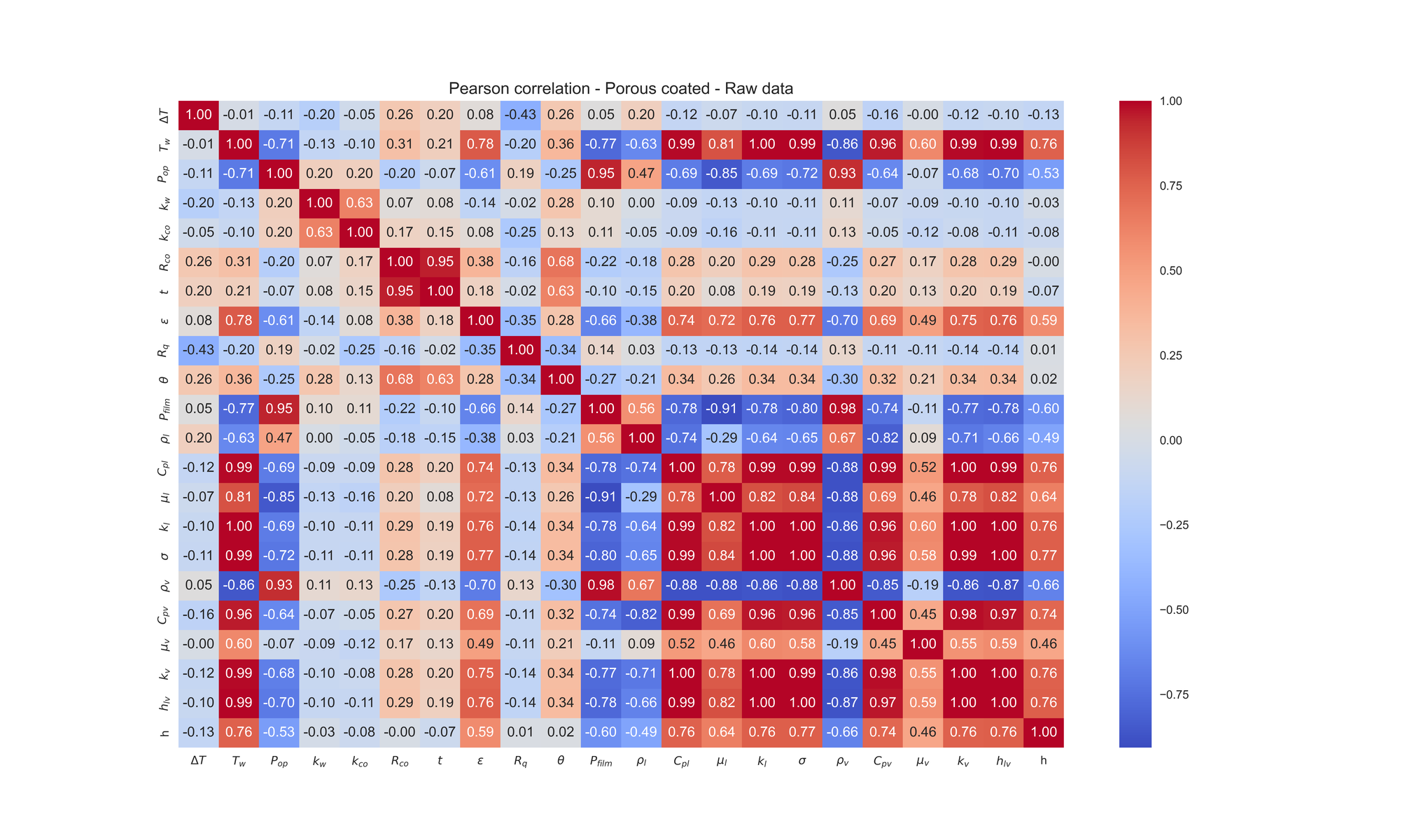}
\caption{Pearson correlation chart for raw data (Porous-coated).}
\label{fig: Pearson correlation chart for raw data (Porous Coated).}
\end{figure}

\begin{figure}[H]
	\centering
  \includegraphics[width=18cm, height=12cm]{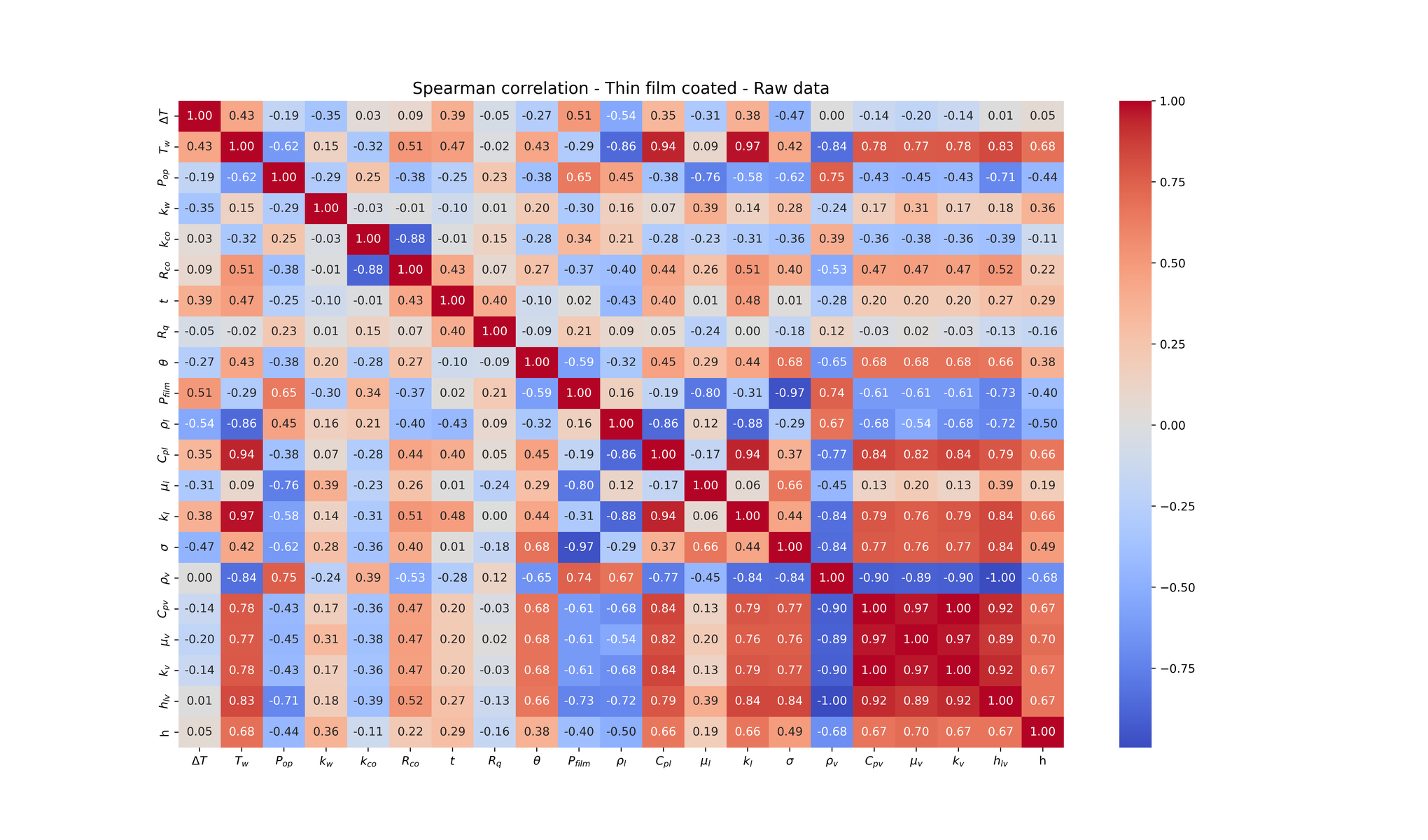}
\caption{Spearman correlation chart for raw data (Thin film-coated).}
\label{fig: Spearman correlation chart for raw data (Surface Coated).}
\end{figure}

\begin{figure}[H]
	\centering
  \includegraphics[width=18cm,height=12cm]{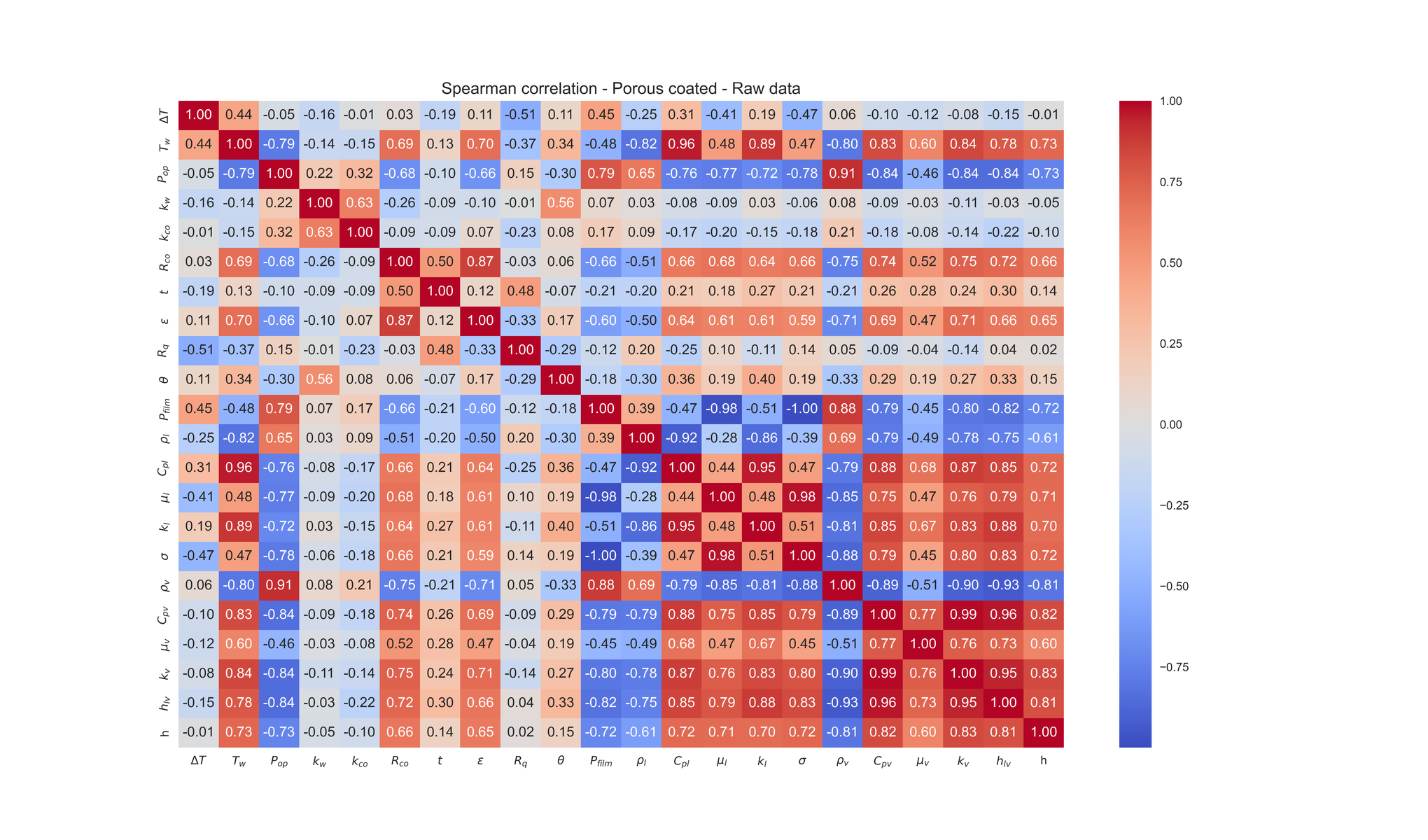}
\caption{Spearman correlation chart for raw data (Porous-coated).}
\label{fig: Spearman correlation chart for raw data (Porous Coated).}
\end{figure}

\begin{figure}[H]
	\centering
  \includegraphics[width=18cm,height=12cm]{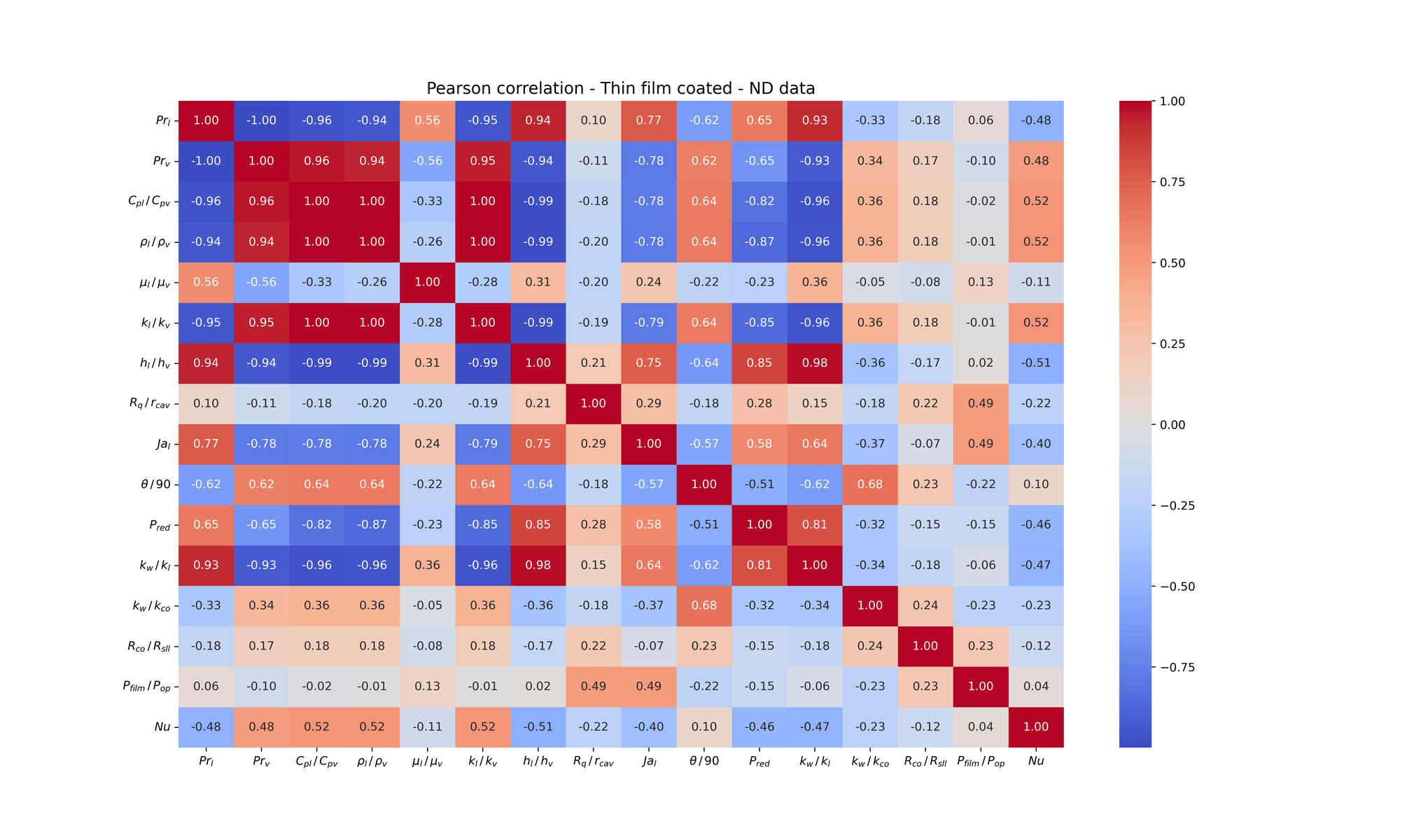}
\caption{Pearson correlation chart for ND data (Thin film-coated).}
\label{fig: Pearson correlation chart for ND data (Surface Coated).}
\end{figure}

\begin{figure}[H]
	\centering
  \includegraphics[width=18cm,height=12cm]{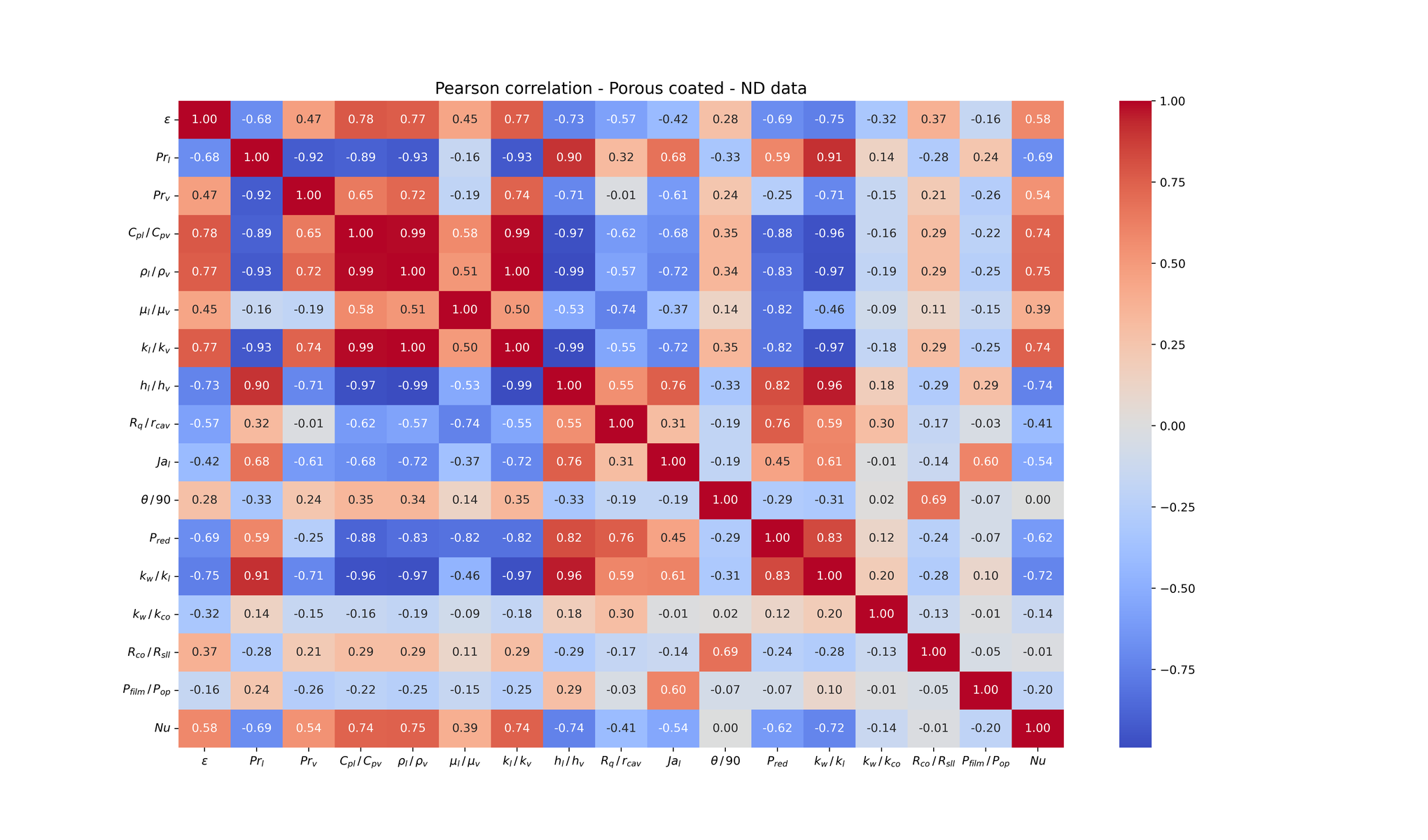}
\caption{Pearson correlation chart for ND data (Porous-coated).}
\label{fig: Pearson correlation chart for ND data (Porous Coated).}
\end{figure}

\begin{figure}[H]
	\centering
  \includegraphics[width=18cm,height=12cm]{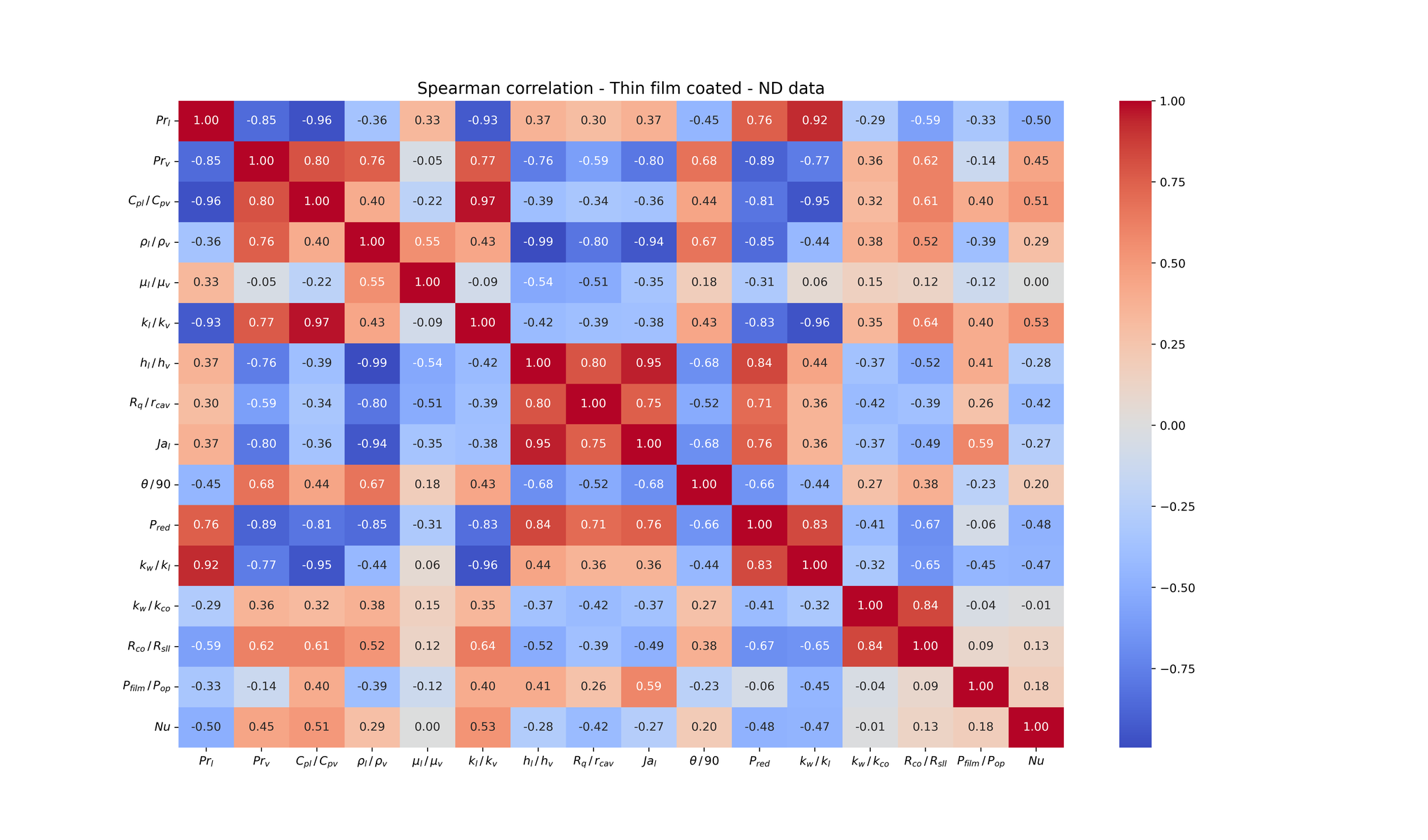}
\caption{Spearman correlation chart for ND data (Thin film-coated).}
\label{fig: Spearman correlation chart for ND data (Surface Coated).}
\end{figure}

\begin{figure}[H]
	\centering
  \includegraphics[width=18cm,height=12cm]{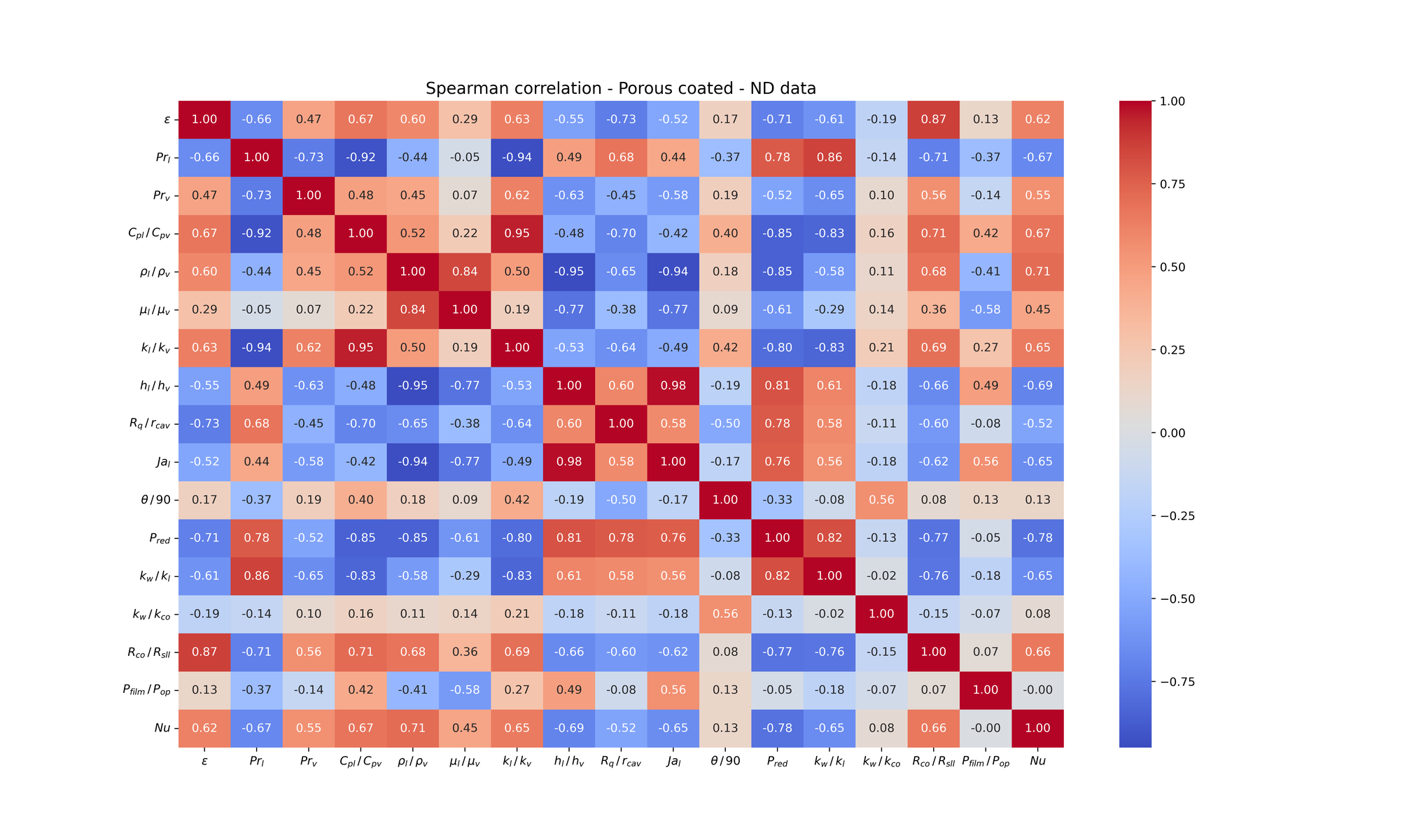}
\caption{Spearman correlation chart for ND data (Porous-coated).}
\label{fig: Spearman correlation chart for ND data (Porous Coated).}
\end{figure}

Figs. \ref{fig: Data distribution of raw data (Surface Coated).}-\ref{fig: Data distribution of ND data (Porous Coated).} show the distribution of the data used in this analysis. Pearson and Spearman correlation coefficients are estimated to identify the dependency between variables. While the Pearson correlation estimates the linear relationship, the Spearman correlation determines the monotonic relationship between variables. The coefficients vary between -1 and 1. Values close to the extremes represent a strong relationship (linear or monotonic), while the values close to zero represent a weak linear or monotonic relationship. Figs. \ref{fig: Pearson correlation chart for raw data (Surface Coated).}-\ref{fig: Spearman correlation chart for raw data (Porous Coated).} and Figs. \ref{fig: Pearson correlation chart for ND data (Surface Coated).}-\ref{fig: Spearman correlation chart for ND data (Porous Coated).} shows the Pearson and Spearman correlation chart for raw and ND data, respectively. Also, it can be seen that multicollinearity between features is negligible, which helps in developing a stable and interpretable model.

\subsection{Data preprocessing}
Data preprocessing plays a crucial role in building an effective and reliable model. The collected data from various sources may contain missing data, duplicates, errors, and outliers. Data cleaning is performed to remove the duplicates and outliers from the dataset, and impute or remove the missing values (approximately 2\% of the dataset). ML models process only the numerical inputs, so categorical features must be encoded to numerical values \cite{Goodfellow-et-al-2016}. In this study, one-hot encoding of the categorical variable ”Fluid” was performed before loading the data into the ML model. Also, the features in the ML model may vary on different scales. This results in features with larger values masking the smaller magnitude features. Here, z-score normalization is employed, such that each feature has a mean of 0 and a standard deviation of 1 \cite{Ahsan2021}. It is calculated by the formula as shown in Eq.\eqref{eq:z-score}: 
\begin{equation} 
\label{eq:z-score}
z = \frac{x - \mu}{\sigma}
\end{equation}

where \( z \) is the transformed value of the data point \( x \). \( \mu \) and \( \sigma \) represent the mean and standard deviation of the data in a particular feature.

Thus, the above data preprocessing steps ensure that quality data is fed into the model, making the model more interpretable and accurate \cite{pyle1999data}.

\subsection{Machine learning models}

ML models perform distinctly for different datasets. Identification of best-performing models for a particular case is crucial. The working principle of best-performing ML models for thin film-coated and porous-coated substrates is discussed below.

\subsubsection{Decision tree regressor}
Decision Trees (DTs) are widely used in supervised learning for both classification and regression tasks. It creates a hierarchical tree structure that divides the dataset into different subsets iteratively based on the input features \cite{Quinlan1986}. The root node contains the entire sample, which is further split into various nodes based on the feature values. The criteria for splitting each node is based on the Mean Squared Error (MSE). Leaf nodes represent the final prediction, which is the mean of the target variable in a specific leaf node. DTs offer better interpretability and can capture non-linear relationships well. However, deeper DTs (with more depth) can overfit the data, leading to poor generalization and high variance.

\subsubsection{Extra trees regressor}
Extra Trees (Extremely Randomized Trees) Regressor combines multiple decision trees, where each tree is developed based on the random subset of features. Also, the threshold to split each node in a tree is done at random \cite{Geurts2006}. The predictions from all the trees are averaged to estimate the final prediction. The randomization introduced in the algorithm helps to reduce the overfitting and variance in the model, thus enhancing the performance.

\subsubsection{Random forest regressor}
The Random Forest (RF) Algorithm works on the principle of Bagging / Bootstrapping. Bagging involves combining multiple decision trees, which are trained on different subsets of data with replacement \cite{Breiman2001}. To split each node in a tree, a random subset of features is used, and MSE criteria is employed. Each tree is trained independently, and the final prediction is the average of the output from all the multiple trees considered. The combination of multiple models and randomization effectively reduces overfitting, decreases bias, and increases performance.

\subsubsection{Gradient boosting regressor}
The gradient boosting algorithm introduced by Jerome H. Friedman is also an ensemble ML algorithm \cite{Friedman2001}. Unlike RF, where each tree is built independently, Gradient boosting employs the boosting technique, which combines multiple decision trees, where each tree is built sequentially to correct the errors/residuals made by the previous tree \cite{Friedman2001}. MSE is the objective function used in this algorithm.  All the predictions from the trees are then added to estimate the final prediction. Gradient Boosting greatly reduces the bias as each tree is built on the residuals from the previous trees.

\subsubsection{Extreme gradient boosting (XGBoost)}
Chen and Guestrin \cite{Chen2016} developed the XGBoost algorithm and addressed the problem of overfitting in the gradient boosting algorithm. XGBoost also employs the boosting algorithm and each tree is trained sequentially based on the residuals from the preceding tree. In addition to MSE, it introduces L1 and L2 regularization in the objective function. By adding the regularization parameters, it reduces the variance in the model and prevents overfitting. Thus, XGBoost yields better model generalization and increased accuracy.

\subsubsection{Light gradient boosting machine (LightGBM)}
For training large datasets with higher dimensional feature space, LightGBM, developed by Microsoft Research, proves particularly useful due to faster training time \cite{NIPS2017_6449f44a}. Gradient-based One-Side Sampling (GOSS), which uses only data points with large gradients to calculate the information gain, and Exclusive Feature Bundling (EFP), which minimizes the number of features by grouping mutually exclusive features, are two novel techniques proposed in this algorithm for faster execution \cite{NIPS2017_6449f44a}. Moreover, this model uses leaf-wise growth of trees rather than level-wise growth and uses the boosting technique, which results in increased accuracy. But, it may lead to overfitting, which is controlled by setting the maximum limit for the depth of a tree.

\subsubsection{CatBoost regressor}
CatBoost algorithm \cite{NEURIPS2018_14491b75} developed by Yandex proposed an innovative technique called ordered boosting to handle categorical variables. This algorithm doesn't require preprocessing of categorial variables like one-hot encoding. Numerical encoding of categorical features is done based on their significance in relation to the output variable \cite{NEURIPS2018_14491b75}. Catboost also uses the boosting technique combined with regularization parameters in the objective function to reduce overfitting and increase prediction accuracy. The trees in the Catboost algorithm are symmetric. 

\subsection{Hyperparameter optimization}

It is essential to perform hyperparameter optimization, where the best set of hyperparameters is identified to enhance the model performance. Hyperparameters specific to the model are fine-tuned through various methods to improve accuracy. In this study, the Random Search optimization approach \cite{bergstra2012random} is used. This method selects random hyperparameters over a range of values and identifies the best combination, which is particularly advantageous in high-dimensional space \cite{bergstra2012random}.

\subsection{k-fold cross validation}

K-fold cross-validation is a commonly employed approach to evaluate the model's reliability and performance. 5-fold or 10-fold cross-validation is widely used to assess the ML models. In this study, 10-fold cross-validation is used, which divides the total training dataset into 10 folds randomly. A total of 10 iterations will occur, and in each iteration, one of the folds will act as a test set, and the remaining nine folds will be used for training. Each fold will serve as a test set only once. The model is assessed based on the mean performance of all the iterations. Thus, hyperparameters are fine-tuned to perform well across all the iterations, mitigating the issues of overfitting, resulting in a low bias and low variance model.

\subsection{Evaluation metrics}

To assess the effectiveness of the models, several evaluation metrics are used to provide insights into the model's performance. These metrics are also used in fine-tuning the hyperparameters of a model. This study uses the commonly employed regression metrics, including Mean Absolute Error (MAE), Coefficient of Determination ($R^2$), Root Mean Squared Error (RMSE), and Mean Absolute Percentage Error (MAPE). 

\subsubsection{Coefficient of determination}
The coefficient of determination ($R^2$) is a statistical measure evaluating how well the variance in the dependent variable is explained by the independent variables. $R^2$, ranging between 0 and 1, represents the goodness of fit for the regression model. A value close to 1 implies a better fit, whereas a value close to 0 represents an underfit model. The $R^2$ value is given by Eq.\eqref{eq:R2}
\begin{equation} 
\label{eq:R2}
R^2 = 1 - \frac{\sum\limits_{i=1}^{n} (y_i - \hat{y}_i)^2}{\sum\limits_{i=1}^{n} (y_i - \bar{y})^2}
\end{equation}
where \( n \) is the number of data points, \( y_i \) represents the actual values, \( \hat{y}_i \) represents the predicted values, and \( \bar{y} \) is the mean of the actual values.

\subsubsection{Mean Absolute Error (MAE)}
MAE is the average of the absolute difference between predicted and actual values in the dataset. It represents the average magnitude of the errors as expressed by Eq.\eqref{eq:MAE}
\begin{equation} 
\label{eq:MAE}
\text{MAE} = \frac{1}{n} \sum_{i=1}^{n} \lvert y_i - \hat{y}_i \rvert
\end{equation}

\subsubsection{Root Mean Squared Error (RMSE)}
RMSE represents the square root of the average of the sum of squared errors between predicted and actual values. It brings the scale of the errors to the same scale as that of the target, which facilitates easier understanding. The formulation is given by Eq.\eqref{eq:MSE}
\begin{equation} 
\label{eq:MSE}
\text{MSE} = \frac{1}{n} \sum_{i=1}^{n} (y_i - \hat{y}_i)^2
\end{equation}

\subsubsection{Mean Absolute Percentage Error (MAPE):}
MAPE represents the average percentage error between the predicted and actual values. This metric is scale-independent and used to compare models across different scales of datasets. MAPE is calculated by the following formula (Eq.\eqref{eq: MAPE})
\begin{equation} 
\label{eq: MAPE}
\text{MAPE} = \frac{1}{n} \sum_{i=1}^{n} \left| \frac{y_i - \hat{y}_i}{y_i} \right| \times 100
\end{equation}

\subsection{SHAP (SHapley Additive exPlanations) 
Technique for Model Interpretation}

The interpretation of the ML models is as significant as the predictive accuracy to ensure its reliability while making predictions. 
SHAP is a powerful tool for understanding ML models by evaluating the contribution of each feature to the model predictions. SHAP, derived from the cooperative game theory, estimates the contribution of each player to the outcome of the game to provide a fair distribution of payoffs to the players \cite{NIPS2017_8a20a862}. This idea is applied to machine learning to calculate the contribution of each feature to the model predictions \cite{NIPS2017_8a20a862}. It provides both global and local interpretations of the predictions. Shapley values calculate the average marginal contribution of each feature across all possible combinations of features, which is expressed by the Eq.\eqref{eq: SHAP}.

\begin{equation} 
\label{eq: SHAP}
\phi_i = \sum_{S \subseteq N \setminus \{i\}} \frac{|S|!(|N| - |S| - 1)!}{|N|!} \left[f(S \cup \{i\}) - f(S)\right]
\end{equation}

where:

\begin{itemize}
    \item \( \phi_i \) is the Shapley value for feature \(i\), representing its contribution to the model predictions.
    \item \( N \) is the set of all features used in the model.
    \item \( S \) is a subset of features not including feature \(i\).
    \item \( f(S) \) is the model’s prediction using only the features in subset \(S\).
    \item \( f(S \cup \{i\}) \) is the model’s prediction using the features in subset \(S\), including feature \(i\).
    \item \( |S| \) represents the number of features in subset \(S\).
    \item \( |N| \) represents the total number of features in the model.
    \item \(\frac{|S|!(|N| - |S| - 1)!}{|N|!}\) represents the weight of a particular permutation of feature \(i\) being added to subset \(S\).
    \item \( f(S \cup \{i\}) - f(S) \) is the marginal contribution of feature \(i\), representing the change in the model's predictions by adding feature \(i\) to subset \(S\).
\end{itemize}

\section{Results and discussions}\label{sec:results}

\subsection{Performance of ML models}
The thin film-coated and porous-coated dataset, after preprocessing and z-score normalization, was divided into training data and testing data with 80\% for model training and 20\% for testing its performance. The training dataset with specific and appropriate initial hyperparameters was fed into various algorithms to estimate the performance. To determine the effective models for thin film-coated and porous-coated data, different regression ML models were evaluated, which include CatBoost, Extra Trees, Extreme Gradient Boosting, Random Forest, Light Gradient Boosting Machine, Decision Tree, Gradient Boosting, K Nearest Neighbors, AdaBoost, Linear Regression, Ridge Regression, Bayesian Ridge, Lasso Regression, Lasso Least Angle Regression, Huber Regressor, Elastic Net, Orthogonal Matching Pursuit, Passive Aggressive Regressor, and Dummy Regressor. All these models were evaluated based on a 10-fold cross-validation approach, and their results are shown in Tables \ref{tab:Comparison of Regression Models for Surface-Coated Substrates} and \ref{tab:Comparison of Regression Models for Porous-Coated Substrates}. All the nineteen ML models available in the scikit-learn library \cite{JMLR:v12:pedregosa11a} were implemented in Python 3.9.16. These models were then further developed and fine-tuned.

\begin{figure}[H]
\centering
\subfloat[]{{\includegraphics[width=9cm]{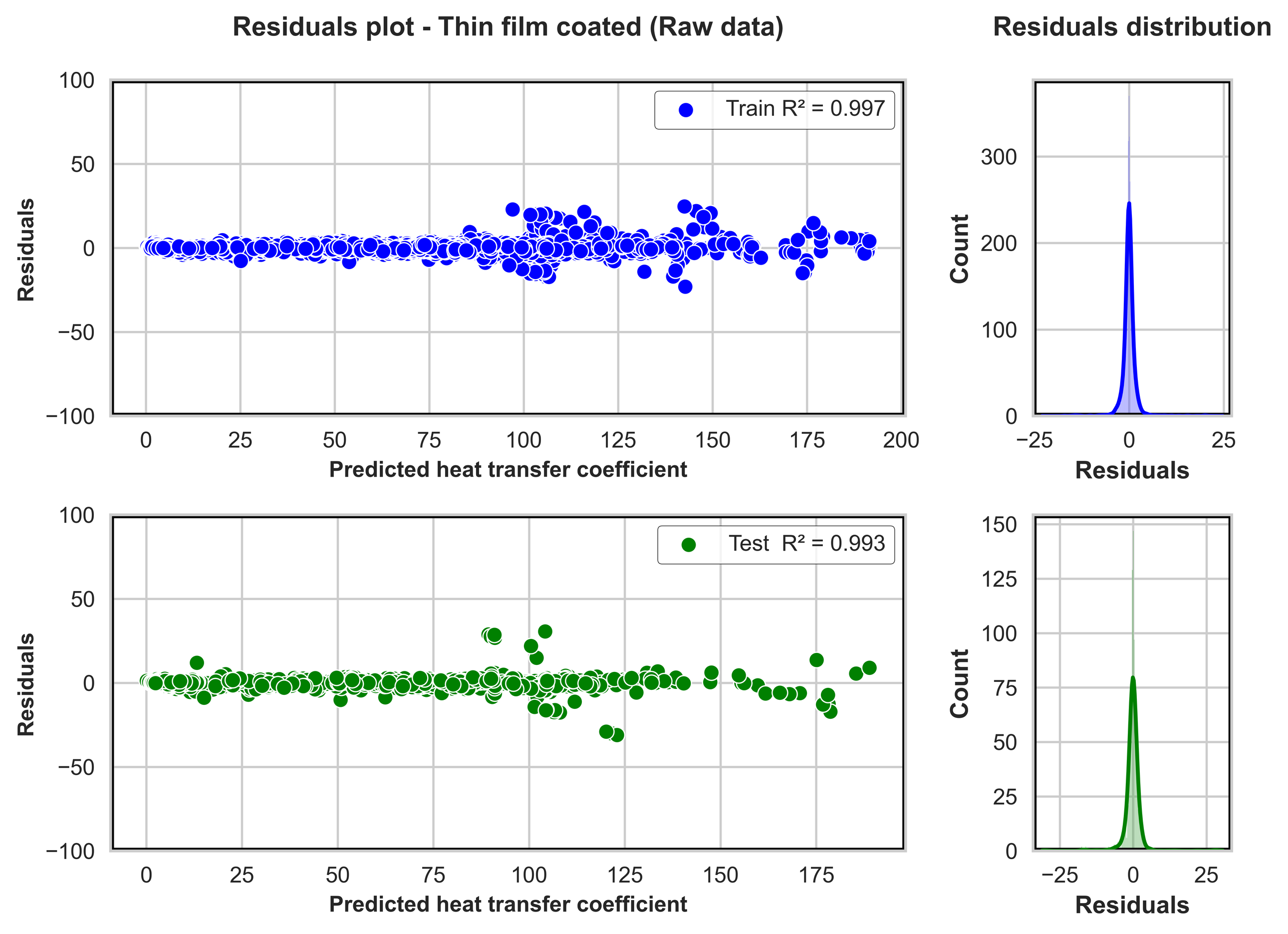}}\label{fig:residual_plot_surface_train_test_raw_data}}
\subfloat[]{{\includegraphics[width=9cm]{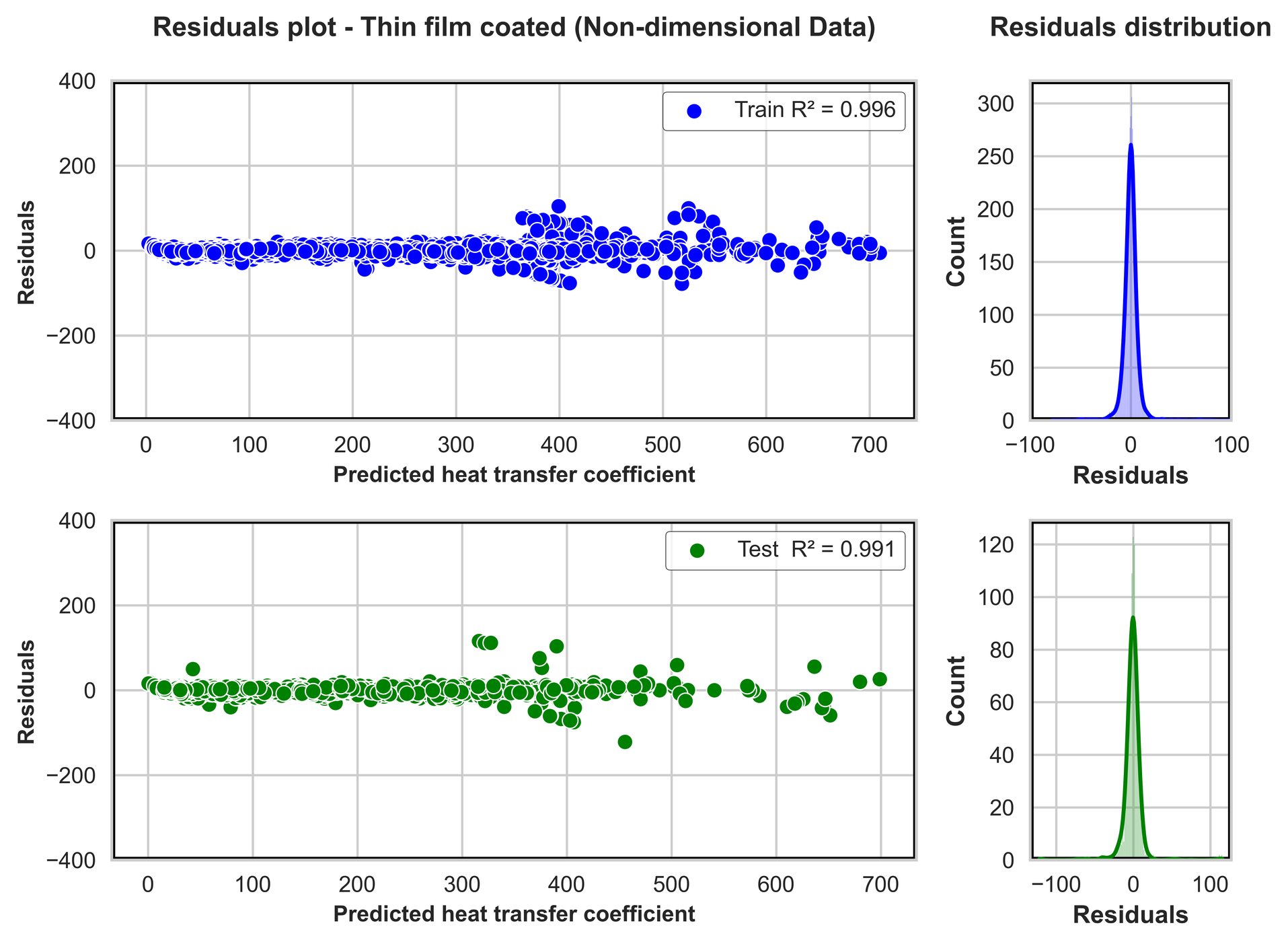}}\label{fig:residual_plot_surface_train_test_nd_data}}\\
\caption{Residual plot of CatBoost model for thin film-coated substrates for (a) Raw data and (b) Non-dimensional data.}
\end{figure}

\begin{figure}[H]
\centering
\subfloat[]{{\includegraphics[width=9cm]{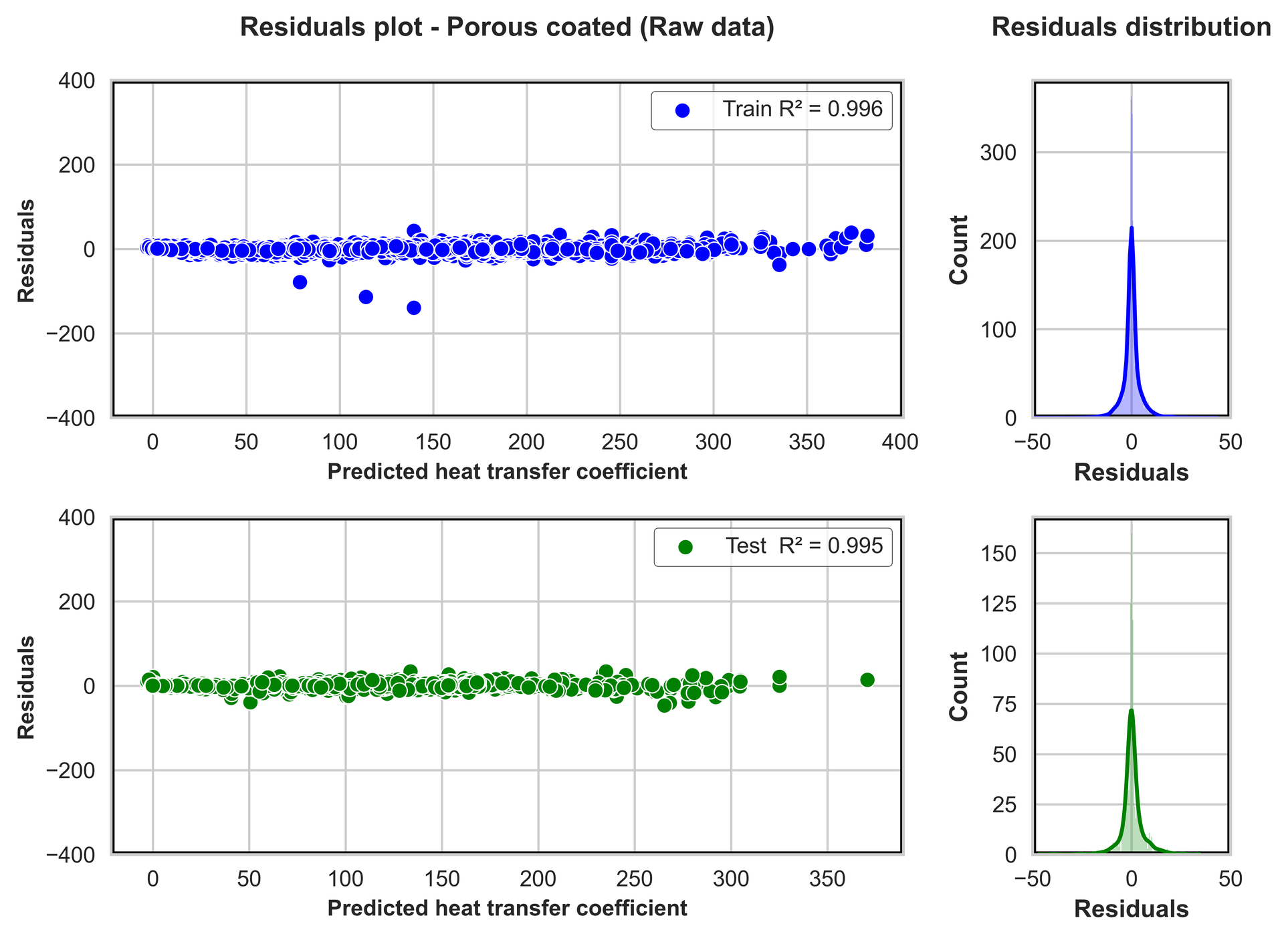}}\label{fig:residual_plot_porous_train_test_raw_data}}
\subfloat[]{{\includegraphics[width=9cm]{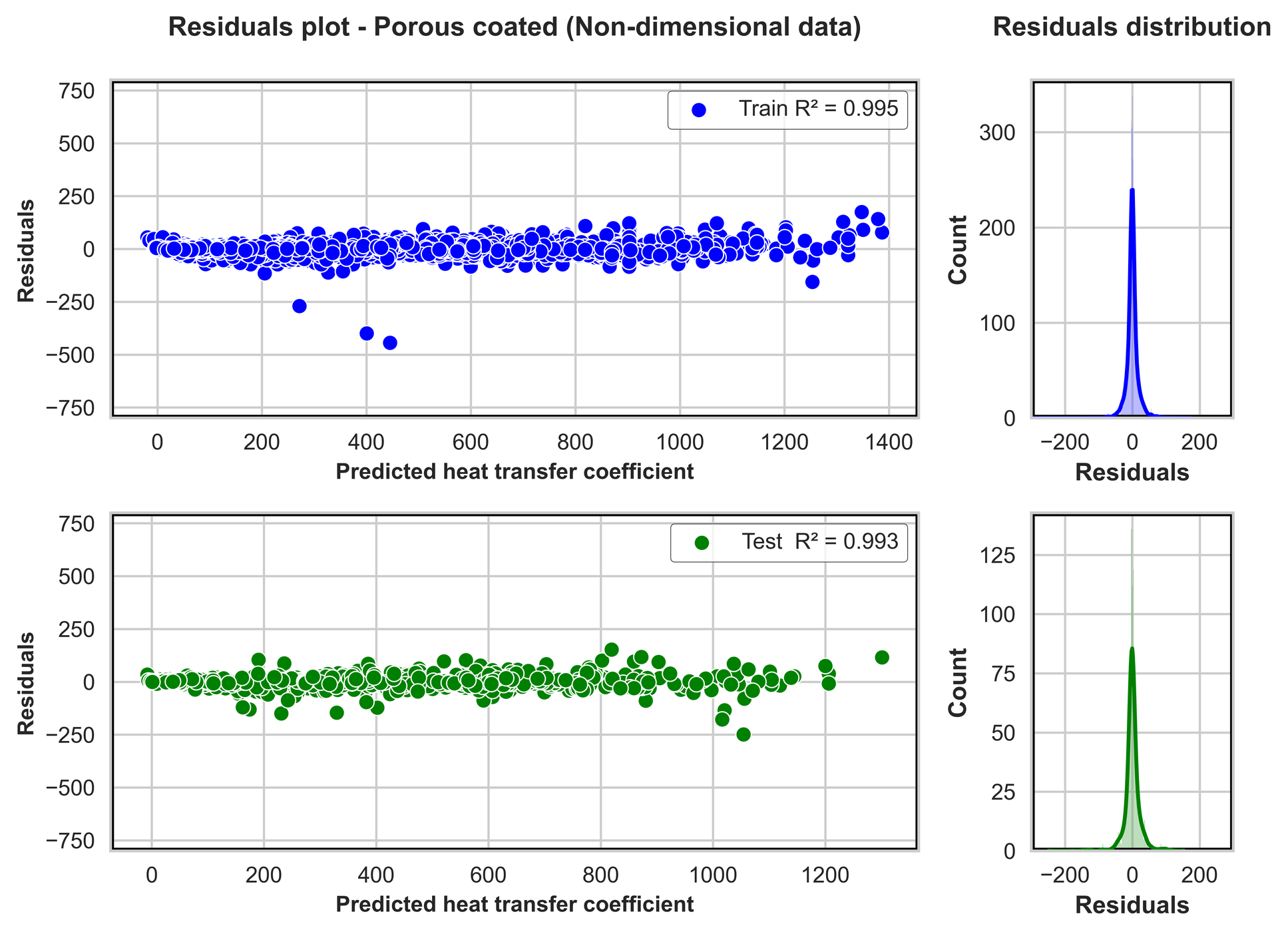}}\label{fig:residual_plot_porous_train_test_nd_data}}\\
\caption{Residual plot of CatBoost model for porous-coated substrates for (a) Raw data and (b) Non-dimensional data.}
\end{figure}

From the assessment of the models based on the performance metrics, CatBoost, Extra Trees, Extreme Gradient Boosting, Random Forest, Light Gradient Boosting Machine, Decision Tree, and Gradient Boosting were found to display better performance for thin film-coated and porous-coated substrates. The above models were then fine-tuned by hyperparameter optimization, and their performance was determined based on the results of cross-validation. Markedly, the CatBoost Regressor showed the best performance across all the considered metrics after hyperparameter optimization and 10-fold cross-validation for both thin film-coated and porous-coated substrates. Figures \ref{fig:residual_plot_surface_train_test_raw_data} and \ref{fig:residual_plot_surface_train_test_nd_data} show the training and testing performance of the optimized CatBoost Model for the raw data and the non-dimensional data of the thin-film coated substrates and Figs. \ref{fig:residual_plot_porous_train_test_raw_data} and  \ref{fig:residual_plot_porous_train_test_nd_data} show the same for the porous-coated substrates. The fine-tuned hyperparameters for the CatBoost model are detailed in Table \ref{tab: Hyperparameters employed in the Catboost Model after hyperparameter optimization}.

\begin{table}[H]
    \centering
    \caption{Comparison of regression models for thin film-coated substrates.}
    \label{tab:Comparison of Regression Models for Surface-Coated Substrates}
    \begin{adjustbox}{max width=\textwidth}
    \renewcommand{\arraystretch}{1.5} 
    \begin{tabular}{llcccccccc}
        \hline
        \multirow{2}{*}{\textbf{Model}} & \multicolumn{4}{c}{\textbf{Raw Data}} & \hspace{1em} & \multicolumn{4}{c}{\textbf{ND Data}} \\ \cline{2-5} \cline{7-10}
        & \textbf{R\textsuperscript{2}} & \textbf{MAE} & \textbf{RMSE} & \textbf{MAPE} & & \textbf{R\textsuperscript{2}} & \textbf{MAE} & \textbf{RMSE} & \textbf{MAPE} \\ \hline
        CatBoost Regressor & 0.993 & 1.539 & 3.525 & 0.073 & & 0.991 & 6.780 & 13.414 & 0.070 \\ 
        Extra Trees Regressor & 0.992 & 0.870 & 3.640 & 0.025 & & 0.991 & 3.734 & 13.680 & 0.027 \\ 
        Extreme Gradient Boosting & 0.991 & 1.569 & 3.972 & 0.061 & & 0.988 & 6.698 & 15.337 & 0.065 \\ 
        Random Forest Regressor & 0.991 & 1.201 & 3.874 & 0.039 & & 0.986 & 5.384 & 16.571 & 0.053 \\ 
        Light Gradient Boosting Machine & 0.991 & 1.957 & 4.116 & 0.104 & & 0.984 & 8.610 & 17.760 & 0.105 \\ 
        Decision Tree Regressor & 0.984 & 1.428 & 5.027 & 0.049 & & 0.975 & 6.014 & 21.558 & 0.055 \\ 
        Gradient Boosting Regressor & 0.968 & 4.798 & 7.597 & 0.306 & & 0.947 & 22.069 & 33.414 & 0.281 \\ 
        K Neighbors Regressor & 0.952 & 3.445 & 9.255 & 0.183 & & 0.943 & 14.222 & 34.487 & 0.169 \\ 
        AdaBoost Regressor & 0.834 & 15.224 & 17.354 & 2.163 & & 0.827 & 48.246 & 60.548 & 0.844 \\ 
        Linear Regression & 0.529 & 19.368 & 29.326 & 1.359 & & 0.533 & 66.364 & 99.805 & 0.818 \\ 
        Ridge Regression & 0.515 & 19.707 & 29.748 & 1.287 & & 0.517 & 69.915 & 101.550 & 0.898 \\ 
        Bayesian Ridge & 0.514 & 19.778 & 29.800 & 1.294 & & 0.516 & 70.059 & 101.595 & 0.903 \\ 
        Lasso Regression & 0.507 & 20.349 & 30.020 & 1.420 & & 0.513 & 71.467 & 101.978 & 0.965 \\ 
        Lasso Least Angle Regression & 0.507 & 20.348 & 30.020 & 1.420 & & 0.513 & 71.501 & 101.987 & 0.965 \\ 
        Huber Regressor & 0.489 & 19.256 & 30.538 & 1.142 & & 0.510 & 68.802 & 102.245 & 0.849 \\ 
        Elastic Net & 0.484 & 21.106 & 30.704 & 1.628 & & 0.499 & 71.463 & 103.473 & 0.937 \\ 
        Orthogonal Matching Pursuit & 0.425 & 22.229 & 32.418 & 1.668 & & 0.478 & 77.338 & 105.589 & 1.220 \\ 
        Passive Aggressive Regressor & 0.319 & 25.149 & 35.149 & 2.018 & & 0.264 & 94.100 & 125.323 & 1.673 \\ 
        Dummy Regressor & -0.006 & 37.395 & 42.872 & 4.681 & & -0.005 & 124.437 & 146.470 & 2.115 \\ \hline
    \end{tabular}
    \end{adjustbox}
\end{table}

\begin{table}[H]
    \centering
    \caption{Comparison of regression models for porous-coated substrates.}
    \label{tab:Comparison of Regression Models for Porous-Coated Substrates}
    \begin{adjustbox}{max width=\textwidth}
    \renewcommand{\arraystretch}{1.5} 
    \begin{tabular}{llllllllll}
        \hline
        \multirow{2}{*}{\textbf{Model}} & \multicolumn{4}{c}{\textbf{Raw Data}} & \hspace{1em} & \multicolumn{4}{c}{\textbf{ND Data}} \\ \cline{2-5} \cline{7-10}
        & \textbf{R\textsuperscript{2}} & \textbf{MAE} & \textbf{RMSE} & \textbf{MAPE} & & \textbf{R\textsuperscript{2}} & \textbf{MAE} & \textbf{RMSE} & \textbf{MAPE} \\ \hline
        CatBoost Regressor & 0.989 & 3.984 & 8.538 & 0.629 & & 0.987 & 16.404 & 34.329 & 0.608 \\ 
        Extreme Gradient Boosting & 0.989 & 3.718 & 8.758 & 0.594 & & 0.985 & 16.031 & 37.053 & 0.542 \\ 
        Extra Trees Regressor & 0.988 & 2.795 & 8.904 & 0.494 & & 0.984 & 12.488 & 37.515 & 0.499 \\ 
        Random Forest Regressor & 0.987 & 3.347 & 9.197 & 0.513 & & 0.983 & 19.296 & 38.782 & 0.655 \\ 
        Light Gradient Boosting Machine & 0.986 & 4.800 & 9.904 & 0.732 & & 0.983 & 14.027 & 38.429 & 0.518 \\ 
        K Neighbors Regressor & 0.978 & 5.357 & 12.512 & 0.554 & & 0.981 & 18.936 & 41.632 & 0.564 \\ 
        Decision Tree Regressor & 0.977 & 4.140 & 12.260 & 0.518 & & 0.975 & 15.753 & 46.959 & 0.493 \\ 
        Gradient Boosting Regressor & 0.941 & 11.928 & 20.966 & 1.322 & & 0.929 & 49.373 & 81.775 & 1.292 \\ 
        AdaBoost Regressor & 0.792 & 33.900 & 39.185 & 17.250 & & 0.779 & 119.696 & 143.966 & 7.451 \\ 
        Linear Regression & 0.723 & 30.186 & 45.267 & 7.795 & & 0.658 & 118.954 & 179.700 & 2.336 \\ 
        Ridge Regression & 0.700 & 30.708 & 47.144 & 7.351 & & 0.641 & 120.140 & 184.097 & 2.869 \\ 
        Bayesian Ridge & 0.699 & 30.678 & 47.189 & 6.920 & & 0.640 & 119.983 & 184.307 & 2.775 \\ 
        Lasso Regression & 0.692 & 30.548 & 47.767 & 3.900 & & 0.636 & 116.883 & 185.457 & 2.086 \\ 
        Lasso Least Angle Regression & 0.692 & 30.549 & 47.767 & 3.901 & & 0.636 & 117.004 & 185.459 & 2.142 \\ 
        Huber Regressor & 0.687 & 29.112 & 48.141 & 4.366 & & 0.633 & 112.884 & 186.326 & 1.513 \\ 
        Elastic Net & 0.667 & 31.805 & 49.693 & 4.440 & & 0.627 & 119.158 & 187.817 & 2.415 \\ 
        Passive Aggressive Regressor & 0.658 & 33.580 & 50.319 & 9.735 & & 0.619 & 117.684 & 189.602 & 1.492 \\ 
        Orthogonal Matching Pursuit & 0.642 & 31.338 & 51.462 & 2.104 & & 0.619 & 122.684 & 189.825 & 2.903 \\ 
        Dummy Regressor & -0.001 & 75.534 & 86.138 & 41.225 & & -0.001 & 268.201 & 307.619 & 15.771 \\ \hline
    \end{tabular}
    \end{adjustbox}
\end{table}

\begin{figure}[H]
\centering
\subfloat[]{{\includegraphics[width=9cm]{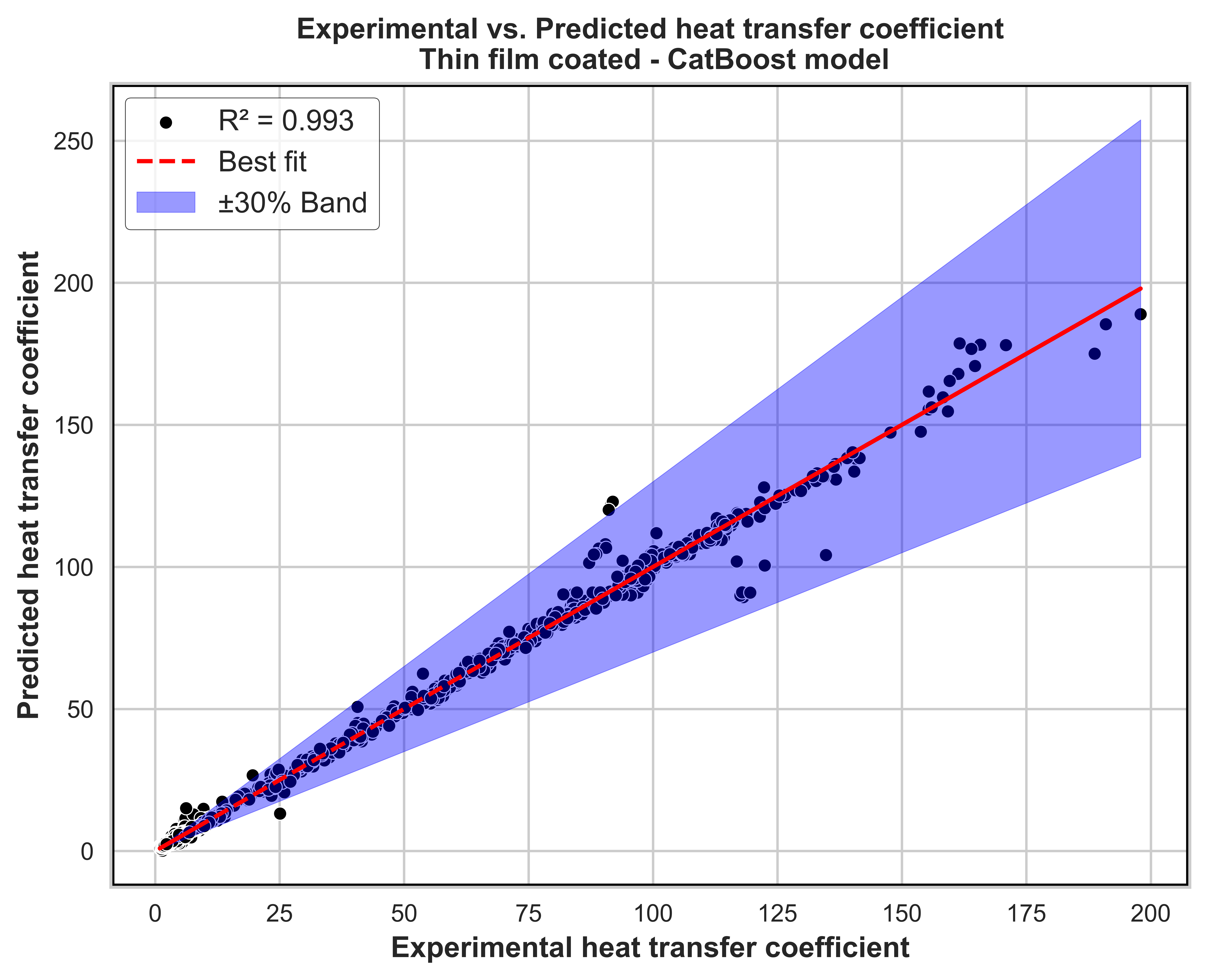}}\label{surface_coated_catboost_model_parity_plot_raw_data}}
\subfloat[]{{\includegraphics[width=9cm]{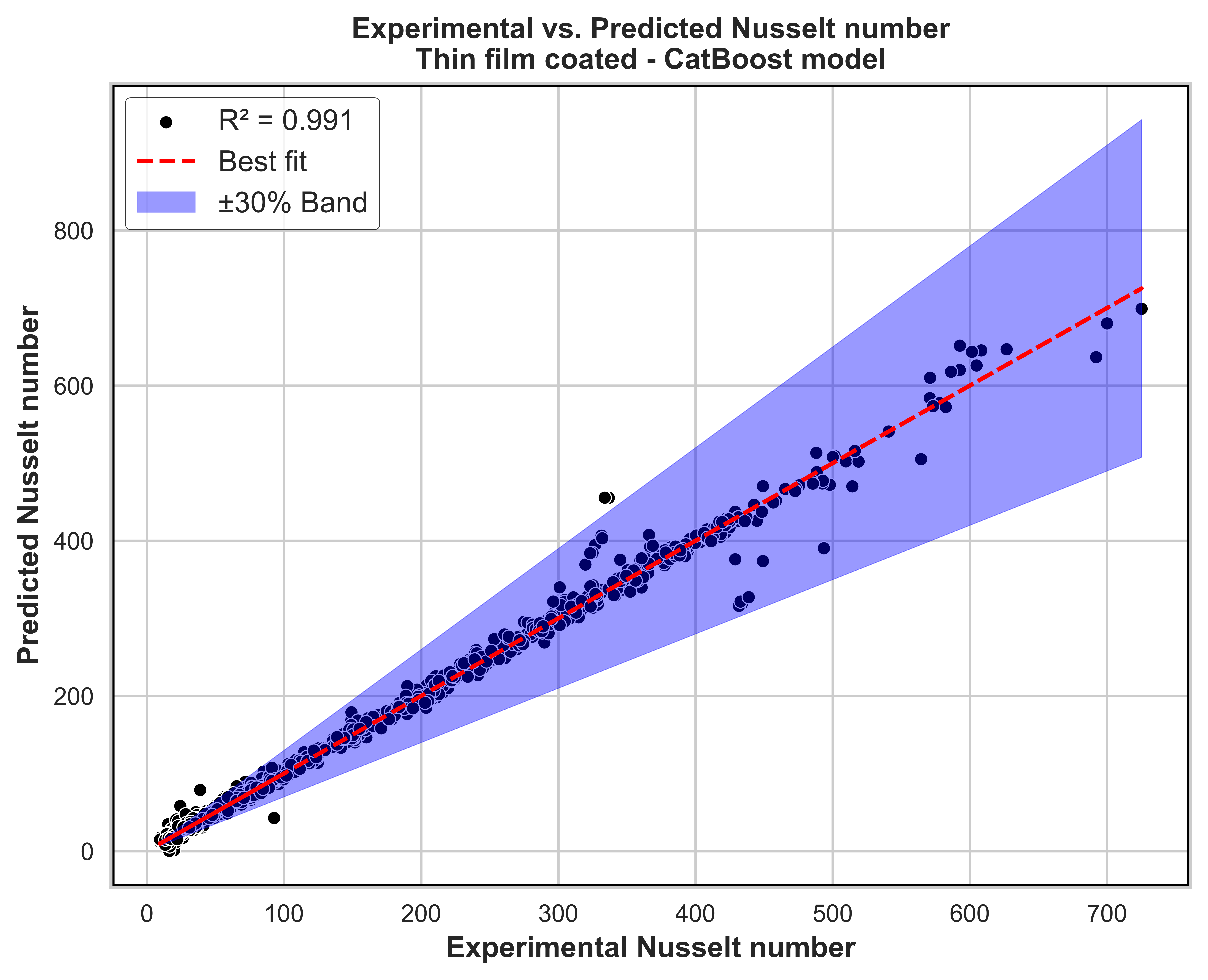}}\label{fig:surface_coated_catboost_model_nd_parity_plot}}\\
\caption{CatBoost model prediction for (a) Heat transfer coefficient and (b) Nusselt number on thin film-coated substrates.}
\end{figure}

\begin{figure}[H]
\centering
\subfloat[]{{\includegraphics[width=9cm]{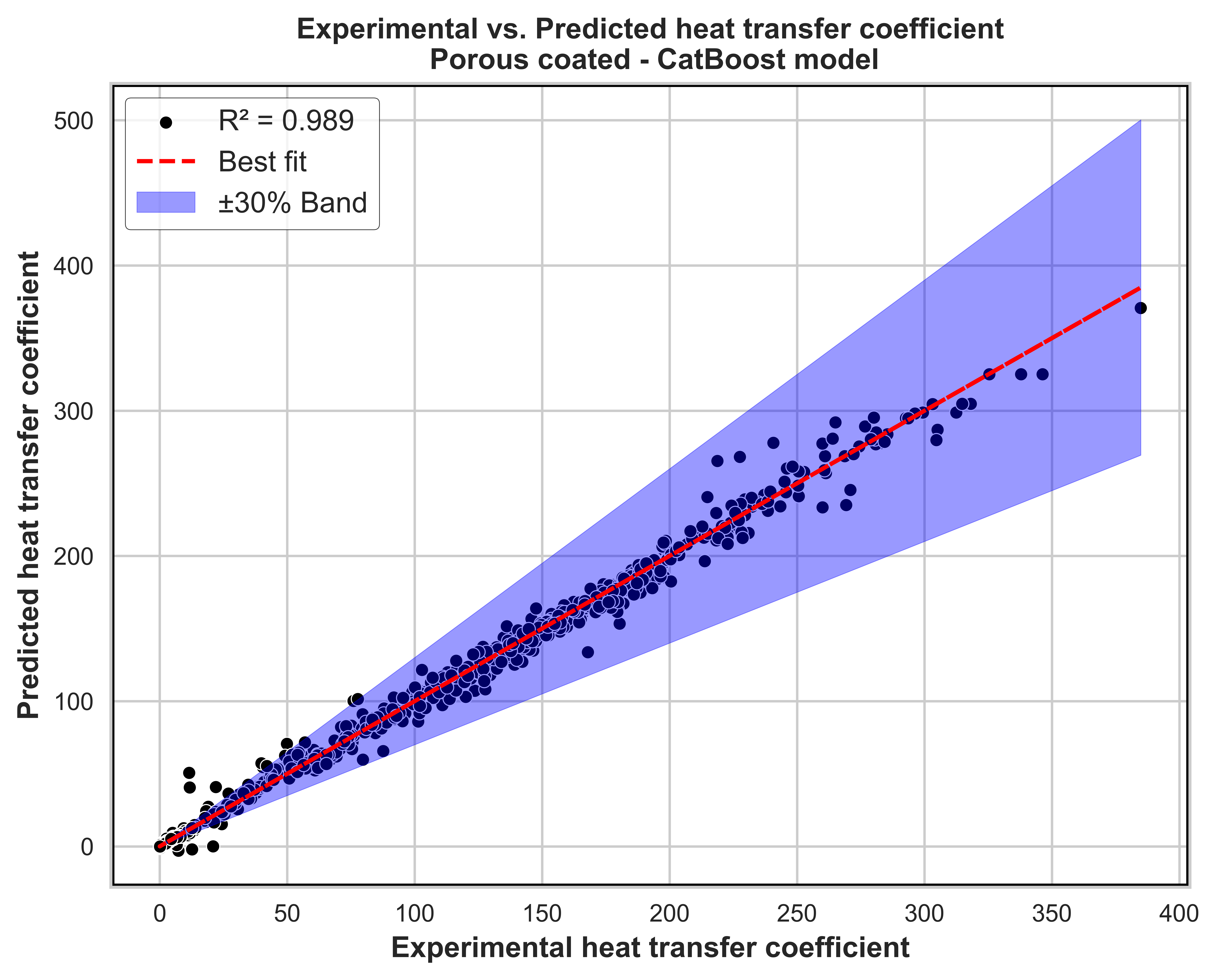}}\label{porous_coated_catboost_model_parity_plot_raw_data}}
\subfloat[]{{\includegraphics[width=9cm]{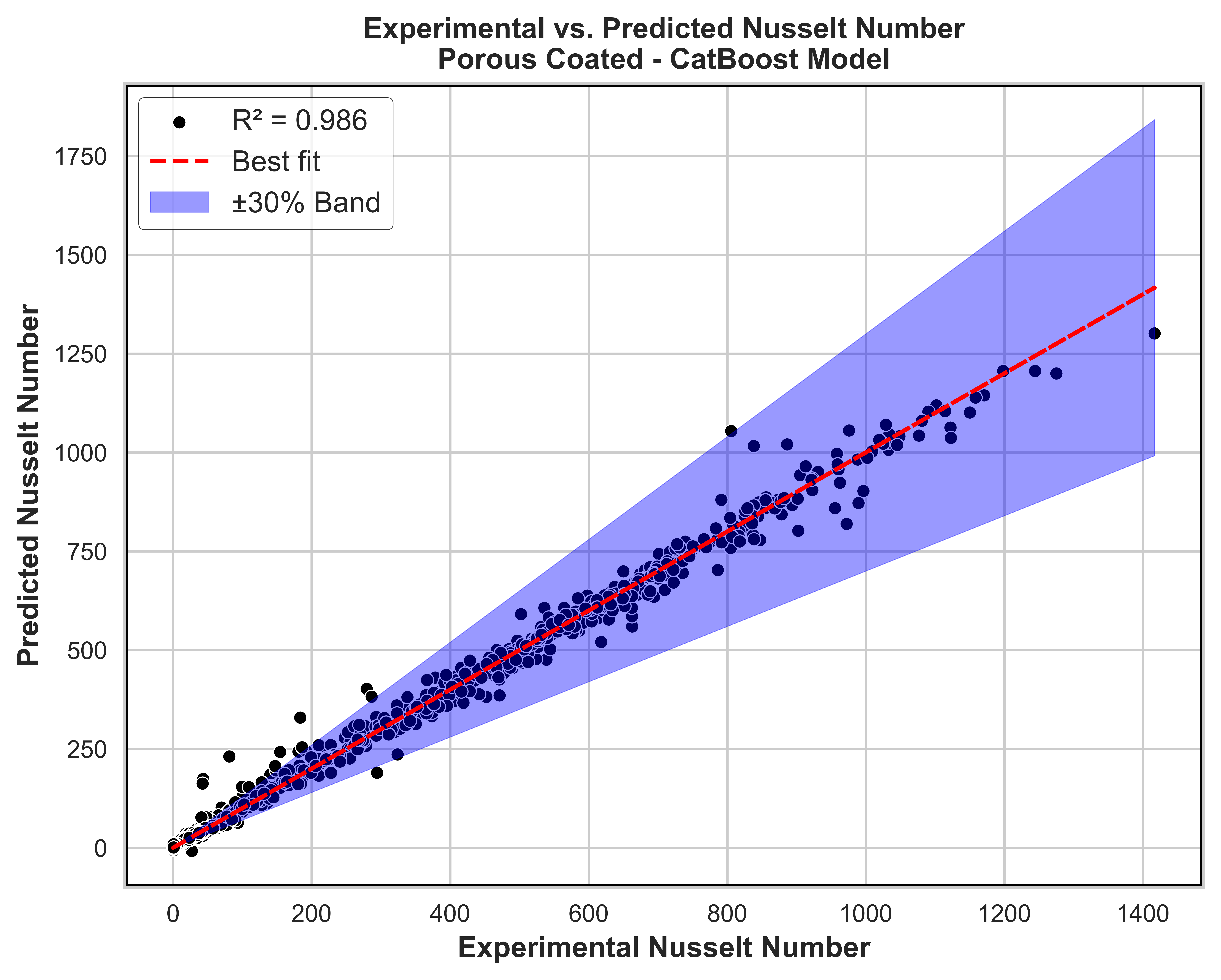}}\label{fig:porous_coated_catboost_model_nd_parity_plot}}\\
\caption{CatBoost model prediction for (a) Heat transfer coefficient and (b) Nusselt number on porous-coated substrates.}
\end{figure}

Table \ref{tab: Machine Learning Model Performance Comparison} demonstrates the predictive performance of the optimized ML models for the overall dataset, and separately for water and fluids other than water, for raw and non-dimensional data, for both substrates. The water dataset includes 3421 data points for thin film-coated and 3087 data points for porous-coated, and the dataset for fluids other than water includes 1823 data points for thin film-coated and 2055 data points for porous-coated. The optimal feature selection of appropriate surface characteristics such as coating resistance, the thermal conductivity of coating and substrate, coating thickness, contact angle, surface roughness, and porosity, in addition to the operating conditions and thermophysical properties, have led to the identification of the underlying interactions between variables. In spite of the high-dimensional dataset of coating substrates, CatBoost emerges as the robust and reliable model for HTC prediction on coated surfaces due to the combined effect of regularization and sequential residual modeling. Figures \ref{surface_coated_catboost_model_parity_plot_raw_data} and \ref{fig:surface_coated_catboost_model_nd_parity_plot} show parity plots of the predicted data versus experimental data for the heat transfer coefficient and the Nusselt number on the thin film-coated substrates, respectively and Figs. \ref{porous_coated_catboost_model_parity_plot_raw_data}, \ref{fig:porous_coated_catboost_model_nd_parity_plot} shows the same on the porous-coated substrates. It can be seen that the CatBoost model predicts the heat transfer characteristics with R2 value around 0.99.

\begin{table}[H]
    \centering
    \caption{Hyperparameters employed in the catboost model after hyperparameter optimization.}
    \label{tab: Hyperparameters employed in the Catboost Model after hyperparameter optimization}
    \begin{adjustbox}{max width=\textwidth}
    \renewcommand{\arraystretch}{1.5} 
    \begin{tabular}{lll}
        \hline
        \textbf{Hyperparameters} & \textbf{Values} & \textbf{Description} \\ \hline
        iterations & 1000 & Number of boosting trees built in the model. \\
        subsample & 0.8 & Fraction of data used for each tree. \\
        depth & 6 & Maximum depth of a tree in the model. \\
        min\_data\_in\_leaf & 1 & Minimum number of samples in the leaf node. \\
        max\_leaves & 64 & Maximum number of leaves in a tree. \\
        learning\_rate & 0.0514 & Rate at which the model learns. \\
        score\_function & Cosine & Measures the quality of a split. \\
        leaf\_estimation\_method & Newton & Adopted method to find the value of each leaf node. \\ \hline
    \end{tabular}
    \end{adjustbox}
\end{table}

\begin{table}[H]
    \centering
    \caption{Performance comparison of machine learning models.}
    \label{tab: Machine Learning Model Performance Comparison}
    \begin{adjustbox}{max width=\textwidth}
    \renewcommand{\arraystretch}{1.5} 
    \begin{tabular}{llllllll}
        \hline
        \multirow{2}{*}{\textbf{Type of surface}} & \multirow{2}{*}{\textbf{Cases}} & \multirow{2}{*}{\textbf{Dataset}} & \multirow{2}{*}{\textbf{Model}} & \multicolumn{4}{c}{\textbf{Optimized Model Metrics}} \\ \cline{5-8}
        & & & & \textbf{R\textsuperscript{2}} & \textbf{MAE} & \textbf{RMSE} & \textbf{MAPE} \\ \hline
        \multirow{6}{*}{\textbf{Thin film-coated}} & \multirow{3}{*}{\textbf{Raw Data}} & Overall dataset & CatBoost & 0.993 & 1.539 & 3.525 & 0.073 \\ 
        & & Water dataset & CatBoost & 0.988 & 2.143 & 4.541 & 0.055 \\ 
        & & Other Fluids dataset & CatBoost & 0.988 & 0.181 & 0.311 & 0.041 \\ 
        \cline{2-8}
        & \multirow{3}{*}{\textbf{ND Data}} & Overall dataset & CatBoost & 0.991 & 6.780 & 13.414 & 0.070 \\ 
        & & Water dataset & CatBoost & 0.988 & 7.978 & 16.766 & 0.058 \\ 
        & & Other Fluids dataset & Random Forest & 0.988 & 2.334 & 4.008 & 0.047 \\ 
        & & \textit{Other Fluids dataset} & \textit{CatBoost} & \textit{0.968} & \textit{4.982} & \textit{7.042} & \textit{0.100} \\
        \hline
        \multirow{6}{*}{\textbf{Porous-coated}} & \multirow{3}{*}{\textbf{Raw Data}} & Overall dataset & CatBoost & 0.989 & 3.984 & 8.538 & 0.629 \\ 
        & & Water dataset & CatBoost & 0.970 & 6.286 & 11.837 & 0.752 \\ 
        & & Other Fluids dataset & CatBoost & 0.994 & 0.145 & 0.222 & 0.071 \\ 
        \cline{2-8}
        & \multirow{3}{*}{\textbf{ND Data}} & Overall dataset & CatBoost & 0.987 & 16.404 & 34.329 & 0.608 \\ 
        & & Water dataset & LightGBM & 0.963 & 26.347 & 48.762 & 0.771 \\ 
        & & \textit{Water dataset} & \textit{CatBoost} & \textit{0.949} & \textit{22.353} & \textit{50.892} & \textit{0.793} \\ 
        & & Other Fluids dataset & CatBoost & 0.995 & 1.545 & 2.362 & 0.079 \\ 
        \hline
    \end{tabular}
    \end{adjustbox}
\end{table}

\subsection{Model interpretation through SHAP analysis}

Tables \ref{tab: Overall Significant Variables Influencing Heat Transfer} and \ref{tab: Significant Variables Influencing Heat Transfer for Water and Other Fluids} show the significant dimensional and non-dimensional variables affecting heat transfer in nucleate pool boiling on coated substrates. The SHAP summary plot shows the trend of a feature, and the influencing variables are ordered in decreasing order of the mean SHAP value. When the feature SHAP value changes from blue on the left to red on the right, it has a positive effect on the target, and if the value changes from red to blue, it has a negative effect. The trend of a feature that is not clearly interpretable is also depicted by the SHAP value plot and the SHAP dependency plot. The following subsections present the observations that can be made from the mean SHAP values of the optimized model for the thin film-coated and porous-coated substrates. 

\begin{table}[H]
\centering
\caption{Overall significant variables influencing heat transfer in thin film-coated and porous-coated substrates.}
\label{tab: Overall Significant Variables Influencing Heat Transfer}
\begin{adjustbox}{max width=\textwidth}
\begin{tabular}{@{}llllll@{}}
\toprule
\multicolumn{2}{c}{\textbf{Thin film-coated}} & \multicolumn{2}{c}{\textbf{Porous-coated}} \\ 
\cmidrule(lr){1-2} \cmidrule(lr){3-4}
\hspace{10pt} \textbf{Raw Data} & \hspace{10pt} \textbf{ND Data} & \hspace{10pt} \textbf{Raw Data} & \hspace{10pt} \textbf{ND Data} \\ 
\midrule
\hspace{10pt} $\theta$ & \hspace{10pt} $\theta / 90^\circ$ & \hspace{10pt} $C_{pl}$ & \hspace{10pt} $R_{co} / R_{sll}$ \\

\hspace{10pt} $R_{co}$ & \hspace{10pt} $R_{co} / R_{sll}$ & \hspace{10pt} $k_{l}$ & \hspace{10pt} $k_{w} / k_{l}$ \\

\hspace{10pt} $T_{w}$ & \hspace{10pt} $R_{q} / r_{cav}$ & \hspace{10pt} $R_{q}$ & \hspace{10pt} $Pr_{l}$ \\

\hspace{10pt} $k_{l}$ & \hspace{10pt} $k_{w} / k_{co}$ & \hspace{10pt} $\theta$ & \hspace{10pt} $\theta / 90^\circ$ \\

\hspace{10pt} $C_{pl}$ & \hspace{10pt} $k_{w} / k_{l}$ & \hspace{10pt} $\varepsilon$ & \hspace{10pt} $R_{q} / r_{cav}$ \\

\hspace{10pt} $R_{q}$ & \hspace{10pt} $Pr_{l}$ & \hspace{10pt} $R_{co}$ & \hspace{10pt} $k_{l} / k_{v}$ \\

\hspace{10pt} $t$ & \hspace{10pt} $C_{pl} / C_{pv}$ & \hspace{10pt} $\rho_{l}$ & \hspace{10pt} $k_{w} / k_{co}$ \\

\hspace{10pt} $k_{co}$ & & \hspace{10pt} $T_{w}$ & \hspace{10pt} $\varepsilon$ \\

 & & \hspace{10pt} $k_{co}$ & \hspace{10pt} $C_{pl} / C_{pv}$ \\
 
\bottomrule
\end{tabular}
\end{adjustbox}
\end{table}

\begin{table}[H]
\centering
\caption{Significant variables influencing heat transfer for water and fluids other than water.}
\label{tab: Significant Variables Influencing Heat Transfer for Water and Other Fluids}
\begin{adjustbox}{max width=\textwidth}
\begin{tabular}{@{}llllllll@{}}
\toprule
\multicolumn{4}{c}{\textbf{Thin film-coated}} & \multicolumn{4}{c}{\textbf{Porous-coated}} \\ 
\cmidrule(lr){1-4} \cmidrule(lr){5-8}
\multicolumn{2}{c}{\textbf{Raw Data}} & \multicolumn{2}{c}{\textbf{ND Data}} & \multicolumn{2}{c}{\textbf{Raw Data}} & \multicolumn{2}{c}{\textbf{ND Data}} \\ 
\cmidrule(lr){1-2} \cmidrule(lr){3-4} \cmidrule(lr){5-6} \cmidrule(lr){7-8}
\hspace{10pt} \textbf{Water} & \hspace{10pt} \textbf{Other Fluids} & \hspace{10pt} \textbf{Water} & \hspace{10pt} \textbf{Other Fluids} & \hspace{10pt} \textbf{Water} & \hspace{10pt} \textbf{Other Fluids} & \hspace{10pt} \textbf{Water} & \hspace{10pt} \textbf{Other Fluids} \\ 
\midrule
\hspace{10pt} $R_{co}$ & \hspace{10pt} $\triangle T$ & \hspace{10pt} $R_{co} / R_{sll}$ & \hspace{10pt} $k_{w} / k_{l}$ & \hspace{10pt} $\theta$ & \hspace{10pt} $T_{w}$ & \hspace{10pt} $\theta / 90^\circ$ & \hspace{10pt} $k_{l} / k_{v}$ \\

\hspace{10pt} $\theta$ & \hspace{10pt} $R_{q}$ & \hspace{10pt} $\theta / 90^\circ$ & \hspace{10pt} $k_{w} / k_{co}$ & \hspace{10pt} $R_{q}$ & \hspace{10pt} $\triangle T$ & \hspace{10pt} $\varepsilon$ & \hspace{10pt} $Ja_l$ \\

\hspace{10pt} $R_{q}$ & \hspace{10pt} $k_{w}$ & \hspace{10pt} $k_{w} / k_{co}$ & \hspace{10pt} $Pr_v$ & \hspace{10pt} $k_{co}$ & \hspace{10pt} $\varepsilon$ & \hspace{10pt} $R_{q} / r_{cav}$ & \hspace{10pt} $R_{co} / R_{sll}$ \\

\hspace{10pt} t & \hspace{10pt} $R_{co}$ & \hspace{10pt} $R_{q} / r_{cav}$ & \hspace{10pt} $R_{co} / R_{sll}$ & \hspace{10pt} t & \hspace{10pt} $R_{q}$ & \hspace{10pt} $Pr_l$ & \hspace{10pt} $R_{q} / r_{cav}$ \\

\hspace{10pt} $k_{co}$ & \hspace{10pt} t & & \hspace{10pt} $R_{q} / r_{cav}$ & \hspace{10pt} $R_{co}$ & \hspace{10pt} $R_{co}$ & \hspace{10pt} $R_{co} / R_{sll}$ & \hspace{10pt} $\varepsilon$ \\

 & & & \hspace{10pt} $\rho_{l} / \rho_{v}$ & \hspace{10pt} $\varepsilon$ & & & \hspace{10pt} $Pr_v$ \\

 & & & \hspace{10pt} $Ja_l$ & & & & \\
\bottomrule
\end{tabular}
\end{adjustbox}
\end{table}

\begin{figure}[H]
    \centering
  \includegraphics[width=18cm]{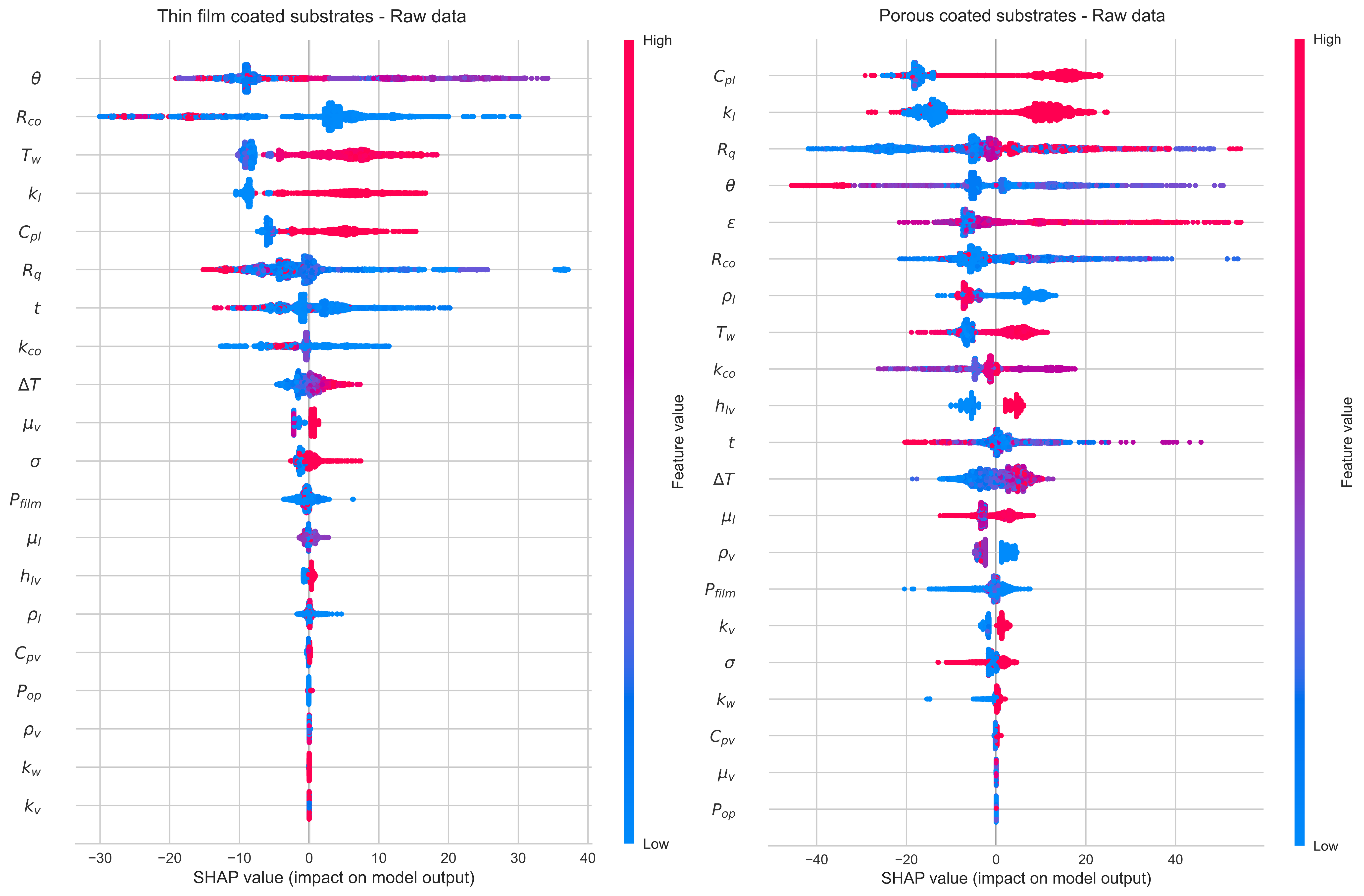}
\caption{SHAP summary plot for thin film-coated and porous-coated substrates on raw data.}
\label{fig: SHAP beeswarm plot for surface and porous coated substrates on Raw Data.}
\end{figure}

\begin{figure}[H]
    \centering
  \includegraphics[width=18cm]{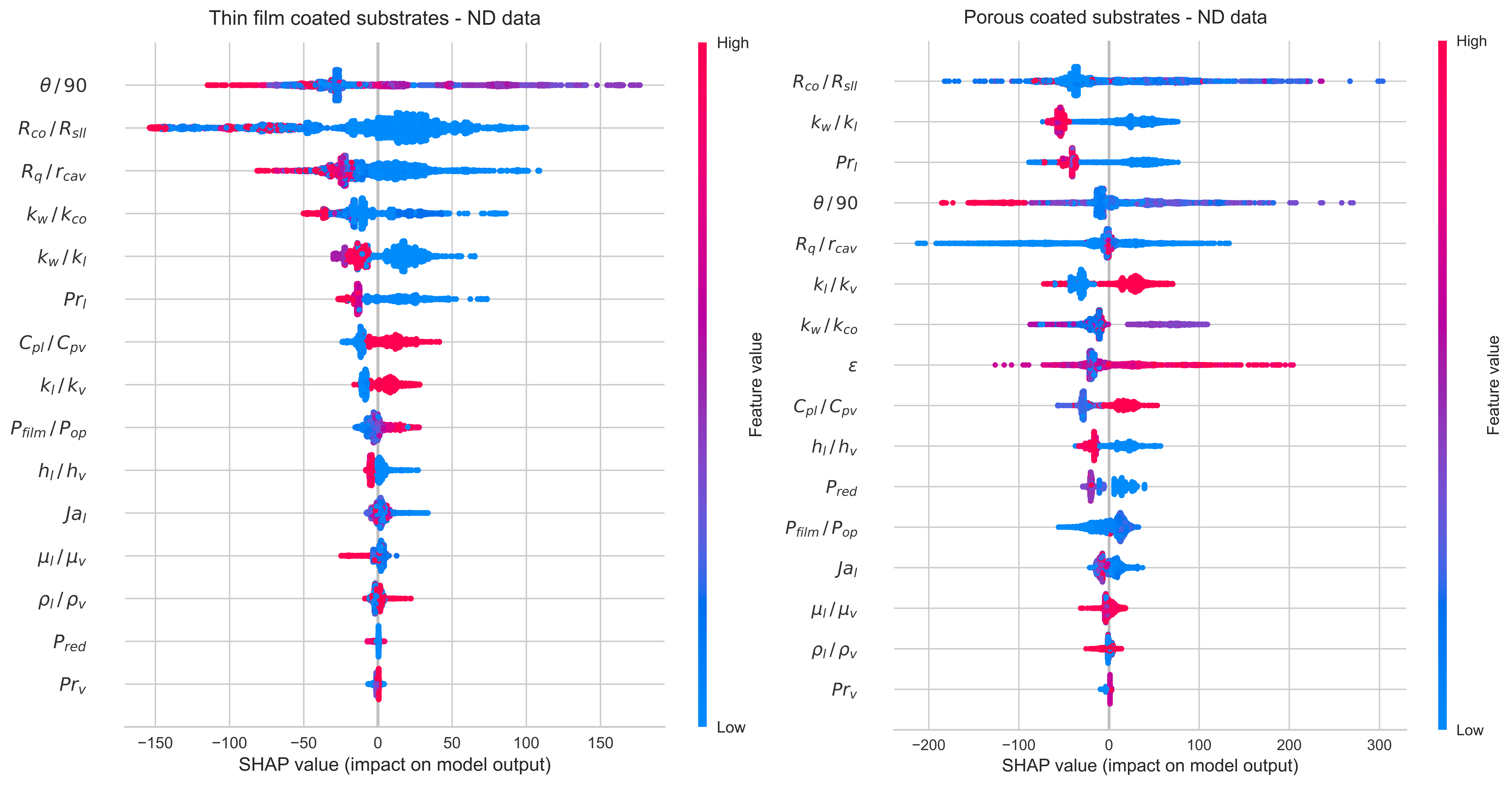}
\caption{SHAP summary plot for thin film-coated and porous-coated substrates on non-dimensional data.}
\label{fig: SHAP beeswarm plot for surface and porous coated substrates on non-dimensional Data.}
\end{figure}

\begin{figure}[H]
\centering
\subfloat[]{{\includegraphics[width=18cm]{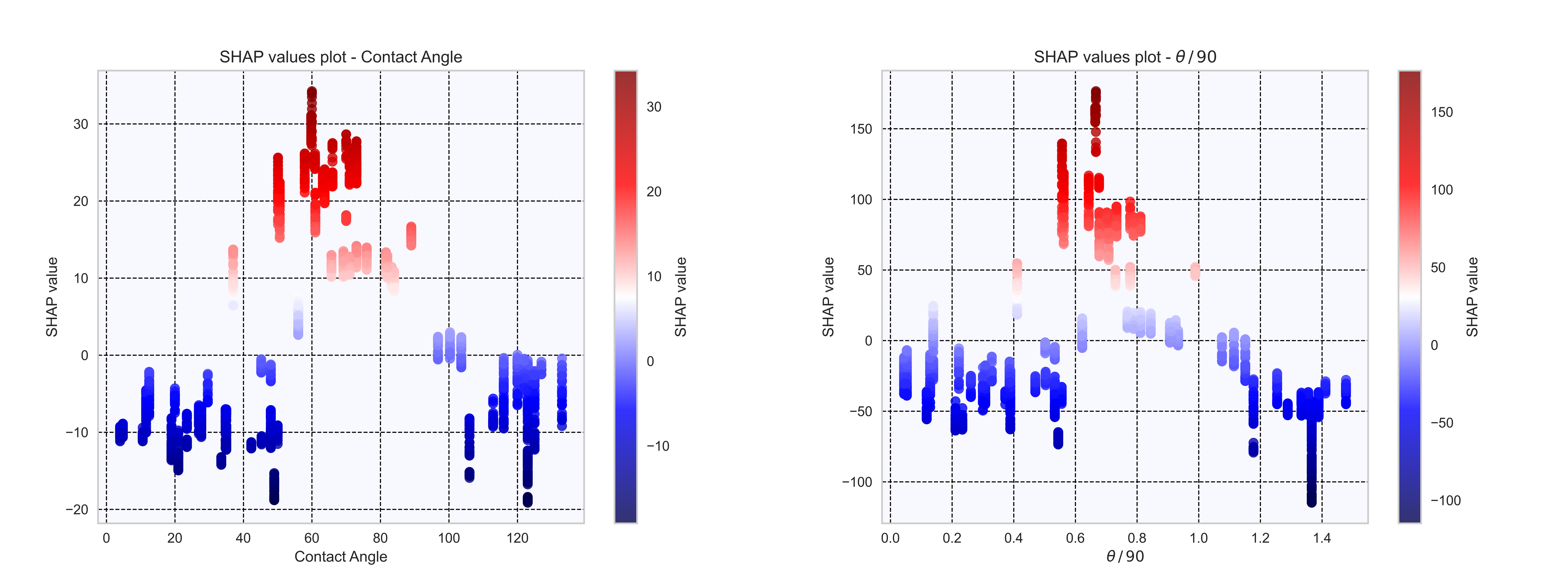}}\label{fig: contact_angle_variation_shap_plot_surface}}\\
\subfloat[]{{\includegraphics[width=18cm]{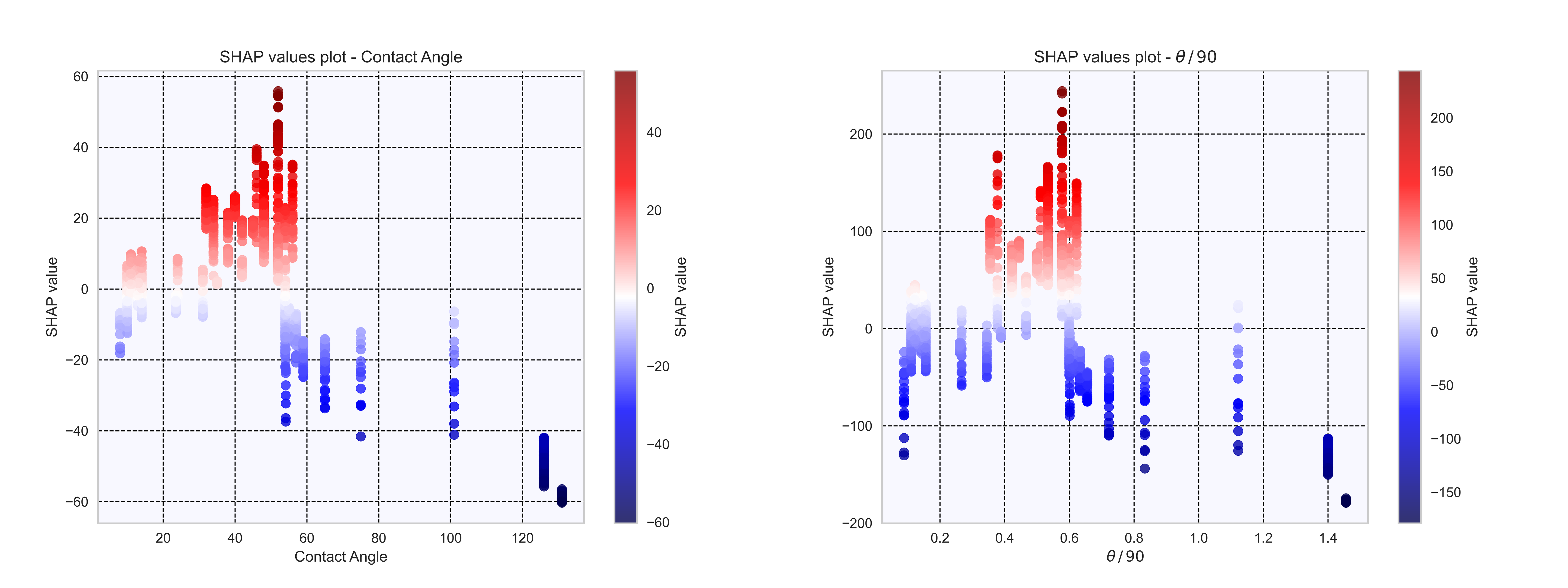}}\label{fig:contact_angle_variation_shap_plot_porous}}\\
\caption{SHAP value plot of contact angle and $\theta/90^\circ$ for (a) thin film-coated substrates and (b) porous-coated substrates.}
\end{figure} 

\begin{figure}[H]
\centering
\subfloat[]{{\includegraphics[width=9cm]{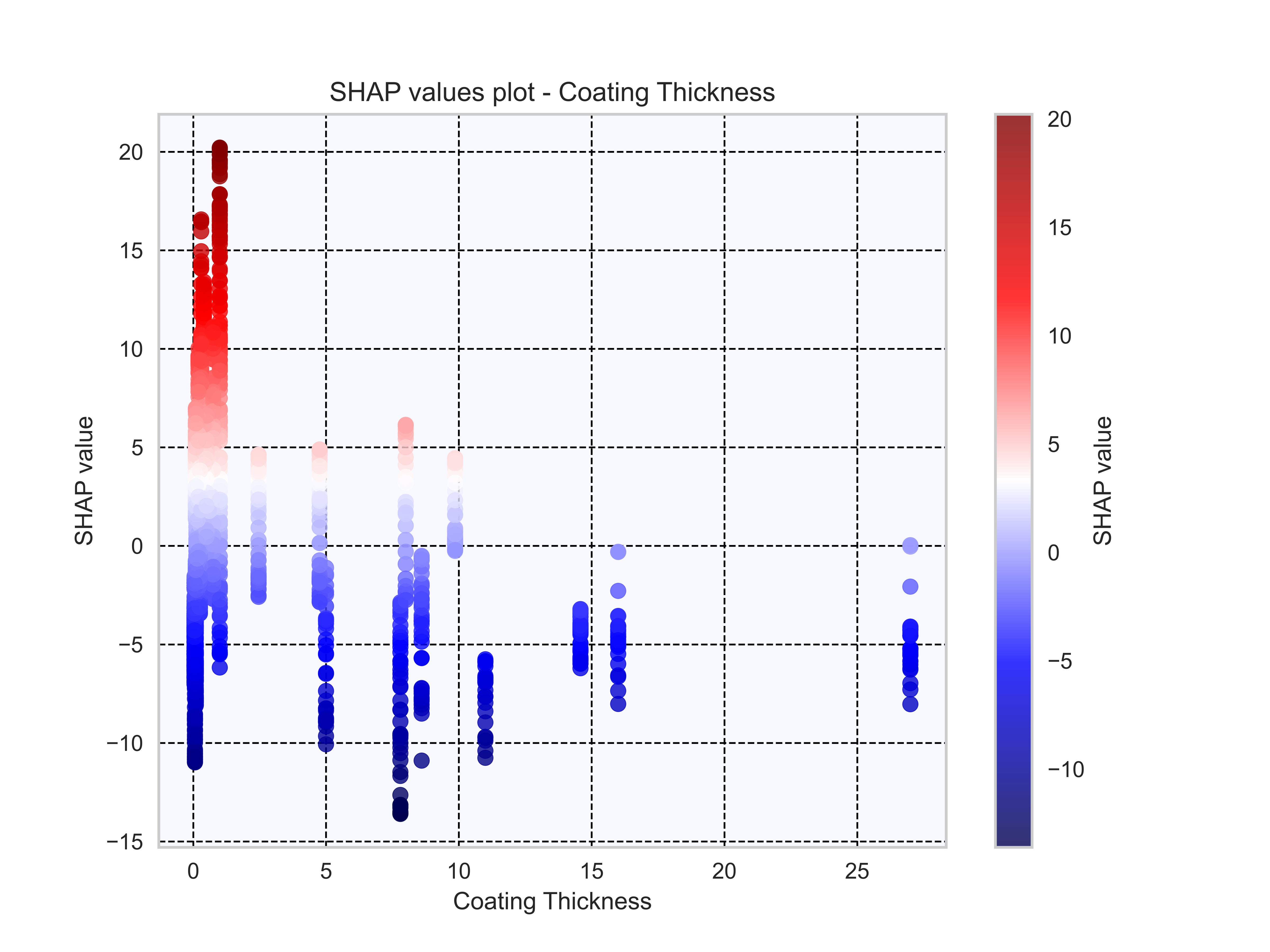}}\label{fig: coating_thickness_surface_raw_data_shap_plot}}
\subfloat[]{{\includegraphics[width=9cm]{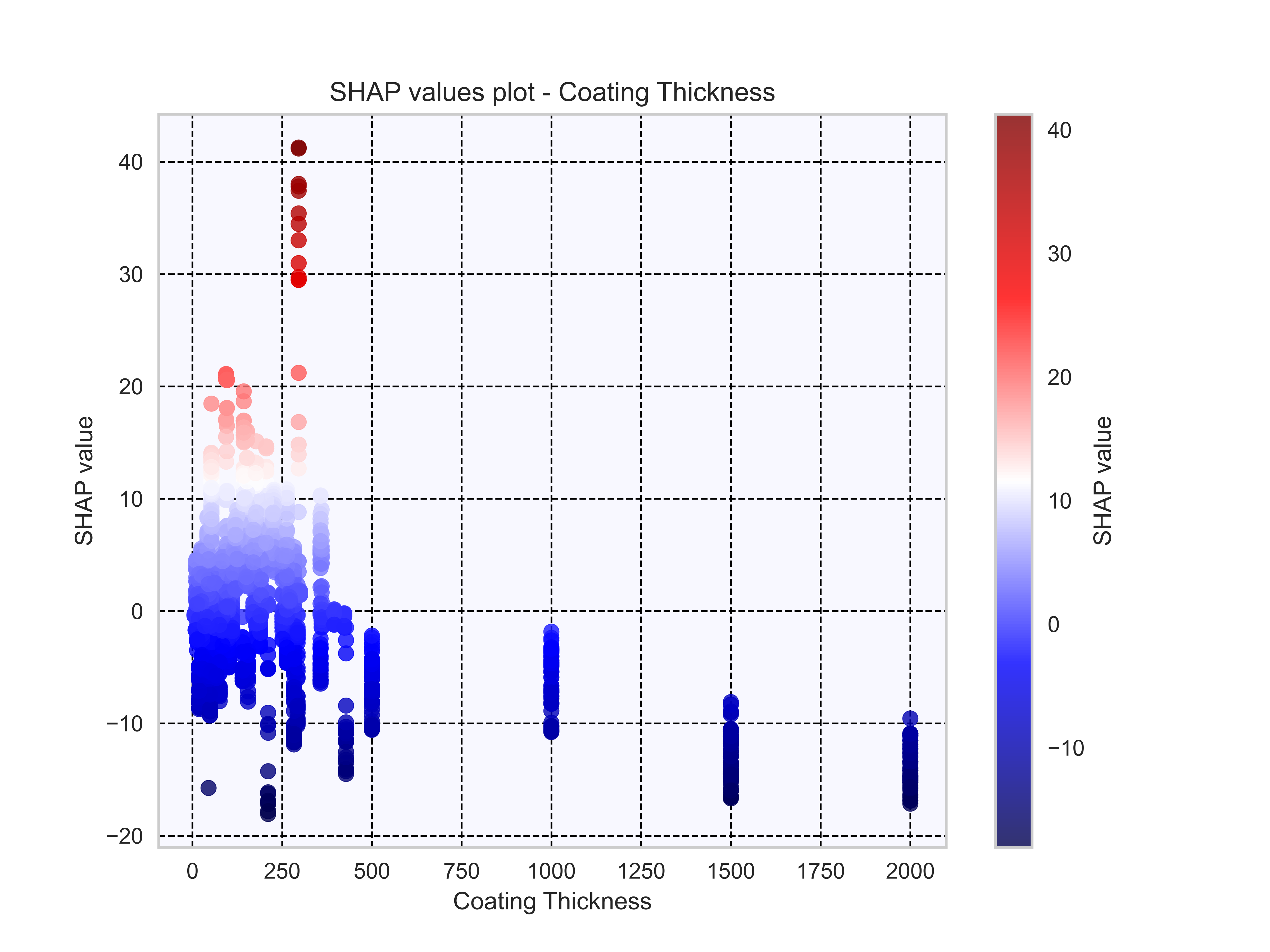}}\label{fig:coating_thickness_porous_raw_data_shap_plot}}\\
\caption{SHAP value plot of coating thickness for (a) thin film-coated substrates and (b) porous-coated substrates.}
\end{figure} 

\clearpage

\begin{figure}[H]
\centering
\subfloat[]{{\includegraphics[width=9cm]{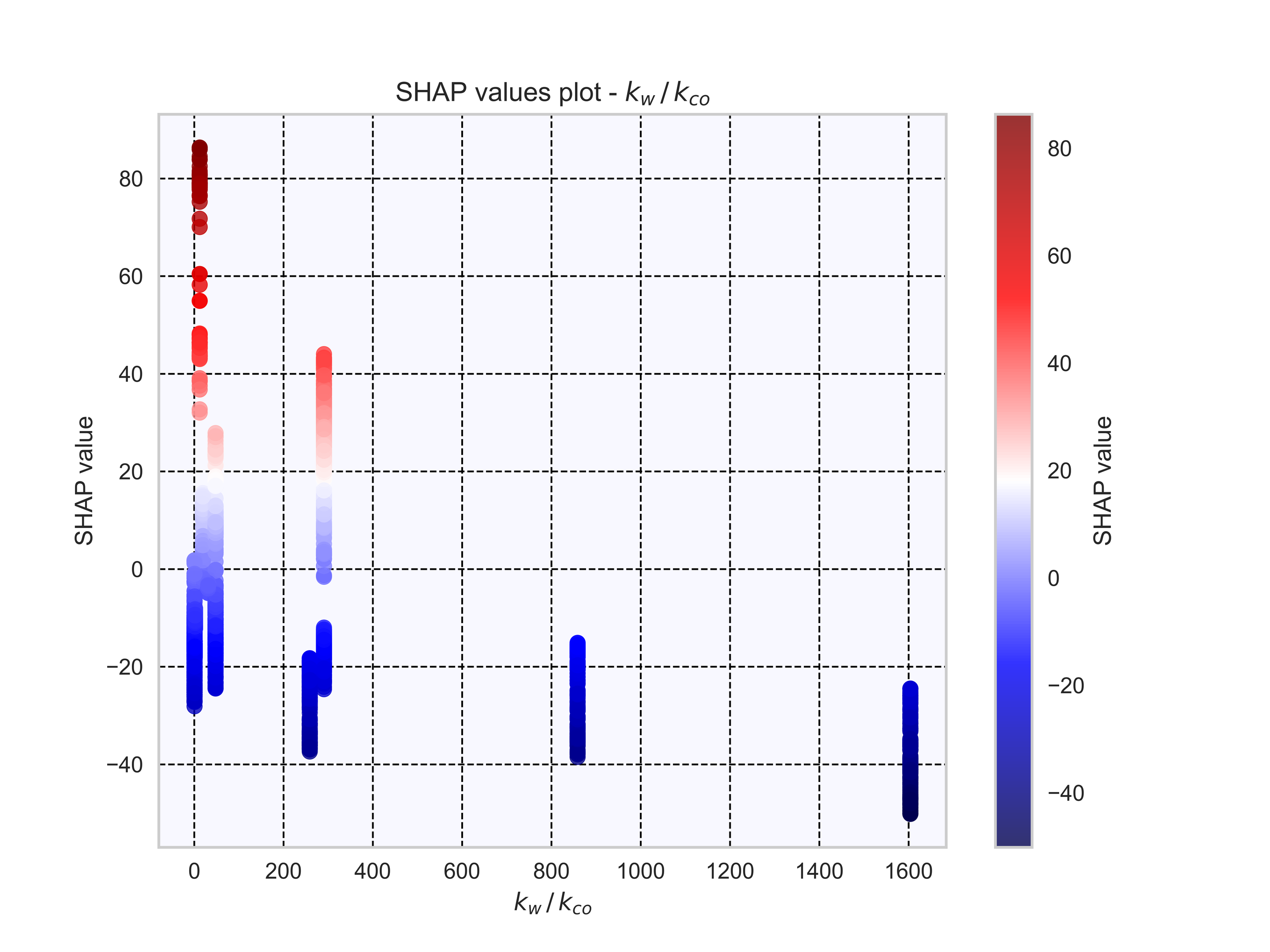}}\label{fig: kw_k_coating_surface_nd_data_shap_plot}}
\subfloat[]{{\includegraphics[width=9cm]{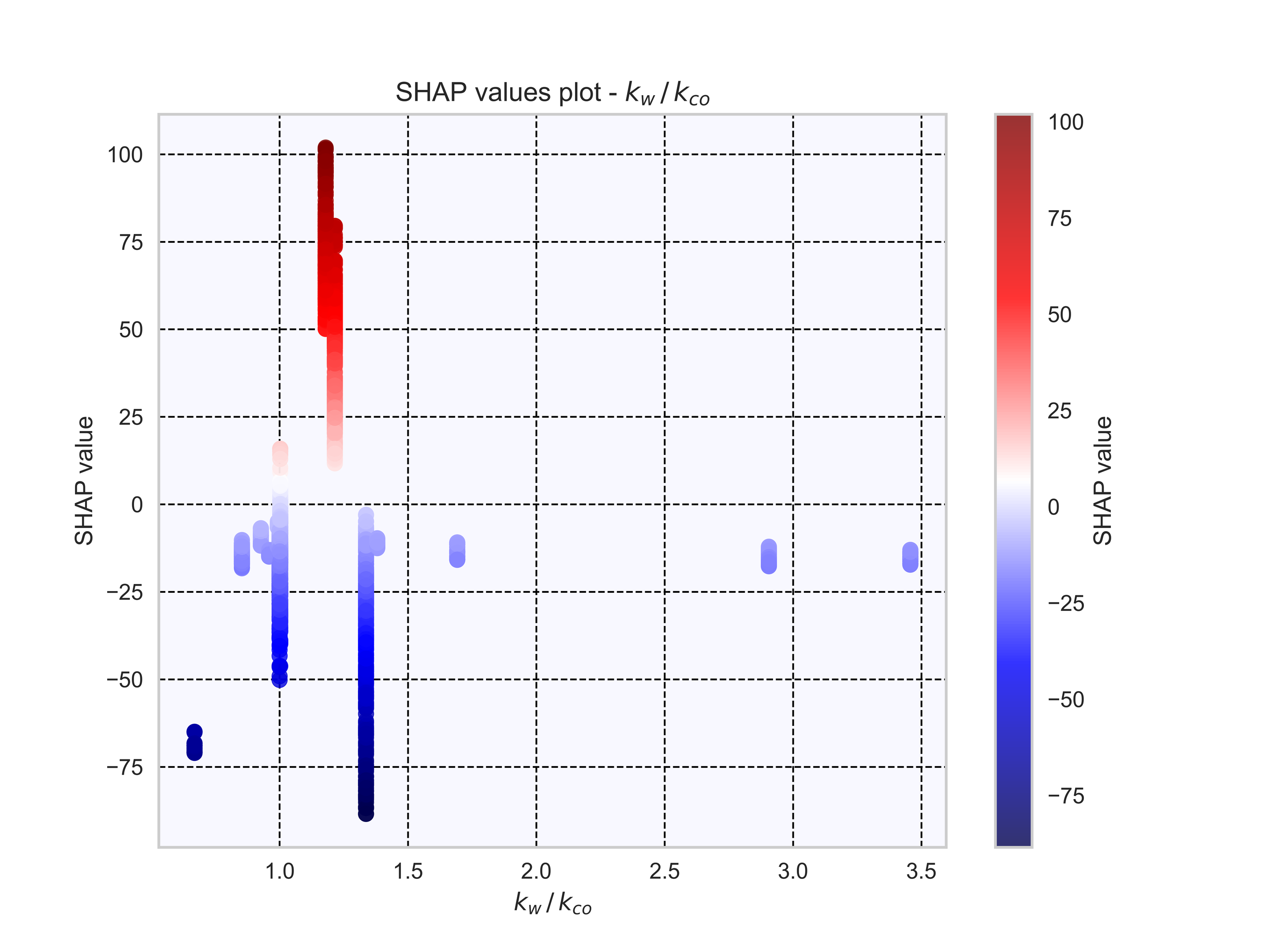}}\label{fig:kw_k_coating_porous_nd_data_shap_plot}}\\
\caption{SHAP value plot of $k_w/k_{co}$ for (a) thin film-coated substrates and (b) porous-coated substrates.}
\end{figure} 

\begin{figure}[H]
\centering
\subfloat[]{{\includegraphics[width=18cm]{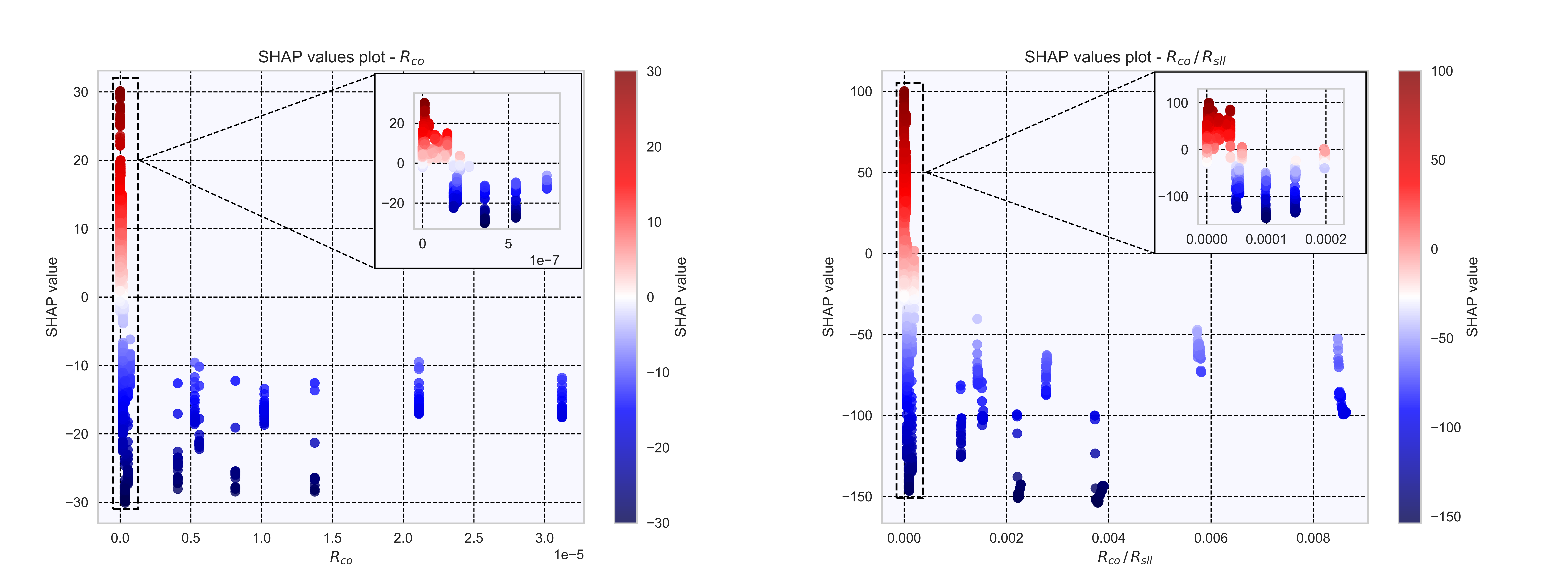}}\label{fig: coating_resistance_variation_shap_plot_surface}}\\
\subfloat[]{{\includegraphics[width=18cm]{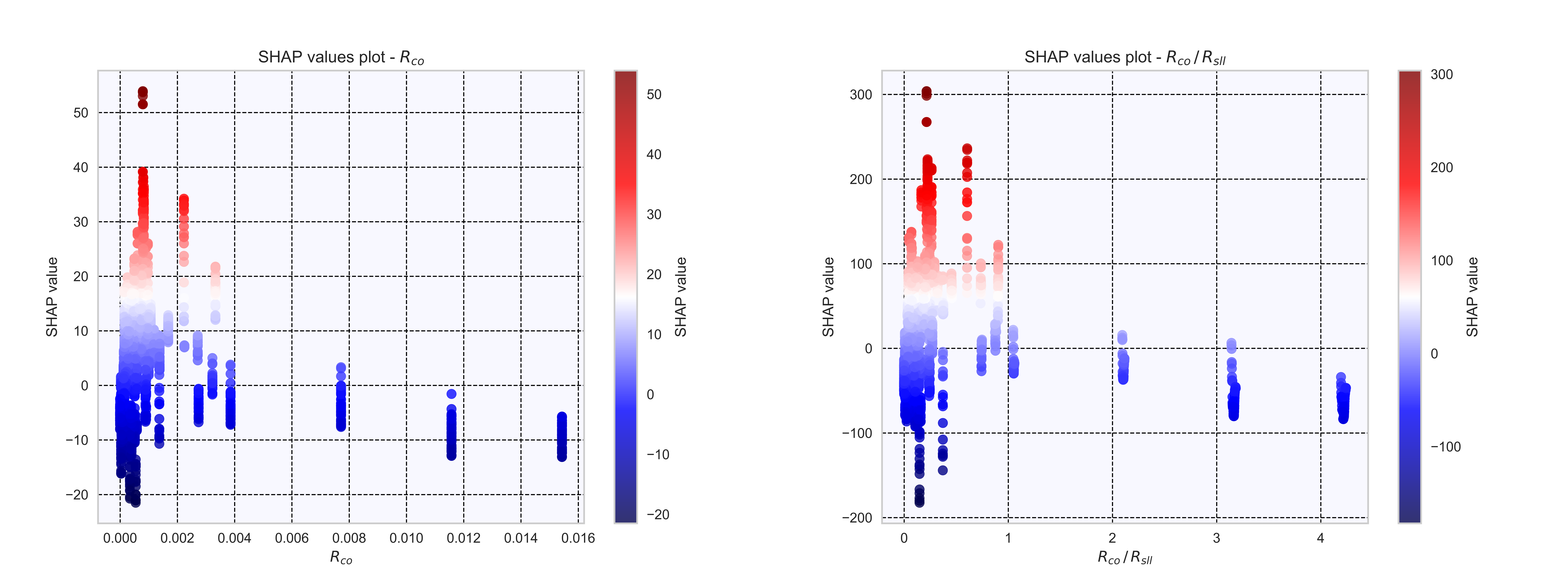}}\label{fig:coating_resistance_variation_shap_plot_porous}}\\
\caption{SHAP value plot of Coating resistance and $R_{co}$/$R_{sll}$ for (a) thin film-coated substrates and (b) for porous-coated substrates.}
\end{figure} 

\clearpage

\subsubsection{For both thin film-coated and porous-coated substrates:}

\begin{itemize}

    \item The SHAP value plot of the contact angle and $\theta/90^\circ$ (Fig. \ref{fig: contact_angle_variation_shap_plot_surface} and \ref{fig:contact_angle_variation_shap_plot_porous}), shows a positive impact on the heat transfer till 60$^\circ$ approximately, and after that, it shows a negative impact.
    The positive impact is perhaps due to increased nucleation with the increase in contact angle, and the negative effect is perhaps due to an increase in the bubble growth time or a reduction in the bubble frequency with the increase in contact angle \cite{MAHMOUD2021101024}.

    \item Higher specific heat liquids retain more thermal energy for the same $\Delta T$, which leads to efficient heat transfer to the bulk liquid. $C_{pl}/C_{pv}$ also shows the same trend. Thus, $C_{pl}$ positively affects the HTC (Figs. \ref{fig: SHAP beeswarm plot for surface and porous coated substrates on Raw Data.} and \ref{fig: SHAP beeswarm plot for surface and porous coated substrates on non-dimensional Data.}).

    \item Higher thermal conductivity liquids such as water exhibit increased heat transfer than lower thermal conductivity liquids such as refrigerants. The present analysis also shows that $k_l$ positively influences the HTC, and $Nu$ increases with a decrease in $k_w/k_l$ (i.e., an increase in $k_l$) and increases with an increase in $k_l/k_v$. This is due to the effective conduction of heat through the liquid microlayer beneath the nucleating bubbles \cite{KUMAR2021100827} for the liquids with high thermal conductivity (Figs. \ref{fig: SHAP beeswarm plot for surface and porous coated substrates on Raw Data.} and \ref{fig: SHAP beeswarm plot for surface and porous coated substrates on non-dimensional Data.}).

    \item The Prandtl number signifies the ratio of momentum diffusivity to thermal diffusivity. From the scatter plots of $Pr_l$, it can be seen that $Pr_l$ varies from 1.33 to 4.93. An increase in $Pr_l$ indicates a reduction in liquid thermal diffusivity, reducing thermal energy transfer. Thus, $Pr_l$ shows a negative impact on the predictions (Fig. \ref{fig: SHAP beeswarm plot for surface and porous coated substrates on non-dimensional Data.}).

    \item As the thickness of the coating increases, $R_{co}$ also increases, and hence, it shows a negative impact on the HTC prediction, which can be seen from the SHAP values plot for coating thickness (Fig. \ref{fig: coating_thickness_surface_raw_data_shap_plot} and \ref{fig:coating_thickness_porous_raw_data_shap_plot}).

    \item When $k_{w}$/$k_{co}$ decreases ((i.e., $k_{co}$ increases), heat transfer increases, which implies that large $k_{co}$ reduces the activation time required for bubble nucleation, increasing the bubble frequency from the surface \cite{AN2021107110}, and hence it has a positive impact on the prediction. This phenomenon is observed when $k_{w}$/$k_{co}$ $>$ 1. In porous-coated substrates, the same phenomenon is observed when $k_{w}$/$k_{co}$ $>$ 1, whereas for $k_{w}$/$k_{co}$ $<$ 1, $k_{w}$ shows prominence than $k_{co}$, so that with increase in $k_{w}$/$k_{co}$, HTC also increases. (Fig. \ref{fig: kw_k_coating_surface_nd_data_shap_plot} and \ref{fig:kw_k_coating_porous_nd_data_shap_plot}).
    
    \item HTC increases with an increase in $T_{w}$ (Fig. \ref{fig: SHAP beeswarm plot for surface and porous coated substrates on Raw Data.}), as the large surface temperature of the surface or the wall superheat increases the nucleation site density and bubble frequency for effective heat transfer \cite{MCHALE2010249,GOEL2017163}.

    \item From the SHAP plot of $R_{co}$ and $R_{co}$/$R_{sll}$ (Fig. \ref{fig: coating_resistance_variation_shap_plot_surface} and \ref{fig:coating_resistance_variation_shap_plot_porous}), it is clearly seen that the HTC and $Nu$ decreases overall with an increase in the thermal resistance of the coating. Increased thermal resistance leads to lower surface conduction heat transfer, and hence HTC decreases \cite{TRISAKSRI20091582}.

\end{itemize}

\begin{figure}[H]
\centering
\subfloat[]{{\includegraphics[width=18cm]{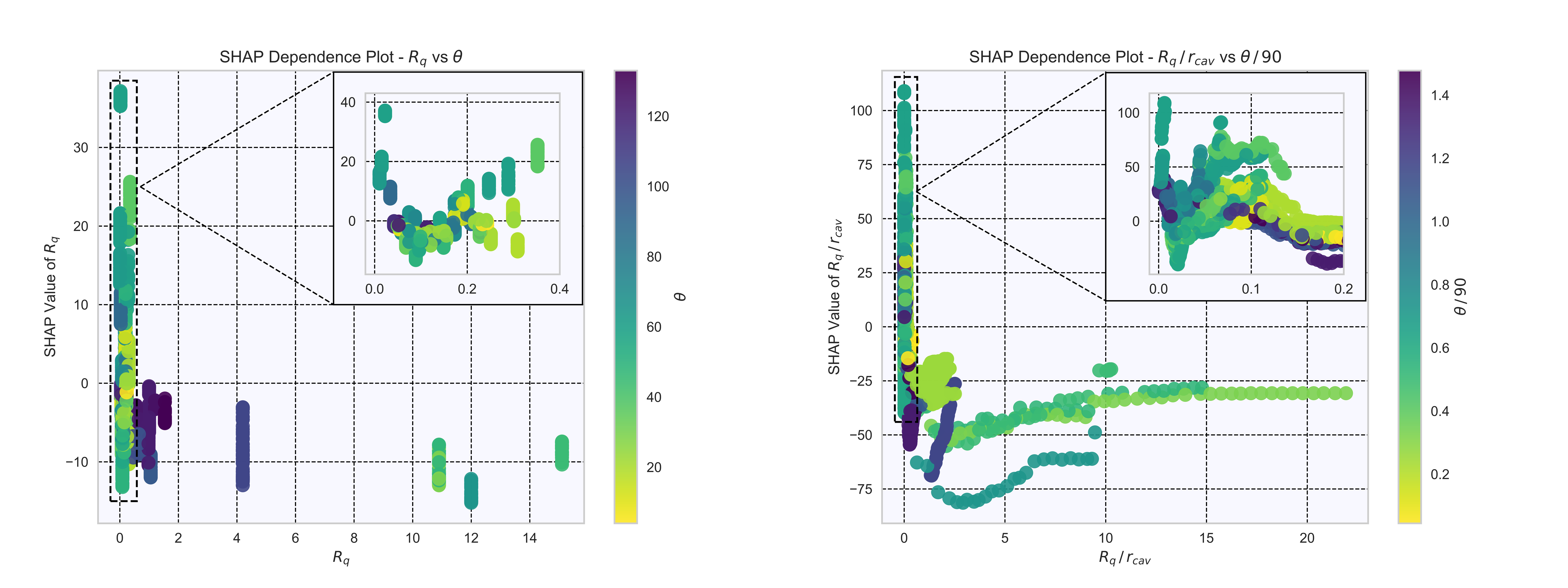}}\label{fig: roughness_variation_with_theta_shap_dp_plot_surface}}\\
\subfloat[]{{\includegraphics[width=18cm]{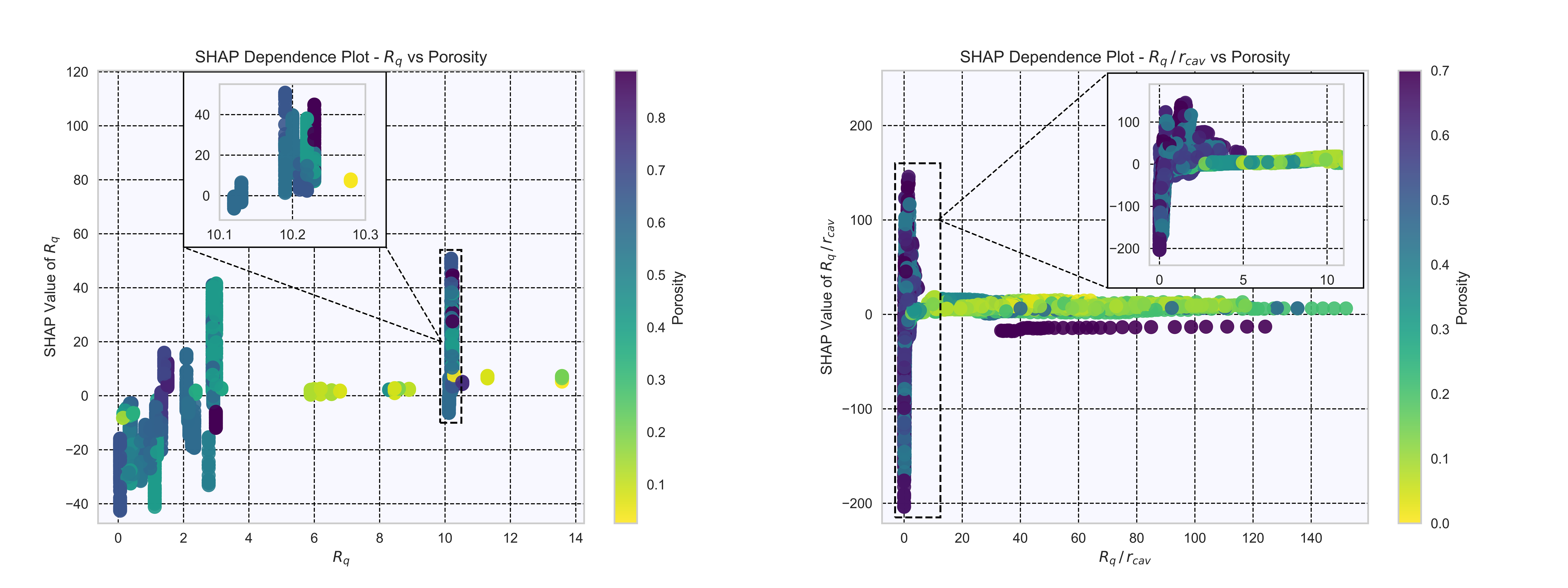}}\label{fig:roughness_variation_with_porosity_shap_dp_plot_porous}}\\
\caption{(a) SHAP dependency plot of $R_q$ vs. $\theta$ and $R_q/r_{cav}$ vs. $\theta/90^\circ$ for thin film-coated substrates. (b) SHAP dependency plot of $R_q$ vs. $\varepsilon$ and $R_q/r_{cav}$ vs. $\varepsilon$ for porous-coated substrates.}
\end{figure} 

\subsubsection{For the thin film-coated substrates:}

\begin{itemize}

    \item From the SHAP dependency plot of $R_q$ and $\theta$ (Fig. \ref{fig: roughness_variation_with_theta_shap_dp_plot_surface}), it is seen that for low surface roughness range between 0.01 $\mu$m and 0.1 $\mu$m, data points with contact angle ranging between $0^\circ$ and $90^\circ$ show increased HTC with contact angle due to enhanced nucleation sites. Above $90^\circ$, HTC decreases due to the bubble agglomeration on the surface \cite{MAHMOUD2021101024}. In this roughness range, contact angle plays a major role. From 0.1 $\mu$m to 0.35 $\mu$m, with an increase in roughness and contact angle ($0^\circ$ $\leq$ $\theta$ $\leq$ $90^\circ$), HTC increases. Here, both roughness and contact angle play a significant role. For 0.35 $\mu$m $\leq$ $R_{q}$ $\leq$ 4.2 $\mu$m, data points with $\theta$ $>$ $90^\circ$, hinder heat transfer due to the accumulation of bubbles on the surface \cite{MAHMOUD2021101024}. Here contact angle shows prominence. After this, even though the roughness values increase ($R_{q}$ $>$ 10.9 $\mu$m), the model shows a negative impact on the HTC irrespective of the contact angle. This is possibly due to the fact that liquid flooding the cavities increases with the increase in the cavity radius, thus decreasing the vapor trapped in the cavity and requiring higher wall superheats for nucleation \cite{collier1994convective}. A similar trend is observed in the dependency plot between $R_q/r_{cav}$ and $\theta/90^\circ$ (Fig. \ref{fig: roughness_variation_with_theta_shap_dp_plot_surface}). 
\end{itemize}

\begin{figure}[H]
    \centering
  \includegraphics[width=9cm]{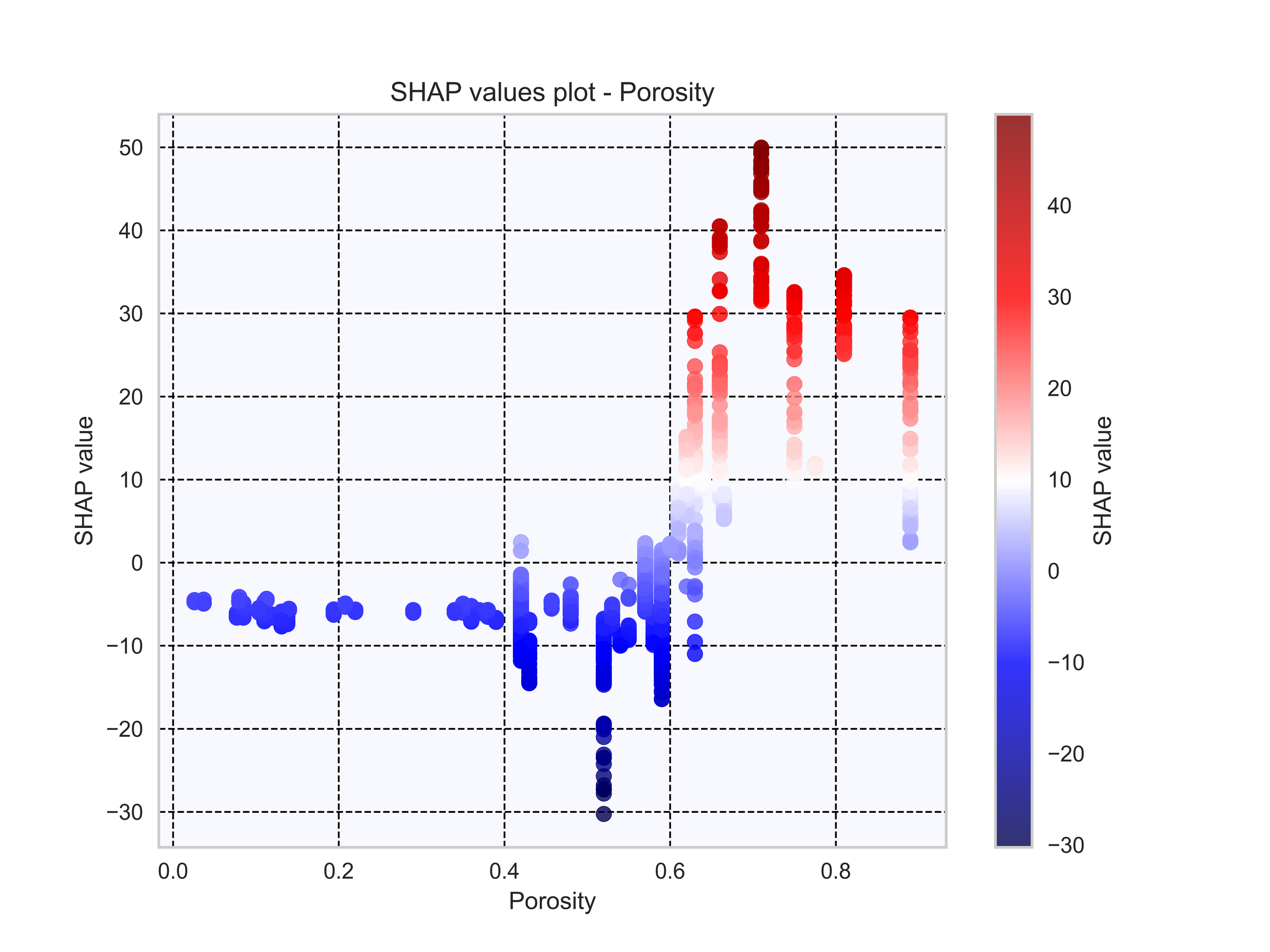}
\caption{SHAP value plot of porosity for porous-coated substrates.}
\label{fig: Porosity_porous_raw_nd_data_shap_plot}
\end{figure}

\subsubsection{For the porous-coated substrates:}

\begin{itemize}

    \item Porosity of the surface provides a large effective surface area, and microcavities play an important role as they act as nucleation sites for bubble incipience and growth. Also, pores act as reentrant channels, which assists in liquid replenishment \cite{Li2007,Surtaev2018,Katarkar2021a}. These combined effects increase HTC with an increase in porosity from 0.03 to 0.71. However, highly porous surfaces lead to bubble agglomeration at the surface, which leads to decreased HTC in the porosity range of 0.71 to 0.89 (Fig. \ref{fig: Porosity_porous_raw_nd_data_shap_plot}).

    \item Generally, surface roughness leads to increased nucleation sites, and hence HTC increases, but if the roughness is very high, it may lead to flooding of the cavities, hindering nucleation and hence heat transfer \cite{collier1994convective}. But for porous surfaces, the variation of $R_{q}$ is also dependent on the porosity. Generally, surface roughness leads to increased nucleation sites, and hence HTC increases. From the dependency plot of $R_{q}$ vs $\varepsilon$, and $R_q/r_{cav}$ vs $\varepsilon$ (Fig. \ref{fig:roughness_variation_with_porosity_shap_dp_plot_porous}), it can be observed that even though the surface roughness is higher (10.13 $\mu$m  $\leq$ $R_{q}$  $\leq$ 13.58 $\mu$m), HTC increases, since porosity of the surface (0.59 $\leq$ $\varepsilon$ $\leq$ 0.66) is large. This is due to the liquid replenishment of the surface, as pores act as reentrant channels \cite{Li2007,Surtaev2018,Katarkar2021a}. It can also be observed that large surface roughness with low porosity shows decreased HTC due to low liquid replenishment. Thus, the porosity of the surface plays an important role in enhancing HTC, and the influence of roughness on the porous-coated substrates differs from that on the thin film-coated substrates.

\end{itemize}

\begin{figure}[H]
    \centering
  \includegraphics[width=18cm]{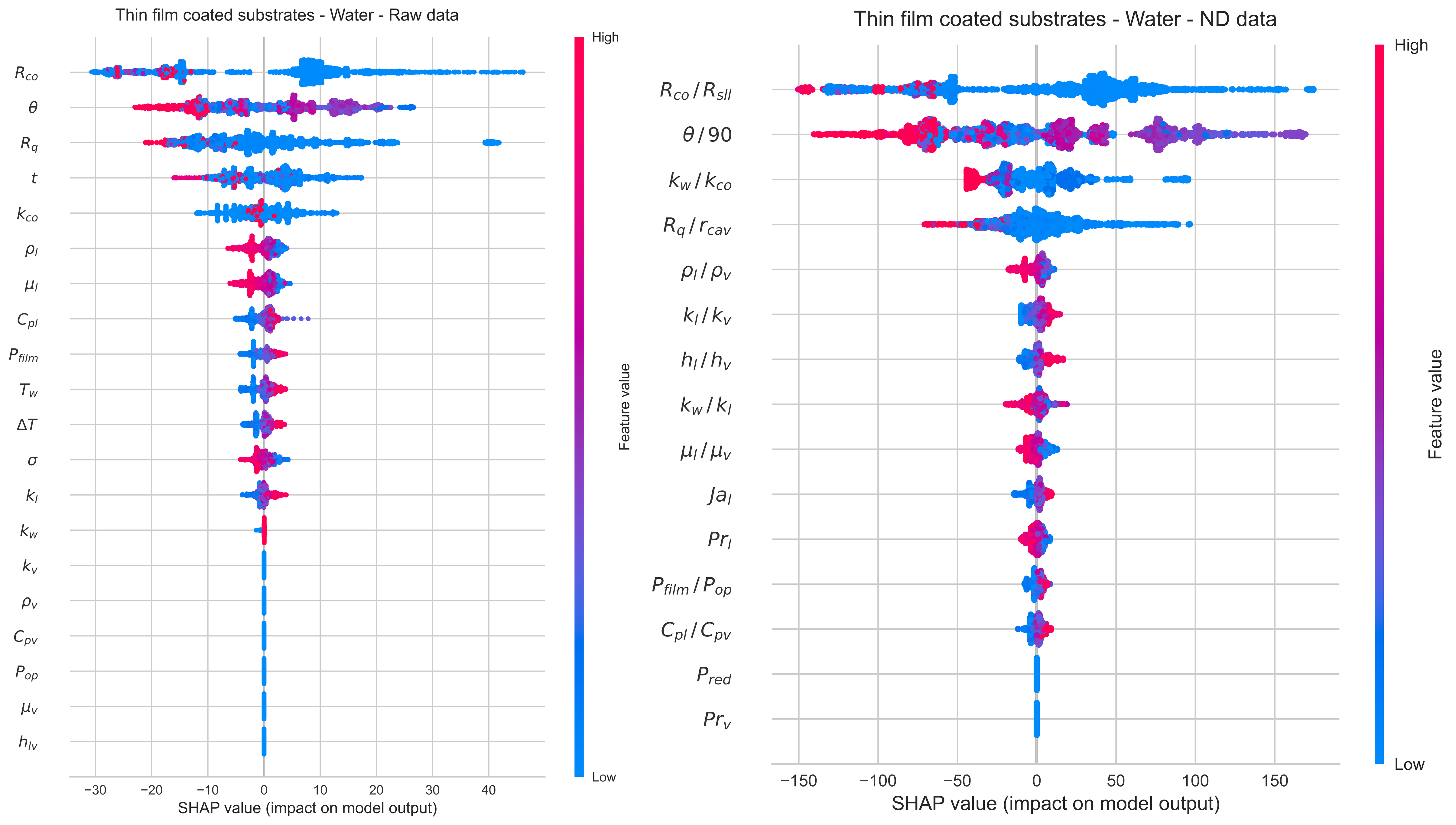}
\caption{SHAP summary plot for thin film-coated substrates on raw data and non-dimensional data with only water.}
\label{fig: SHAP beeswarm plot for surface coated substrates on raw and non-dimensional data with only water.}
\end{figure}

\begin{figure}[H]
    \centering
  \includegraphics[width=18cm]{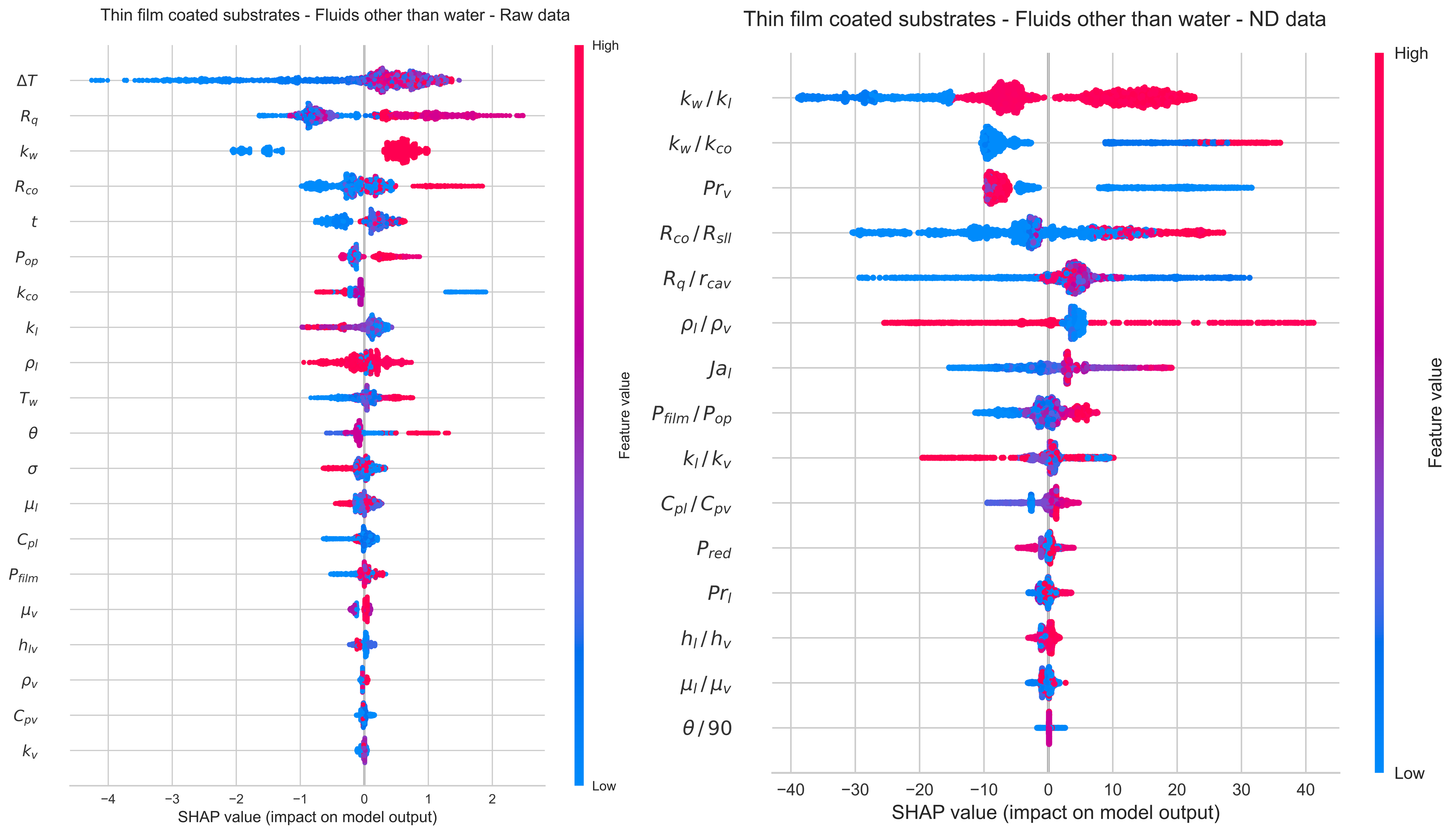}
\caption{SHAP summary plot for thin film-coated substrates on raw data and non-dimensional data with fluids other than water.}
\label{fig: SHAP beeswarm plot for surface coated substrates on raw and non-dimensional data with fluids other than water.}
\end{figure}

\begin{figure}[H]
    \centering
  \includegraphics[width=18cm]{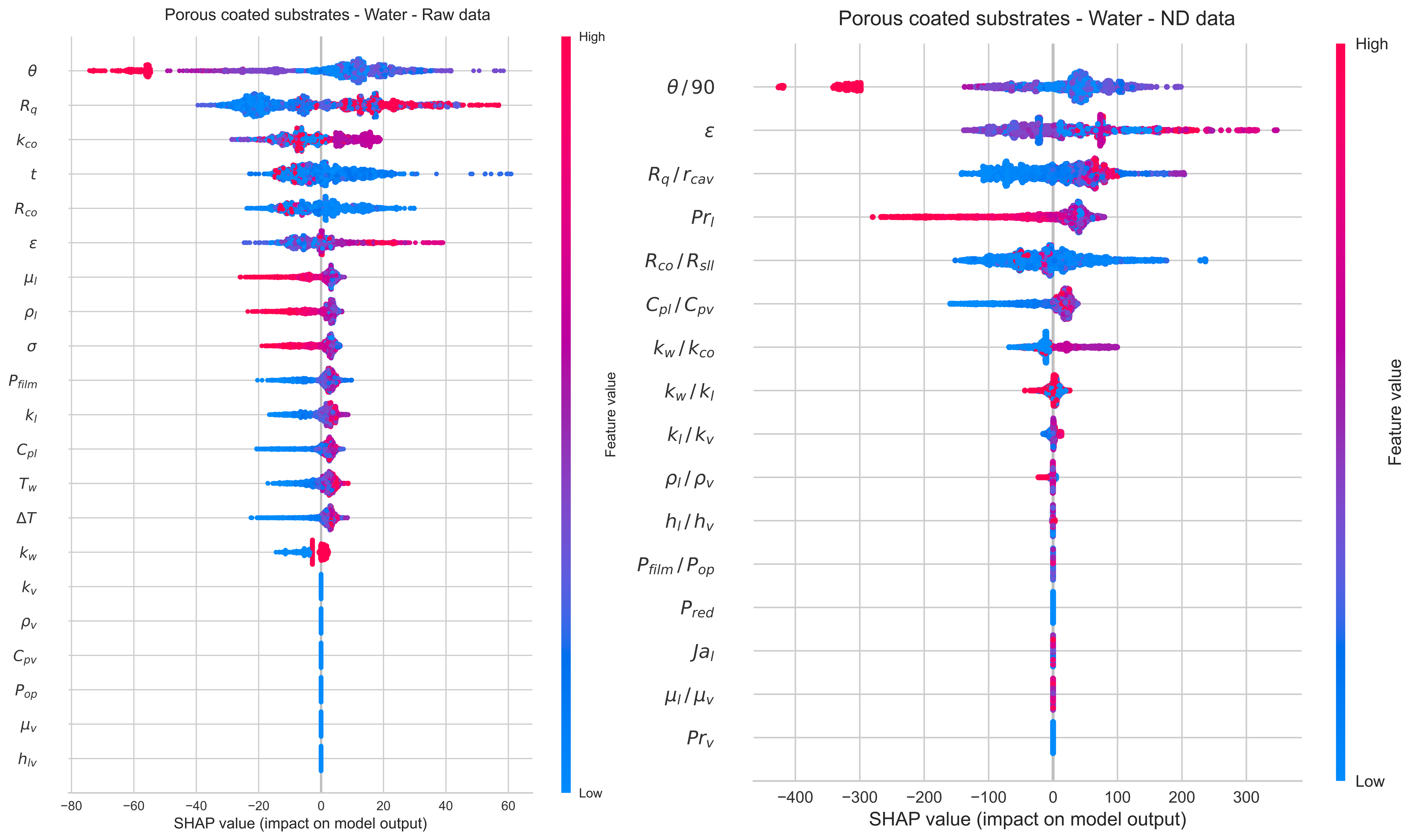}
\caption{SHAP summary plot for porous-coated substrates on raw data and non-dimensional data with only water.}
\label{fig: SHAP beeswarm plot for porous coated substrates on raw and non-dimensional data with only water.}
\end{figure}

\begin{figure}[H]
    \centering
  \includegraphics[width=18cm]{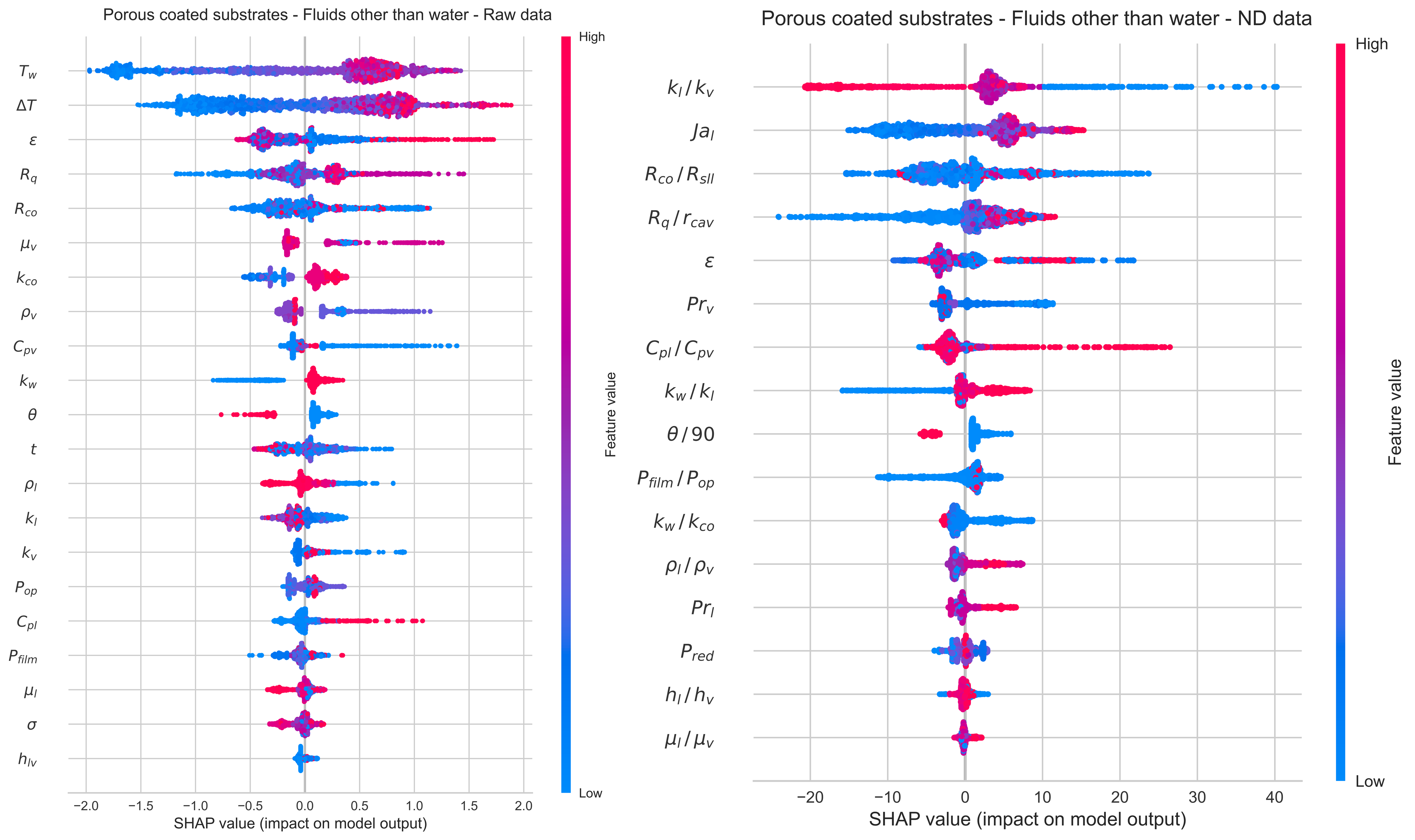}
\caption{SHAP summary plot for porous-coated substrates on raw data and non-dimensional data with fluids other than water.}
\label{fig: SHAP beeswarm plot for porous coated substrates on raw and non-dimensional data with fluids other than water.}
\end{figure}

\clearpage

\subsubsection{From the segregation of the data into water and other fluids:}
\begin{itemize}

    \item From the SHAP summary plot (Figs. \ref{fig: SHAP beeswarm plot for surface coated substrates on raw and non-dimensional data with only water.}-\ref{fig: SHAP beeswarm plot for porous coated substrates on raw and non-dimensional data with fluids other than
    water.}), it can be observed that $\theta$ and $\theta/90^\circ$ play a significant role in water, while for other fluids, its effect is negligible. This is due to the highly wetting nature of the refrigerants, or fluids other than water. Whereas $R_q$ and $R_q/r_{cav}$ show significant effects on the heat transfer for water and other fluids for both types of surfaces. Thus, the effect of surface roughness is pivotal in pool boiling with both types of fluids.

    \item $R_{co}$, $R_{co}$/$R_{sll}$, $t$, and $k_{co}$ consistently show major significance for both thin film-coated and porous-coated substrates irrespective of the fluid used. This shows that the coating resistance, coating thickness, and thermal conductivity of the coating are the major surface characteristics, in addition to surface roughness and wettability, affecting boiling heat transfer on coated surfaces. Furthermore, porosity is a major surface characteristic for porous-coated substrates, which shows prominence for both types of fluids (Figs. \ref{fig: SHAP beeswarm plot for surface coated substrates on raw and non-dimensional data with only water.}-\ref{fig: SHAP beeswarm plot for porous coated substrates on raw and non-dimensional data with fluids other than water.}).

    \item In addition to the thermal conductivity of the coating, substrate thermal conductivity plays a major role for both types of fluids on thin film-coated substrates. Whereas for porous-coated substrates, $k_{co}$ is more significant than $k_{w}$ (Fig. \ref{fig: SHAP beeswarm plot for surface coated substrates on raw and non-dimensional data with only water.}-\ref{fig: SHAP beeswarm plot for porous coated substrates on raw and non-dimensional data with fluids other than water.}).

    \item Markedly, for fluids other than water, $\Delta T$ and $Ja_l$ show a major effect on the pool boiling heat transfer. This shows that the boiling heat transfer coefficient is highly sensitive to wall superheat for fluids other than water (Fig. \ref{fig: SHAP beeswarm plot for surface coated substrates on raw and non-dimensional data with fluids other than
    water.}, \ref{fig: SHAP beeswarm plot for porous coated substrates on raw and non-dimensional data with fluids other than
    water.}).
    
\end{itemize}

From the SHAP interpretation, the variation of the influence of different parameters playing a pivotal role in pool boiling heat transfer has been observed. Thus, it can be inferred that the CatBoost model identifies the key parameters influencing nucleate boiling heat transfer on the coated surfaces.

\subsection{Assessment of empirical correlations}
The different correlations available in the literature are enlisted below, with detailed descriptions presented in Table \ref{tab:Review of the existing correlations}. These correlations were assessed for the thin film-coated and porous-coated data and their performance in terms of MAE, RMSE, MAPE, and error bands of 10\%, 20\% and 30\% are presented in Tables \ref{tab:Comparison_of_Existing_Correlations_Surface_Coated} and \ref{tab:Comparison_of_Existing_Correlations_Porous_Coated}.

\begin{enumerate}
    \item Rosenhow Correlation \cite{Rohsenow1952}:
\begin{flalign*}
& \begin{aligned}
&\triangle T_{\text{sat\_rosen}} = \left( \frac{h_{lv}}{C_{pl}} \right) \cdot C_{sf} \cdot \left( \left( \frac{q}{\mu_l \cdot h_{lv}} \right) \cdot \left( \frac{\sigma}{g \cdot \left(\rho_l - \rho_v\right)} \right)^{0.5} \right)^{0.33} \cdot \left( \text{Pr}_l \right)^{m+1}\\
&h_{\text{rosen}}= \frac{q}{\triangle T_{sat\_rosen}}
\end{aligned} &
\end{flalign*}

    \item Labuntsov Correlation \cite{labuntsov1973heat}:
\begin{flalign*}
& \begin{aligned}
h_{\text{labuntsov}} = & 0.075 \left( 1 + \left( 10 \left( \frac{\rho_v}{\rho_l - \rho_v} \right)^{0.67} \right) \right) \quad \times \left( \left( \frac{\rho_l \cdot k_l^2}{\sigma \cdot \mu_l \cdot T_l} \right)^{0.33} \right) \left( q \right)^{0.67}
\end{aligned} &
\end{flalign*}

    \item Kruzhilin Correlation \cite{kruzhilin1947free}:
\begin{flalign*}
& \begin{aligned}
h_{\text{kruzhilin}} = \left( 0.082 \cdot \frac{k_l}{L_c} \right) \left( \left( \frac{h_{lv} \cdot q}{g \cdot T_l \cdot k_l} \cdot \frac{\rho_v}{\rho_l - \rho_v} \right)^{0.7} \right) \left( \frac{T_l \cdot C_{pl} \cdot \sigma \cdot \rho_l}{h_{lv}^2 \cdot \rho_v^2 \cdot \left( \frac{\sigma}{(\rho_l - \rho_v) \cdot g} \right)^{0.5}} \right)^{0.33} \left( \text{Pr}_l^{-0.45} \right)
\end{aligned} &
\end{flalign*}

    \item Kichigin and Tobilevich Correlation \cite{kichigin1955generalization}:
\begin{flalign*}
& \begin{aligned}
h_{\text{kichi\_tobil}} = \left( \frac{k_l}{L_c} \right) \left( 3.25 \times 10^{-4} \right) \left( \text{Re} \right)^{0.6} \left( \text{Pr}_l \right)^{0.6} \left( \left( \frac{g \cdot L_c^3}{\nu_l^2} \right)^{0.125} \right) \cdot \left( \frac{P_{op}}{\left( g \cdot \sigma \cdot (\rho_l - \rho_v) \right)^{0.5}} \right)^{0.7}
\end{aligned} &
\end{flalign*}

    \item Forster-Zuber Correlation \cite{Forster1955}:
\begin{flalign*}
& \begin{aligned}
h_{\text{forster\_zuber}} = 0.00122 \left( \frac{k_l^{0.79} \cdot C_{pl}^{0.45} \cdot \rho_l^{0.49}}{\sigma^{0.5} \cdot \mu_l^{0.29} \cdot h_{lv}^{0.24} \cdot \rho_v^{0.24}} \right) \left( \triangle T \right)^{0.24} (P_{\text{film}} - P_{op})^{0.75}
\end{aligned} &
\end{flalign*}

    \item Borishansky correlation \cite{borishanskii1969correlation}:
\begin{flalign*}
& \begin{aligned}
&A^* = 0.1011 \cdot \left( P_c^{0.69} \right)\\
&F = 1.8 \cdot \left( P_{\text{red}}^{0.17} \right) + 4 \cdot \left( P_{\text{red}}^{1.2} \right) + 10 \cdot \left( P_{\text{r}}^{10} \right)\\
&h_{\text{borishansky}} = \left( A^* \right)^{3.33} \cdot \left( \triangle T \right)^{2.33} \cdot \left( F\right)^{3.33}
\end{aligned} &
\end{flalign*}

    \item Kutateladze and Borishanski Correlation \cite{kutateladze1966concise}:
\begin{flalign*}
& \begin{aligned}
h_{\text{kuta\_boris}} = \left( 0.44 \cdot \frac{k_l}{L_c} \right) \left( \left( \frac{1 \times 10^{-4} \cdot q \cdot P_{op}}{g \cdot h_{lv} \cdot \rho_v \cdot \mu_l} \cdot \frac{\rho_l}{\rho_l - \rho_v} \right)^{0.7} \right) \left( \text{Pr}_l^{0.35} \right)
\end{aligned} &
\end{flalign*}

    \item Modified Kutateladze Correlation \cite{kutateladze1990heat}:
\begin{flalign*}
& \begin{aligned}
& h_{\text{modif\_kuta}} = \left( 3.37 \times 10^{-9} \cdot \frac{k_l}{L_c} \cdot \left( \frac{h_{lv}}{C_{pl} \cdot q} \right)^{-2} \cdot M^{*-1} \right)^{\frac{1}{3}}\\
& M^* = \frac{g \cdot \sigma}{(\rho_l - \rho_v) \cdot \left( \frac{P_{op}}{\rho_v} \right)^2} \\
\end{aligned} &
\end{flalign*}

    \item Pioro Correlation \cite{Pioro1999}:
\begin{flalign*}
& \begin{aligned}
h_{\text{pioro}} = C_{s} \cdot \frac{k_l}{L_c} \cdot \left( \frac{q}{{h_{lv} \cdot \sqrt{\rho_v}} \cdot {\left( \sigma \cdot g \cdot (\rho_l - \rho_v) \right)^{0.25}}} \right)^{\frac{2}{3}} \cdot \text{Pr}_l^{n}
\end{aligned} &
\end{flalign*}

    \item Cooper Correlation \cite{COOPER1984157}:
\begin{flalign*}
& \begin{aligned}
h_{\text{cooper}} = 55 \cdot \left( \text{P}_{\text{r}}^{0.12 - (0.2 \cdot \log_{10}(\text{R}_{\text{q}}))} \right) \cdot \left (- \log_{10}(\text{P}_{\text{r}})\right)^{-0.55} \cdot \left( \text{M}^{-0.5} \right) \cdot \left( \text{q}^{0.67} \right)
\end{aligned} &
\end{flalign*}

    \item Cornwell–Houston correlation \cite{Cornwell1994}:
\begin{flalign*}
& \begin{aligned}
&h_{\text{cornwell}} = 9.7 \cdot \frac{k_l}{L_c} \cdot F_p \cdot \text{P}_{\text{c}}^{0.5} \cdot (Re)^{0.67} \cdot (\text{Pr}_l)^{0.4} \\
&F_p = 1.8 \cdot \text{P}_{\text{r}}^{0.17} + 4 \cdot \text{P}_{\text{r}}^{1.2} + 10 \cdot \text{P}_{\text{r}}^{10}
\end{aligned} &
\end{flalign*}

    \item Ribatski and Jabardo Correlation \cite{Ribatski2003}:
\begin{flalign*}
& \begin{aligned}
&h_{\text{ribatski}} = 100 \cdot (\text{q}^{m}) \cdot (\text{P}_{\text{r}}^{0.45}) \cdot (-\log(\text{P}_{\text{r}}))^{-0.8} \cdot (\text{R}_{\text{q}}^{0.2}) \cdot (\text{M}^{-0.5}) \\
&m = 0.9 -  0.3 \cdot (\text{P}_{\text{r}}^{0.2}) 
\end{aligned} &
\end{flalign*}

\end{enumerate}

\begin{table}[H]
\centering
\caption{Review of the existing correlations.}
\label{tab:Review of the existing correlations}
\begin{adjustbox}{max width=\textwidth}
\begin{tabular}{lp{10cm}}

\hline
\textbf{Author} & \textbf{Remarks} \\
\hline

Rosenhow \cite{Rohsenow1952} & Fluids: Water, Ethanol, iso-Propanol, n-Butanol, R-11, R-12, R-113, Carbon tetrachloride, Propane, n-pentane, Benzene, n-Heptane, Acetone, 30\% and 50\% Potassium carbonate. \\&Substrates: Copper, Aluminum, Brass, Chromium, Platinum wires, Stainless Steel, Zinc, Nickel, Inconel. \\
\midrule

Labuntsov \cite{labuntsov1973heat} & Can be used for a wide variety of fluids. \\
\midrule

Kruzhilin \cite{kruzhilin1947free}  & Fluids: Water and Refrigerants. \\&Substrates: Horizontal flat plates of various materials. \\
\midrule

Kichigin and Tobilevich \cite{kichigin1955generalization} & Fluids: Water and concentrated solutions. \\&Substrates: Steel Tubes. \\
\midrule

Forster-Zuber \cite{Forster1955} & Fluids: Water, Ethanol, n-Pentane, Benzene. \\&Substrates: Horizontal flat plates of various materials. Used bubble radius and the bubble growth velocity to formulate the correlation. \\
\midrule

Borishansky \cite{borishanskii1969correlation} & Fluids: Water, Ethanol, and other fluids. \\&Substrates: Horizontal tubes and flat plates. \\
\midrule

Kutateladze and Borishanski \cite{kutateladze1966concise} & Can be used for a wide variety of fluids and high heat flux conditions. \\
\midrule

Kutateladze \cite{kutateladze1990heat} & Can be used for a wide variety of fluids. \\
\midrule

Pioro \cite{Pioro1999} & Fluids: Water, Ethanol, iso-propanol, n-butanol, R-11, R-12, R-113, Carbon tetrachloride, Propane, n-pentane, Benzene, n-heptane, Acetone, 30\% and 50\% Potassium carbonate. \\&Substrates: Copper, Aluminum, Brass, Chromium, Platinum wires, Stainless Steel, Zinc, Nickel, Inconel. Modified fluid surface parameter of Rohsenhow Correlation. \\
\midrule

Cooper \cite{COOPER1984157} & 5641 data points. \\& Fluids: Water, R12, R113, R114, Ethanol, Benzene, Propane, Cryogens - Nitrogen, Oxygen, Hydrogen, Helium, Neon. \\&Substrates: Copper, Stainless steel, Platinum wires, Nickel, Aluminum, Brass, Sodium-Potassium alloy. \\
\midrule

Cornwell–Houston \cite{Cornwell1994} & Fluids: Water, R113, R11, R12, R113, R114, R115, R22, Nonane, Pentane, Propane, Hexane, Ethane, Benzene, Methanol, Ethanol, Isobutanol, p-Xylene. \\&Substrates: Horizontal tubes and tube bundles of various materials. \\
\midrule

Ribatski and Jabardo \cite{Ribatski2003} & 2600 data points. \\&Fluids: R11, R12, R123, R22, R134a. \\&Substrates: Cylindrical surfaces - Copper, Brass, and Stainless steel. \\

\hline
\end{tabular}
\end{adjustbox}
\end{table}

The correlations by Rosenhow \cite{Rohsenow1952}, Labunstov \cite{labuntsov1973heat}, Kruzhilin \cite{kruzhilin1947free}, Kichigin \& Tobilevich \cite{kichigin1955generalization}, Kutateladze \cite{kutateladze1990heat}, and Pioro \cite{Pioro1999} exhibit an R$^2$ value of around 0.5 for thin film-coated substrates. Whereas for porous-coated substrates, the above correlations performed poorly, owing to the inability to capture the complex relationship among the porous data.

\begin{table}[H]
    \centering
    \caption{Comparison of existing correlations for thin film-coated substrates.}
    \label{tab:Comparison_of_Existing_Correlations_Surface_Coated}
    \begin{adjustbox}{max width=\textwidth}
    \renewcommand{\arraystretch}{1.5} 
    \begin{tabular}{llllllll}
        \hline
        \multirow{2}{*}{\textbf{Models}} & \multirow{2}{*}{\textbf{R\textsuperscript{2}}} & \multirow{2}{*}{\textbf{MAE}} & \multirow{2}{*}{\textbf{RMSE}} & \multicolumn{4}{c}{\textbf{\% deviation of data within}} \\ \cline{5-8}
        & & & & \textbf{±10\%} & \textbf{±20\%} & \textbf{±30\%} & \textbf{±40\%} \\ \hline
        Rosenhow & 0.533 & 18.460 & 29.303 & 8.429 & 16.743 & 26.945 & 35.831 \\ 
        Labuntsov & 0.526 & 18.982 & 29.523 & 8.352 & 17.601 & 27.193 & 36.918 \\ 
        Kruzhilin & 0.500 & 19.055 & 30.350 & 5.606 & 15.084 & 28.394 & 43.326 \\ 
        Kichigin and Tobilevich & 0.495 & 18.822 & 30.490 & 15.027 & 28.814 & 40.446 & 49.771 \\ 
        Borishansky & -0.754 & 36.818 & 56.814 & 1.602 & 3.318 & 5.187 & 8.352 \\ 
        Kutateladze and Borishanski & 0.227 & 24.716 & 37.723 & 6.846 & 10.889 & 14.455 & 23.722 \\ 
        Modified Kutateladze & 0.554 & 17.465 & 28.651 & 16.285 & 33.047 & 45.290 & 54.462 \\ 
        Pioro & 0.620 & 16.431 & 26.440 & 10.831 & 19.375 & 26.297 & 41.762 \\ 
        Cooper & -0.379 & 33.317 & 50.376 & 1.735 & 6.388 & 13.330 & 27.021 \\ 
        Cornwell-Houston & -35785.495 & 5692.910 & 8116.106 & 0.000 & 0.000 & 0.000 & 0.000 \\ 
        Ribatski and Jabardo & -0.170 & 30.977 & 46.412 & 7.094 & 7.990 & 9.115 & 12.243 \\ 
        \hline
    \end{tabular}
    \end{adjustbox}
\end{table}

\begin{table}[H]
    \centering
    \caption{Comparison of existing correlations for porous-coated substrates.}
    \label{tab:Comparison_of_Existing_Correlations_Porous_Coated}
    \begin{adjustbox}{max width=\textwidth}
    \renewcommand{\arraystretch}{1.5} 
    \begin{tabular}{llllllll}
        \hline
        \multirow{2}{*}{\textbf{Models}} & \multirow{2}{*}{\textbf{R\textsuperscript{2}}} & \multirow{2}{*}{\textbf{MAE}} & \multirow{2}{*}{\textbf{RMSE}} & \multicolumn{4}{c}{\textbf{\% deviation of data within}} \\ \cline{5-8}
        & & & & \textbf{±10\%} & \textbf{±20\%} & \textbf{±30\%} & \textbf{±40\%} \\ \hline
        Rosenhow & 0.037 & 55.853 & 84.339 & 5.445 & 13.516 & 20.517 & 28.899 \\ 
        Labuntsov & 0.030 & 57.524 & 84.653 & 1.050 & 3.792 & 6.301 & 10.463 \\ 
        Kruzhilin & 0.009 & 56.812 & 85.564 & 6.107 & 14.022 & 22.443 & 32.905 \\ 
        Kichigin and Tobilevich & -0.008 & 56.933 & 86.276 & 6.243 & 15.364 & 23.823 & 31.058 \\ 
        Borishansky & -0.873 & 81.034 & 117.634 & 1.984 & 4.959 & 7.371 & 10.093 \\ 
        Kutateladze and Borishanski & -0.213 & 64.083 & 94.655 & 2.334 & 5.795 & 12.155 & 18.242 \\ 
        Modified Kutateladze & 0.054 & 54.961 & 83.587 & 8.246 & 19.020 & 27.285 & 35.045 \\ 
        Pioro & 0.122 & 52.932 & 80.524 & 4.765 & 8.285 & 14.644 & 24.271 \\ 
        Cooper & 0.326 & 45.487 & 70.567 & 8.674 & 18.047 & 28.433 & 36.970 \\ 
        Cornwell-Houston & -11582.644 & 6601.501 & 9249.813 & 0.000 & 0.000 & 0.000 & 0.000 \\ 
        Ribatski and Jabardo & -0.127 & 62.462 & 91.234 & 7.468 & 13.205 & 17.620 & 25.438 \\
        \hline
    \end{tabular}
    \end{adjustbox}
\end{table}

\subsection{Proposed empirical correlation}

Boiling is primarily impacted by the surface characteristics. The above studies did not consider the comprehensive surface parameters in their correlations. In this study, the important surface parameters from SHAP analysis - $R_{co}$/$R_{sll}$, $R_q/r_{cav}$, $k_w/k_l$, $\varepsilon$, and $\theta/90^\circ$ - were incorporated. $P_{red}$, and M/M$_{w}$ were also included in the analysis to take into account the operating conditions and the type of working fluid. The existing empirical correlations were modified with the addition of these parameters, and optimized coefficients were found by curve-fitting on the experimental data. Amongst all the correlations, Kruzhilin correlation with these additional parameters displayed the highest performance with R$^2$ values of 0.9 and 0.81 for thin film-coated and porous-coated substrates, respectively. Eq.(\ref{Proposed Correlation for Surface Coated substrates}) \& Eq.(\ref{Proposed Correlation for Porous Coated substrates}) present the proposed empirical correlations from this analysis.

For thin film-coated substrates:
\begin{equation}
    \label{Proposed Correlation for Surface Coated substrates}
    \begin{aligned}
    & h_{\text{thin film}} = \left( \frac{R_{c}}{R_{sll}} \right)^{-0.032} \cdot \left( \frac{k_{w}}{k_{l}} \right)^{0.008} \cdot \left( \frac{R_{q}}{r_{cav}} \right)^{-0.133} \cdot \left( \frac{\theta}{90} \right)^{0.058} \cdot (P_{\text{red}})^{0.042} \cdot \left( \frac{M}{M_{w}} \right)^{0.058} \cdot \\
    & \quad \left. \left( 0.082 \cdot \frac{k_{l}}{L_{c}} \right) \cdot \left( \left( \frac{h_{lv} \cdot q}{g \cdot T_{l} \cdot k_{l}} \right) \cdot \left( \frac{\rho_{v}}{\rho_{l} - \rho_{v}} \right) \right)^{0.7} \right. \cdot \left( \frac{T_{l} \cdot C_{pl} \cdot \sigma \cdot \rho_{l}}{h_{lv}^{2} \cdot \rho_{v}^{2} \cdot \left( \frac{\sigma}{(\rho_{l} - \rho_{v}) \cdot g} \right)^{0.5} } \right)^{0.33} \cdot \left( \text{Pr}_{l} \right)^{-0.45}
    \end{aligned}
\end{equation}

For porous-coated substrates:
\begin{equation}
    \label{Proposed Correlation for Porous Coated substrates}
    \begin{aligned}
    & h_{\text{porous}} = \left( \frac{R_{c}}{R_{sll}} \right)^{-0.028} \cdot \left( \frac{k_{w}}{k_{l}} \right)^{0.361} \cdot \left( \frac{R_{q}}{r_{cav}} \right)^{0.069} \cdot \left( \frac{\theta}{90} \right)^{-0.086} \cdot \varepsilon^{0.257} \cdot (P_{\text{red}})^{0.205} \cdot \left( \frac{M}{M_{w}} \right)^{-1.431} \cdot \\
    & \quad \left. \left( 0.082 \cdot \frac{k_{l}}{L_{c}} \right) \cdot \left( \left( \frac{h_{lv} \cdot q}{g \cdot T_{l} \cdot k_{l}} \right) \cdot \left( \frac{\rho_{v}}{\rho_{l} - \rho_{v}} \right) \right)^{0.7} \right.\cdot \left( \frac{T_{l} \cdot C_{pl} \cdot \sigma \cdot \rho_{l}}{h_{lv}^{2} \cdot \rho_{v}^{2} \cdot \left( \frac{\sigma}{(\rho_{l} - \rho_{v}) \cdot g} \right)^{0.5} } \right)^{0.33} \cdot \left( \text{Pr}_{l} \right)^{-0.45}
    \end{aligned}
\end{equation}

\begin{table}[H]
    \centering
    \caption{Performance comparison of proposed correlation and machine learning model.}
    \label{tab:Comparison_of_Proposed_Correlations_and_Machine_Learning_Model}
    \begin{adjustbox}{max width=\textwidth}
    \renewcommand{\arraystretch}{1.5} 
    \begin{tabular}{lllllllll}
        \hline
        \multirow{2}{*}{\textbf{Surface Type}} & \multirow{2}{*}{\textbf{Models}} & \multirow{2}{*}{\textbf{R\textsuperscript{2}}} & \multirow{2}{*}{\textbf{MAE}} & \multirow{2}{*}{\textbf{RMSE}} & \multicolumn{4}{c}{\textbf{\% deviation of data within}} \\ \cline{6-9}
        & & & & & \textbf{±10\%} & \textbf{±20\%} & \textbf{±30\%} & \textbf{±40\%} \\ \hline
        \multirow{3}{*}{Thin film-coated} & ML model - Catboost & 0.993 & 1.539 & 3.525 & 83.448 & 93.783 & 97.502 & 98.551 \\ 
         & Kruzhilin Correlation & 0.500 & 19.055 & 30.350 & 5.606 & 15.084 & 28.394 & 43.326 \\ 
         & Proposed Correlation & 0.903 & 7.523 & 13.394 & 30.378 & 53.909 & 69.546 & 78.318 \\  \hline
        \multirow{3}{*}{Porous-coated} & ML model - Catboost & 0.989 & 3.984 & 8.538 & 68.961 & 82.399 & 89.187 & 91.735 \\ 
         & Kruzhilin Correlation & 0.009 & 56.812 & 85.564 & 6.107 & 14.022 & 22.443 & 32.905 \\ 
         & Proposed Correlation & 0.812 & 22.979 & 37.304 & 19.701 & 37.651 & 55.115 & 72.773 \\
         \hline
    \end{tabular}
    \end{adjustbox}
\end{table}

\begin{figure}[H]
	\centering
  \includegraphics[width=18cm]{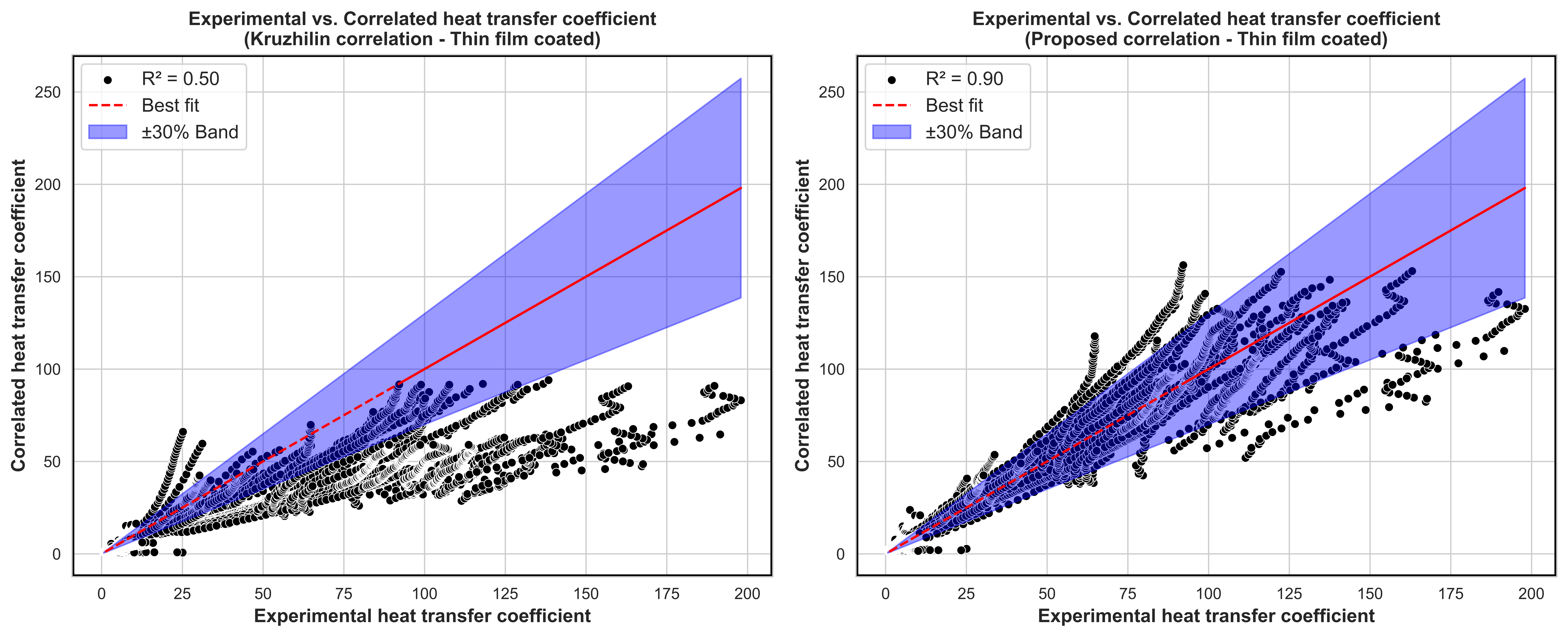}
\caption{Performance comparison of Kruzhilin correlation and proposed correlation for thin film-coated substrates.}
\label{fig: correlation_before_after_modification_surface.}
\end{figure}

\begin{figure}[H]
	\centering
  \includegraphics[width=18cm]{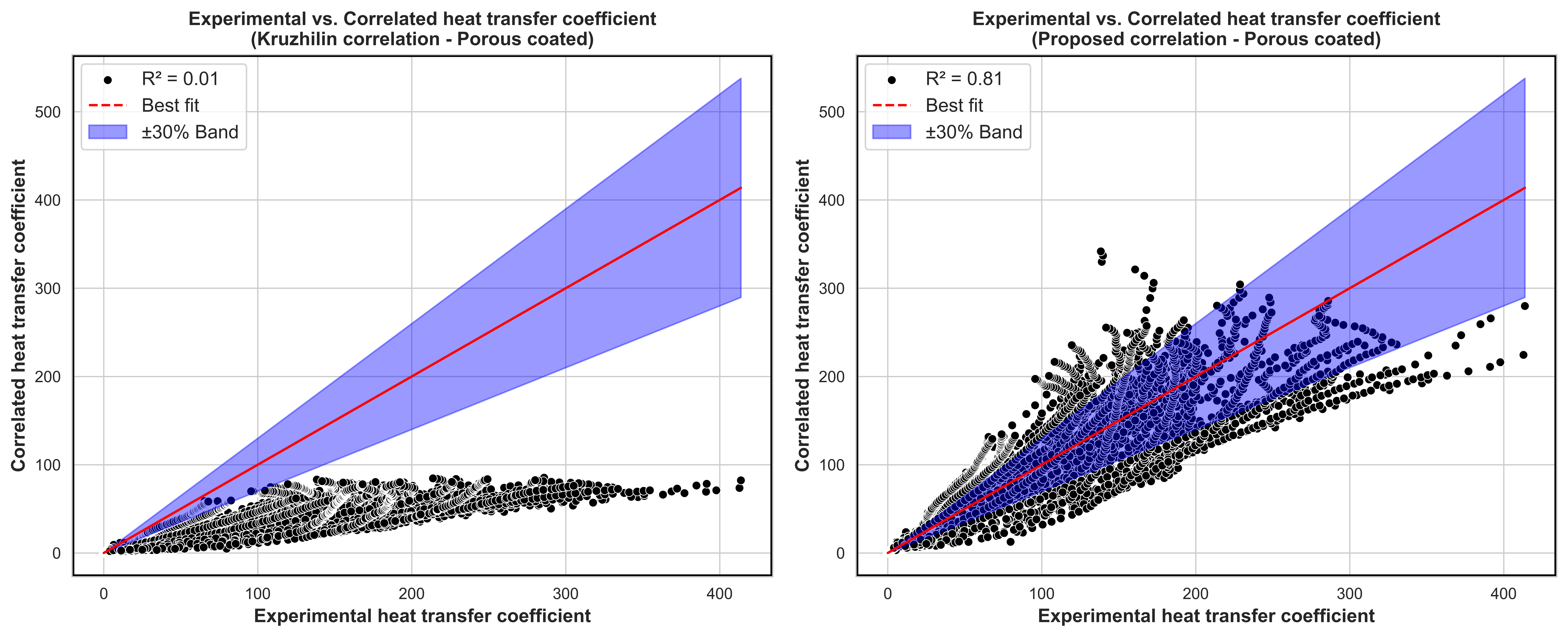}
\caption{Performance comparison of Kruzhilin correlation and proposed correlation for porous-coated substrates.}
\label{fig: correlation_before_after_modification_porous.}
\end{figure}

Table \ref{tab:Comparison_of_Proposed_Correlations_and_Machine_Learning_Model} illustrates the enhanced performance of the proposed correlations (modified Kruzhilin correlations) for thin-film coated and porous-coated substrates. Parity plots shown in Figs. \ref{fig: correlation_before_after_modification_surface.} and \ref{fig: correlation_before_after_modification_porous.} demonstrate the improved performance of the proposed correlations, Eq.(\ref{Proposed Correlation for Surface Coated substrates})and Eq.(\ref{Proposed Correlation for Porous Coated substrates}), relative to the original  Kruzhilin correlation.


\section{Conclusions}\label{sec:conclusions}

A machine learning model to accurately model and predict the nucleate boiling HTC on the thin film-coated and porous-coated substrates has been identified from the present study. Collectively, 10,386 data points have been consolidated from various studies, including diverse operating conditions and surface characteristics of the coated substrates. The most significant parameters influencing the prediction of HTC have also been identified, and the trends in the variation of their influence are in line with the existing studies proving that the ML model correctly identifies the underlying physics of the problem. Furthermore, with the inclusion of major critical parameters from SHAP analysis, new empirical correlations have been proposed that exhibit improved prediction accuracy against the existing correlations. The major conclusions from the study are as follows:

\begin{enumerate}
    \item Among the examined nineteen machine learning algorithms, the CatBoost model consistently exhibited superior performance for thin film-coated and porous-coated substrates with R$^2$ value of around 0.99 across the raw and non-dimensional datasets. The cross-validation results signify that the model doesn’t overfit the collected data. Thus, the proposed model can be effectively used in industrial applications under different parameter ranges for both the thin film-coated and porous-coated surfaces.

    \item Resistance and thickness of the coating, thermal conductivities of coating and substrate, surface roughness, and contact angle clearly highlight the impact of surface features in HTC prediction through SHAP interpretation. On the whole, the parameters - $k_{l}$, $C_{pl}$, $\theta$, $R_{q}$, $R_{co}$, $k_{co}$, $t$, $T_{w}$, $\varepsilon$ and non-dimensional parameters - $C_{pl}/C_{pv}$, $R_{co}$/$R_{sll}$, $k_{w}/k_{co}$, $k_w/k_l$, $\theta/90^\circ$, $R_{q}$/$r_{cav}$, $Pr_l$ consistently influence the HTC prediction for both the thin film-coated and porous-coated surfaces.

    \item For the thin film-coated substrates, for lower surface roughness between 0.01 $\mu$m and 0.1 $\mu$m, the effect of contact angle shows dominance. For roughness between 0.1 $\mu$m to 0.35 $\mu$m, and the contact angle less than 90$^\circ$, HTC increases with an increase in both contact angle and surface roughness. For 0.35 $\mu$m $\leq$ $R_{q}$ $\leq$ 4.2 $\mu$m, large contact angles ($\theta$ $>$ 90$^\circ$) decreases HTC. However, for large roughness values ($R_{q}$ $>$ 10.9 $\mu$m), HTC decreases irrespective of the contact angle.
    
    \item For porous-coated substrates, the effect of surface roughness variation shows unique characteristics. The variation of surface roughness is also dependent on the porosity. For large surface roughness, HTC increases as the pores act as reentrant channels, effectively enhancing liquid replenishment. Large surface roughness with low porosity shows decreased HTC due to low liquid replenishment. However, for porosity greater than 0.71, HTC reduces, regardless of surface roughness levels, as large pores perhaps result in bubble coalescence on the surface.

    \item For substrates with water as the working fluid, the contact angle shows a significant effect on the model prediction. With fluids other than water, the effect of contact angle is negligible as a result of their highly wetting nature caused by lower surface tension. However, the impact of surface roughness is more prominent for substrates with both types of working fluids.

    \item The existing empirical correlations for nucleate boiling heat transfer have been assessed. The proposed empirical correlations with the inclusion of major influencing non-dimensional surface parameters - $R_{co}$/$R_{sll}$, $R_q/r_{cav}$, $k_w/k_l$, $\varepsilon$, and $\theta/90^\circ$ - show better predictive capability than the existing correlations.
\end{enumerate}

\bibliography{Arxiv_Thin_film_and_Porous_coated.bib}
\bibliographystyle{elsarticle-num-names}


\addcontentsline{toc}{chapter}{References}
\clearpage

\end{document}